\pdfoutput = 1
\documentclass[12pt,a4paper]{article}
\usepackage{amsmath}
\usepackage{cite}
\usepackage{hyperref}
\usepackage{graphicx}
\usepackage{booktabs}

\usepackage{xspace}
\newcommand{\Dq}{\ensuremath{D_q}\xspace}
\newcommand{\Dqm}{\ensuremath{D^-_q}\xspace}
\newcommand{\Dqp}{\ensuremath{D^+_q}\xspace}
\newcommand{\Dd}{\ensuremath{D_d}\xspace}
\newcommand{\Ddm}{\ensuremath{D^-_d}\xspace}
\newcommand{\Ddp}{\ensuremath{D^+_d}\xspace}
\newcommand{\Ds}{\ensuremath{D_s}\xspace}
\newcommand{\Dsm}{\ensuremath{D^-_s}\xspace}
\newcommand{\Dsp}{\ensuremath{D^+_s}\xspace}
\newcommand{\Dz}{\ensuremath{D^0}\xspace}
\newcommand{\Dbar}{\kern 0.18em\overline{\kern -0.18em D}\xspace}
\newcommand{\Dzb}{\ensuremath{\Dbar^0}\xspace}
\newcommand{\Du}{\ensuremath{D_u}\xspace}

\newcommand{\B}{\ensuremath{B}\xspace}
\newcommand{\Bbar}{\kern 0.18em\overline{\kern -0.18em B}{}\xspace}
\newcommand{\Bu }{\ensuremath{\B^+}\xspace}

\newcommand{\uBu}{\ensuremath{\B^{\pm}}\xspace}
\newcommand{\Bd }{\ensuremath{\B^0_d}\xspace}
\newcommand{\Bdb}{\ensuremath{\Bbar^0_d}\xspace}
\newcommand{\uBd}{\ensuremath{\B_d^{\vphantom{+}}}\xspace}
\newcommand{\Bs }{\ensuremath{\B^0_s}\xspace}
\newcommand{\Bsb}{\ensuremath{\Bbar^0_s}\xspace}
\newcommand{\uBs}{\ensuremath{\B_s^{\vphantom{+}}}\xspace}
\newcommand{\Bq }{\ensuremath{\B^0_q}\xspace}
\newcommand{\Bqb}{\ensuremath{\Bbar^0_q}\xspace}
\newcommand{\uBq}{\ensuremath{\B_q^{\vphantom{+}}}\xspace}

\newcommand{\BtoDD}{\mbox{\ensuremath{\B\to D\Dbar}}\xspace}

\newcommand{\BqtoDqDq}{\mbox{\ensuremath{\Bq\to \Dqm\Dqp}}\xspace}
\newcommand{\uBqtoDqDq}{\mbox{\ensuremath{\uBq\to \Dqm\Dqp}}\xspace}
\newcommand{\BqTtoDqDq}{\mbox{\ensuremath{\Bq(t)\to \Dqm\Dqp}}\xspace}
\newcommand{\BqbTtoDqDq}{\mbox{\ensuremath{\Bqb(t)\to \Dqm\Dqp}}\xspace}

\newcommand{\BdtoDdDd}{\mbox{\ensuremath{\Bd\to \Ddm\Ddp}}\xspace}
\newcommand{\uBdtoDdDd}{\mbox{\ensuremath{\uBd\to \Ddm\Ddp}}\xspace}
\newcommand{\BstoDsDs}{\mbox{\ensuremath{\Bs\to \Dsm\Dsp}}\xspace}
\newcommand{\uBstoDsDs}{\mbox{\ensuremath{\uBs\to \Dsm\Dsp}}\xspace}

\newcommand{\BdtoDdDs}{\mbox{\ensuremath{\Bd\to\Ddm\Dsp}}\xspace}
\newcommand{\uBdtoDdDs}{\mbox{\ensuremath{\uBd\to\Ddm\Dsp}}\xspace}
\newcommand{\BstoDsDd}{\mbox{\ensuremath{\Bs\to\Dsm\Ddp}}\xspace}
\newcommand{\uBstoDsDd}{\mbox{\ensuremath{\uBs\to\Dsm\Ddp}}\xspace}

\newcommand{\BdtoDsDs}{\mbox{\ensuremath{\Bd\to\Dsm\Dsp}}\xspace}
\newcommand{\uBdtoDsDs}{\mbox{\ensuremath{\uBd\to\Dsm\Dsp}}\xspace}
\newcommand{\BstoDdDd}{\mbox{\ensuremath{\Bs\to\Ddm\Ddp}}\xspace}
\newcommand{\uBstoDdDd}{\mbox{\ensuremath{\uBs\to\Ddm\Ddp}}\xspace}

\newcommand{\ButoDzDd}{\mbox{\ensuremath{\Bu\to\Dzb\Ddp}}\xspace}
\newcommand{\uButoDzDd}{\mbox{\ensuremath{\uBu\to\Du^{\vphantom{\pm}}\Dd^{\pm}}}\xspace}
\newcommand{\ButoDzDs}{\mbox{\ensuremath{\Bu\to\Dzb\Dsp}}\xspace}
\newcommand{\uButoDzDs}{\mbox{\ensuremath{\uBu\to\Du^{\vphantom{\pm}}\Ds^{\pm}}}\xspace}

\newcommand{\BqSL}{\mbox{\ensuremath{\Bq\to\Dqm \ell^+\nu_{\ell}}}\xspace}
\newcommand{\BdSL}{\mbox{\ensuremath{\Bd\to\Ddm \ell^+\nu_{\ell}}}\xspace}
\newcommand{\BsSL}{\mbox{\ensuremath{\Bs\to\Dsm \ell^+\nu_{\ell}}}\xspace}
\newcommand{\BsSLstar}{\mbox{\ensuremath{\Bs\to D_s^{*-} \ell^+\nu_{\ell}}}\xspace}

\begin{document}

\begin{titlepage}

\vspace*{-0.0truecm}

\begin{flushright}
Nikhef-2015-019\\
\end{flushright}

\vspace*{1.3truecm}

\begin{center}
\boldmath
{\Large{\bf Anatomy of \BtoDD Decays}}
\unboldmath
\end{center}

\vspace{0.9truecm}

\begin{center}
{\bf Lennaert Bel\,${}^a$, Kristof De Bruyn\,${}^a$, Robert Fleischer\,${}^{a,b}$, \\
Mick Mulder\,${}^a$ and  Niels Tuning\,${}^a$}

\vspace{0.5truecm}

${}^a${\sl Nikhef, Science Park 105, NL-1098 XG Amsterdam, Netherlands}

${}^b${\sl  Department of Physics and Astronomy, Vrije Universiteit Amsterdam,\\
NL-1081 HV Amsterdam, Netherlands}

\end{center}

\vspace{1.4cm}
\begin{abstract}
\vspace{0.2cm}\noindent
The decays \BdtoDdDd and \BstoDsDs probe the CP-violating
mixing phases $\phi_d$ and $\phi_s$, respectively. The theoretical uncertainty of
the corresponding determinations is limited by contributions from penguin topologies, which can be included
with the help of the $U$-spin symmetry of the strong interaction. We analyse
the currently available data for $B^0_{d,s}\to D_{d,s}^-D_{d,s}^+$ decays
and those with similar dynamics to constrain the involved non-perturbative
parameters. Using further information from semileptonic \BdSL decays, we perform a test of the factorisation
approximation and take non-factorisable $SU(3)$-breaking corrections into
account. 
The branching ratios of the \BdtoDdDd, \BstoDsDd and 
\BstoDsDs, \BdtoDdDs decays show an interesting pattern which 
can be accommodated through significantly enhanced exchange and penguin annihilation topologies.
This feature is also supported by data for the \BstoDdDd channel. Moreover, there are
indications of potentially enhanced penguin contributions in the \BdtoDdDd and 
\BstoDsDs decays, which would make it mandatory to control these effects in the
future measurements of $\phi_d$ and $\phi_s$. We discuss scenarios for high-precision measurements
in the era of Belle II and the LHCb upgrade.
\end{abstract}

\vspace*{0.5truecm}
\vfill
\noindent
May 2015
\vspace*{0.5truecm}

\end{titlepage}

\tableofcontents

\thispagestyle{empty}
\vbox{}
\newpage

\setcounter{page}{1}

\section{Introduction}\label{sec:intro}
CP-violating effects offer important tools to search for new physics (NP) beyond the 
Standard Model (SM). In this endeavour, \Bq--\Bqb mixing ($q=d,s$)
is a key player. This phenomenon does not arise at the tree level in the SM and 
may induce interference effects between oscillation and decay processes, resulting 
in ``mixing-induced" CP violation. The BaBar and Belle experiments at the $e^+e^-$
$B$-factories and the LHCb experiment at the Large Hadron Collider (LHC) have
already performed high precision measurements of the \Bd--\Bdb and 
\Bs--\Bsb mixing phases $\phi_d$ and $\phi_s$, respectively 
\cite{Agashe:2014kda,Amhis:2014hma}. In the era of the
Belle II \cite{Abe:2010gxa} and LHCb upgrade \cite{LHCb-implications},
the experimental analysis will be pushed towards new frontiers of precision. 

In this paper, we present an analysis of the decays \BdtoDdDd and \BstoDsDs
which are related to each other through the $U$-spin symmetry of strong
interactions \cite{RF-psiK,RF-BDD-07}. With the help of this flavour symmetry,
penguin effects can be included in the determination of $\phi_d$ and $\phi_s$
from the mixing-induced CP asymmetries of these decays. The theoretical
precision is limited by non-factorisable $U$-spin-breaking effects.  The impact
of these contributions can be probed in a clean way by comparing branching ratio
measurements of the non-leptonic decays with data from semileptonic \BdSL and
\BsSL decays.  The use of the latter two modes is a new element in this
strategy.  We also explore the role of exchange and penguin annihilation
topologies, which govern the decays \BdtoDsDs and \BstoDdDd \cite{GRP}. 
These modes are also related to each other by the $U$-spin symmetry of strong interactions. 

The analysis of the \BdtoDdDd, \BstoDsDs system complements
the determination of  $\phi_d$ and $\phi_s$ from the decays $\Bd\to J/\psi K_{\rm S}^0$ and
$\Bs\to J/\psi \phi$, respectively, where penguin effects have to be included as well 
\cite{RF-psiK,RF-ang,RF-B99,CPS,FFJM,FFM,GR-09,MJ,LWX,DeBF-pen,FNW}. The dynamics of the 
\BtoDD modes differs from those of the $B\to J/\psi X$ channels. In the latter case, 
the QCD penguins require a colour-singlet exchange and are suppressed by the 
Okubo--Zweig--Iizuka (OZI) rule \cite{Okubo,Zweig,Iizuka}, 
while this feature does not apply to the electroweak (EW) penguins, which are 
colour-allowed and hence contribute significantly
to the decay amplitudes \cite{RF-pen}. On the other hand, the
QCD penguins are not OZI suppressed in the \BdtoDdDd, \BstoDsDs
system, whereas the EW penguins contribute only in colour-suppressed form. The EW penguin sector
offers an interesting avenue for NP to enter weak meson decays \cite{FM-NP,BFRS,BFRS_II,FJPZ}, such as 
in models with extra $Z'$ bosons \cite{Zp, Zp_II}, and would then lead to discrepancies in the determined 
values of $\phi_d$ and $\phi_s$ should the $Z'$ bosons have CP-violating flavour-changing 
couplings to quarks. 
  
The outline of this paper is as follows: in Section~\ref{sec:obs}, we discuss the decay amplitude
structure of the \BdtoDdDd, \BstoDsDs decays and their observables,
while we turn to the picture emerging from the current data in Section~\ref{sec:dat}.
There, we include additional decay modes, which have dynamics similar to the 
\BdtoDdDd, \BstoDsDs system, to address the importance of 
exchange and penguin annihilation topologies, and probe non-factorisable effects by means of
the differential \BdSL rate.
We perform a global analysis of the
penguin parameters $a$ and $\theta$, which allows us to extract $\phi_d$ and $\phi_s$ from measurements
of CP violation in the \BdtoDdDd and \BstoDsDs modes, respectively. 
The current uncertainties of these measurements are unfortunately still very large. 
In Section~\ref{sec:prosp}, we focus on the era of the Belle II and LHCb upgrade, and explore 
the prospects by discussing different scenarios.
Finally, we summarise our conclusions in Section~\ref{sec:concl}. In an appendix, we give a 
summary of the various parameters and observables used in our analysis.

\begin{figure}[tp]
 \centering
\includegraphics[width=0.49\textwidth]{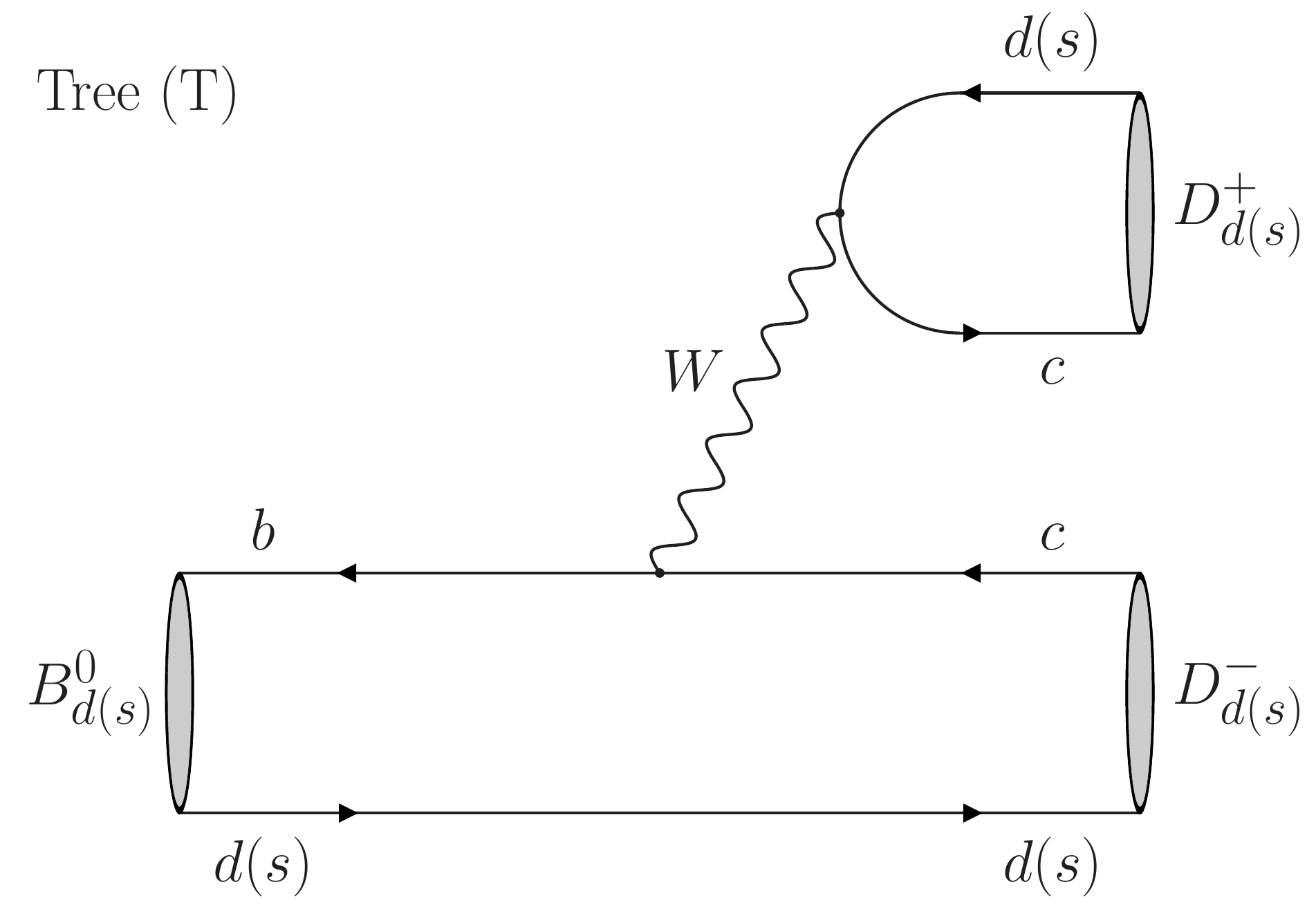}
\includegraphics[width=0.49\textwidth]{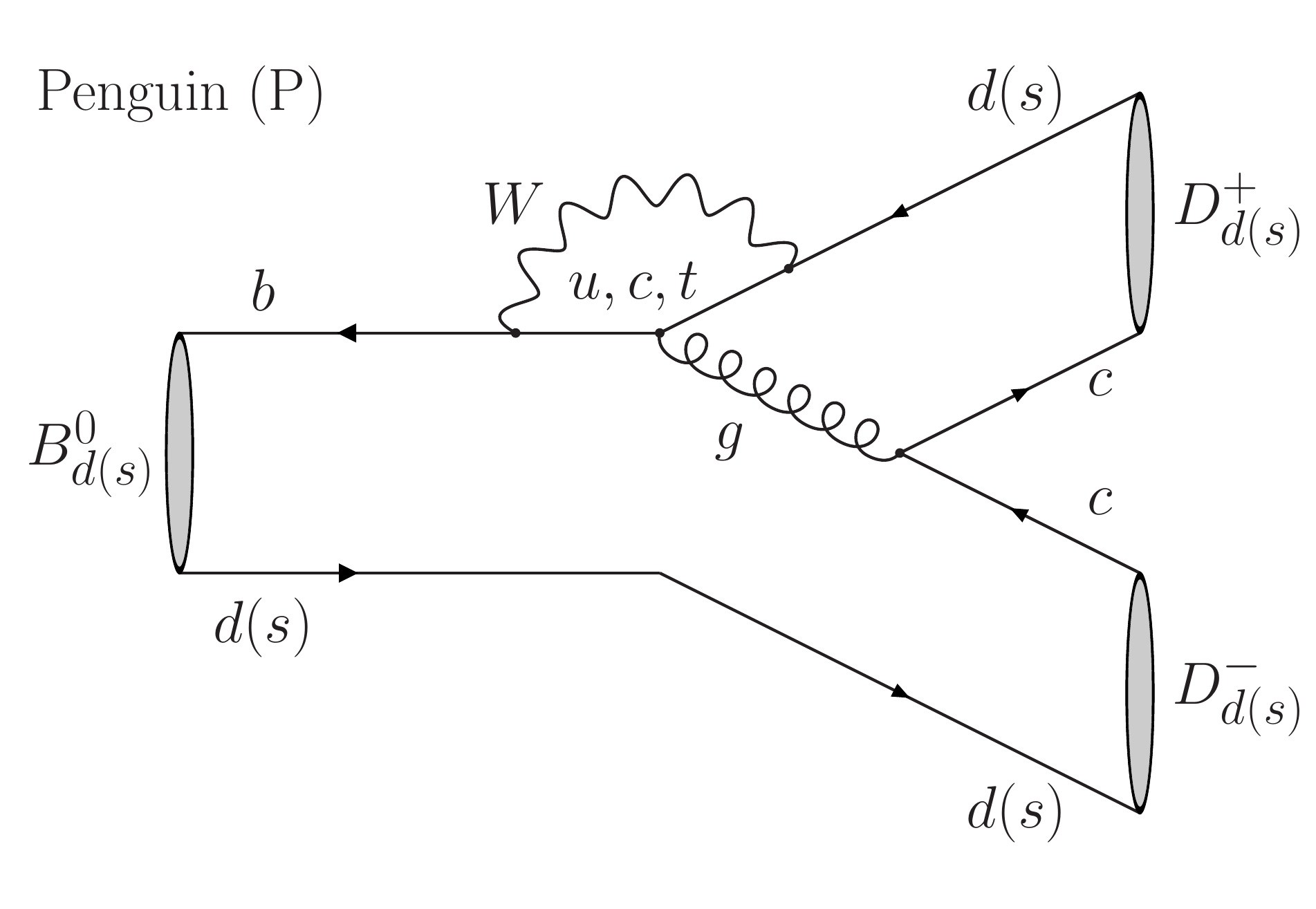}
\includegraphics[width=0.49\textwidth]{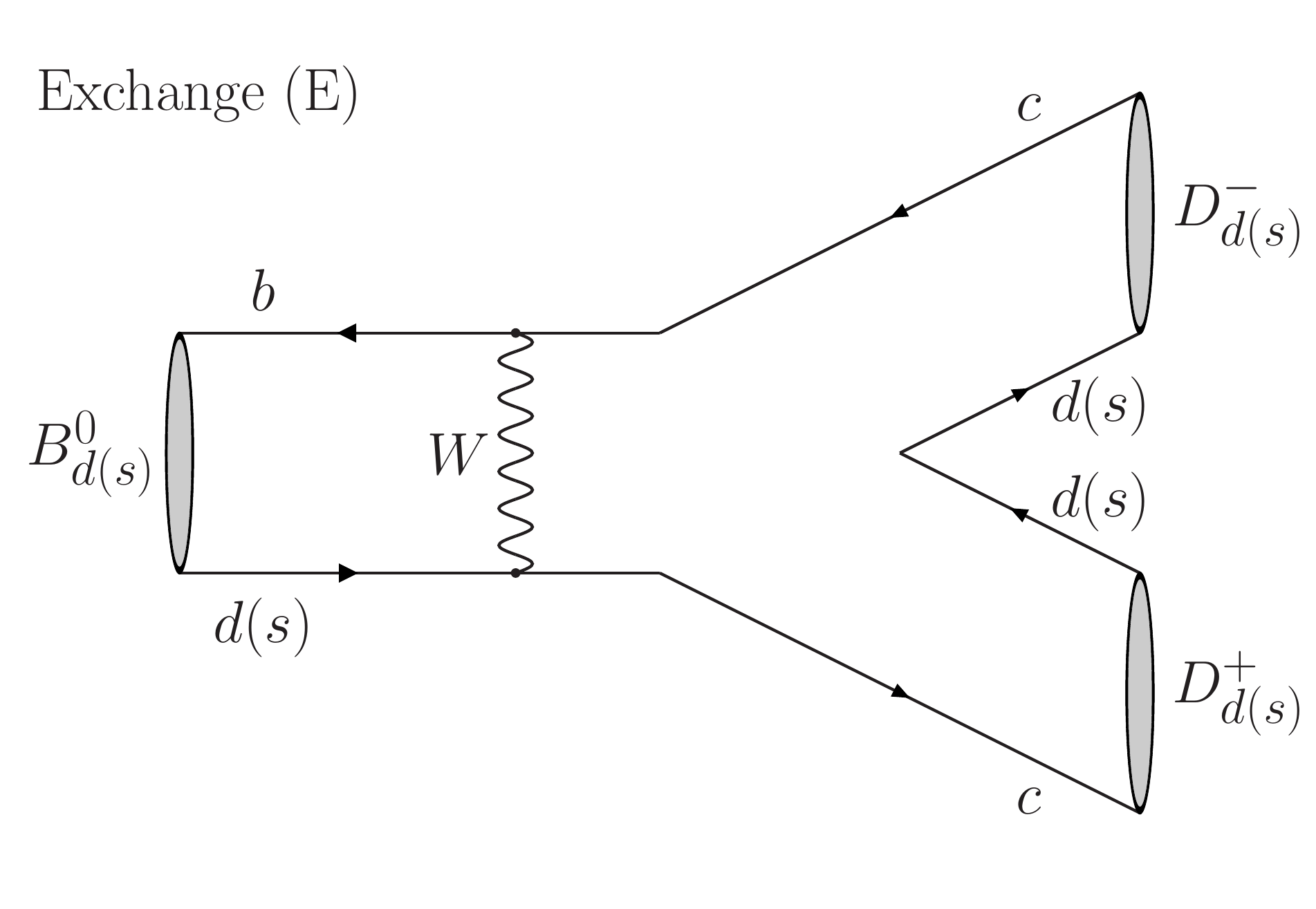}
\includegraphics[width=0.49\textwidth]{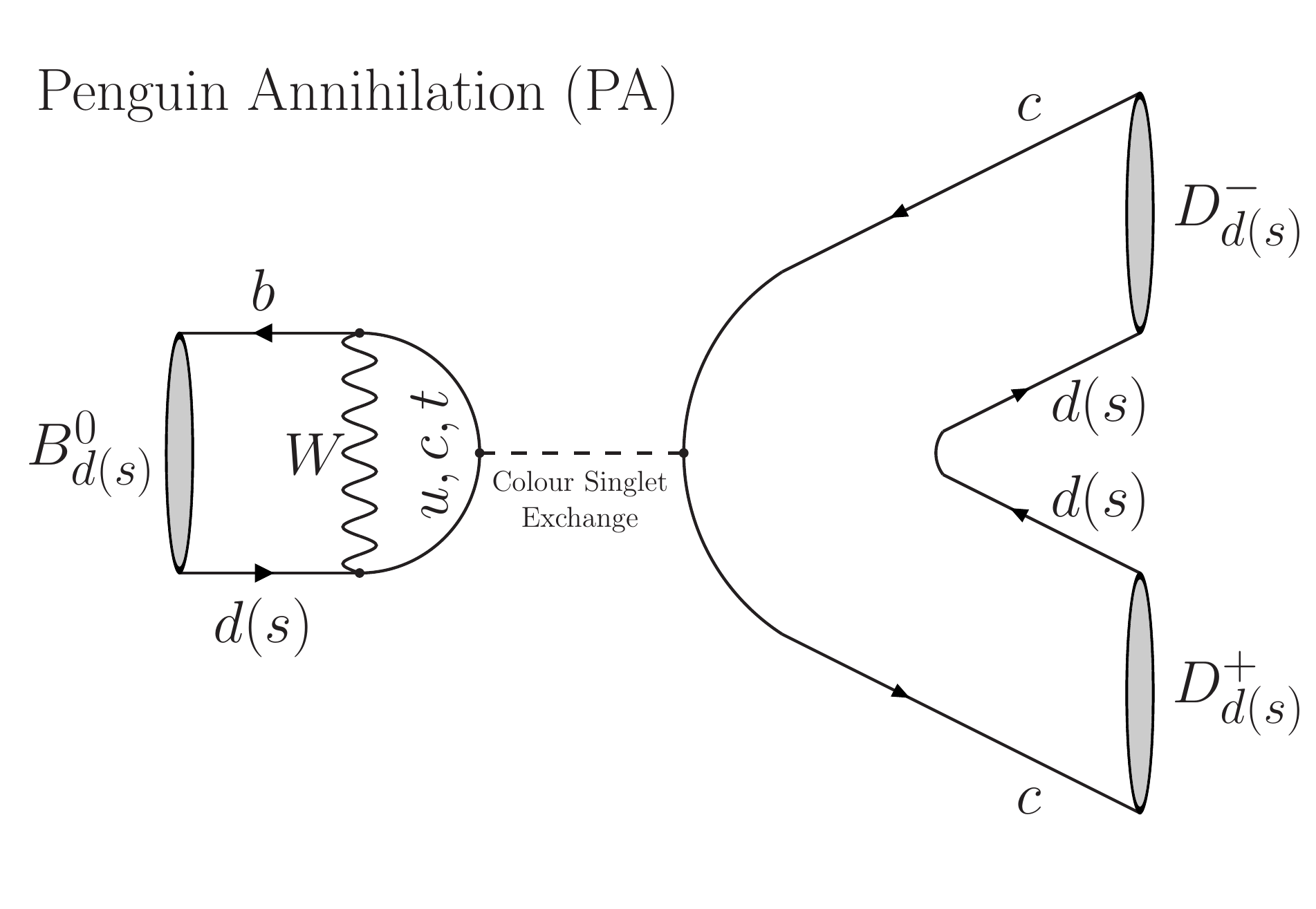}
\caption{Illustration of topologies contributing to the $B^0_{d(s)}\to D_{d(s)}^+D_{d(s)}^-$ 
decays.}\label{fig:TP}
\end{figure}

%
%
%
\section{Decay Amplitudes and Observables}\label{sec:obs}
\subsection{Amplitude Structure}
The \BdtoDdDd mode is caused by $\bar b \to \bar c c \bar d$ quark-level transitions,
and in the SM receives contributions from the decay topologies illustrated in Fig.~\ref{fig:TP}. 
The decay amplitude takes the following form \cite{RF-psiK}:
\begin{equation}\label{BDD-ampl}
A\left(\BdtoDdDd\right) = - \lambda \mathcal{A}
\left[1- a e^{i\theta}e^{i\gamma}\right]\:,
\end{equation}
where $\gamma$ serves as a CP-violating weak phase and is the usual 
angle of the unitarity triangle (UT) of the Cabibbo--Kobayashi--Maskawa (CKM) 
matrix \cite{cab,KM}, while
\begin{equation}\label{A-def}
\mathcal{A} \equiv \lambda^2 A \left[T+E+\left\{P^{(c)}+PA^{(c)}\right\}-\left\{P^{(t)}+
PA^{(t)}\right\}\right]
\end{equation}
and
\begin{equation}\label{a-pen}
ae^{i\theta} \equiv R_b\left[\frac{\left\{P^{(u)}+PA^{(u)}\right\}-\left\{P^{(t)}+PA^{(t)}\right\}}{T+E
+\left\{P^{(c)}+PA^{(c)}\right\}-\left\{P^{(t)}+PA^{(t)}\right\}}\right]
\end{equation}
are CP-conserving hadronic parameters. Here $T$ and $P^{(q)}$ denote the strong amplitudes
of the (colour-allowed) tree and penguin topologies (with internal $q$-quark exchanges), respectively, 
which can be expressed in terms of hadronic matrix elements of the corresponding low-energy 
effective Hamiltonian. We have also included the amplitudes describing exchange $E$
and penguin annihilation $PA^{(q)}$ topologies, which are naively expected to play a minor role
\cite{GHLR}. However, we find that the current data imply sizeable contributions for $E+PA^{(c)}-PA^{(t)}$
with respect to $T+P^{(c)}-P^{(t)}$.  The parameter
\begin{equation}
R_b \equiv \left(1-\frac{\lambda^2}{2}\right)\frac{1}{\lambda}\left|\frac{V_{ub}}{V_{cb}}\right| = 0.390 \pm 0.031
\end{equation}
measures the side of the UT originating from the origin of the complex plane with the angle $\gamma$
between the real axis, while $\lambda \equiv |V_{us}|=0.22548 \pm 0.00068$ is the Wolfenstein
parameter of the CKM matrix \cite{wolf}, and $A\equiv |V_{cb}|/\lambda^2= 0.806 \pm 0.017$; the 
numerical values refer to the analysis of Ref.~\cite{Charles:2015gya}.

The $U$-spin partner \BstoDsDs of the \BdtoDdDd  channel
originates from the $\bar b \to \bar c c \bar s$ quark-level processes. Its SM transition 
amplitude can be written as
\begin{equation}\label{BsDsDs-ampl}
A\left(\BstoDsDs\right)= \left(1-\frac{\lambda^2}{2}\right)\mathcal{A}'
\left[1+\epsilon a'e^{i\theta'}e^{i\gamma}\right],
\end{equation}
where the hadronic parameters $\mathcal{A}'$ and $a'e^{i\theta'}$ are given by expressions
which are analogous to those in Eqs.~(\ref{A-def}) and (\ref{a-pen}), respectively. The key difference
in the structure of the \BstoDsDs decay amplitude with respect to Eq.~(\ref{BDD-ampl})
is the suppression of the $a'e^{i\theta'}e^{i\gamma}$ term by the tiny CKM parameter
\begin{equation}
\epsilon \equiv \frac{\lambda^2}{1-\lambda^2} = 0.0536 \pm 0.0003\:.
\end{equation}
Moreover, the overall factor of $\lambda$ is absent, thereby enhancing the decay rate
with respect to \BdtoDdDd. 
Therefore, sizeable penguin effects in \BdtoDdDd could be probed and subsequently used
to estimate the penguin effects in \BstoDsDs, applying $U$-spin symmetry.

The $U$-spin symmetry of strong interactions implies the following relations between the
hadronic parameters:
\begin{equation}\label{a-theta-rel}
ae^{i\theta} = a'e^{i\theta'}\:,
\end{equation}
\begin{equation}\label{A-rel}
\mathcal{A}= \mathcal{A}'\:.
\end{equation}
It is important to emphasise that hadronic form factors and decay constants cancel within factorisation
in $ae^{i\theta}$ and  $a'e^{i\theta'}$, since these quantities are defined as ratios of hadronic amplitudes,
as can be seen in Eq.~(\ref{a-pen}). Consequently, factorisable $U$-spin-breaking corrections to the
relation in Eq.~(\ref{a-theta-rel}) vanish \cite{RF-psiK,RF-BDD-07}. On the other hand, the $U$-spin
relation in Eq.~(\ref{A-rel}) is affected by $SU(3)$-breaking effects\footnote{%
$SU(3)$ symmetry refers to the symmetry group interchanging $u$, $d$ and $s$ quarks. The
isospin, $U$-spin and $V$-spin subgroups refer to interchanging $u\leftrightarrow d$, $d\leftrightarrow s$,
and $s\leftrightarrow u$, respectively. Throughout the paper the mention of $SU(3)$ refers to the
$U$-spin subgroup, unless specified otherwise.}
in $\uBq\to \Dq$ form factors and
$\Dq$ decay constants ($q=d,s$). We discuss these effects in more detail later.

\subsection{CP-Violating Asymmetries}
Due to \Bq--\Bqb oscillations ($q=d,s$), an initially present \Bq-meson state evolves
in time into a linear combination of \Bq and \Bqb states. CP violation in
the \BqtoDqDq decays, which are characterised by CP-even final states,
is probed through the following time-dependent rate asymmetries \cite{RF-rev}:
\begin{align}
&\frac{|A(\BqTtoDqDq)|^2-
|A(\BqbTtoDqDq)|^2}{|A(\BqTtoDqDq)|^2+
|A(\BqbTtoDqDq)|^2}\notag\\
&\qquad=\frac{{\cal }{\cal A}_{\rm CP}^{\rm dir}(\uBqtoDqDq)
\cos(\Delta M_qt)+{\cal A}_{\rm CP}^{\rm mix}(\uBqtoDqDq)
\sin(\Delta M_qt)}{\cosh(\Delta\Gamma_qt/2)+
{\cal A}_{\Delta\Gamma}(\uBqtoDqDq)\sinh(\Delta\Gamma_qt/2)}\:,\label{ACP-def}
\end{align} 
where \mbox{$\Delta M_q\equiv M^{(q)}_{\rm H}-M^{(q)}_{\rm L}$} and 
\mbox{$\Delta\Gamma_q\equiv \Gamma_{\rm L}^{(q)}-\Gamma_{\rm H}^{(q)}$} denote
the mass and decay width difference between the two $\uBq$ mass eigenstates,  respectively.
The three CP observables are given by%
\footnote{Whenever information from both 
$\Bq\to f$ and $\Bqb\to f$ decays is needed to determine an 
observable, as is the case for CP asymmetries or untagged branching ratios, 
we use the notation \uBd and \uBs.}
\begin{align}
 {\cal A}_{\rm CP}^{\rm dir}(\uBqtoDqDq)&=\frac{2 b_q \sin\rho_q\sin\gamma}{1-
 2b_q\cos\rho_q\cos\gamma+b_q^2}\:,\label{AD}\\
 {\cal A}_{\rm CP}^{\rm mix}(\uBqtoDqDq)&=\phantom{-}\eta_q\left[
 \frac{\sin\phi_q-2 b_q \cos\rho_q\sin(\phi_q+\gamma)+b_q^2\sin(\phi_q+2\gamma)}{1-
 2b_q\cos\rho_q\cos\gamma+b_q^2}\right]\:,\label{AM}\\
\mathcal{A}_{\Delta\Gamma}(\uBqtoDqDq)&=-\eta_q\left[
 \frac{\cos\phi_q-2 b_q \cos\rho_q\cos(\phi_q+\gamma)+b_q^2\cos(\phi_q+2\gamma)}{1-
 2b_q\cos\rho_f\cos\gamma+b_q^2}\right]\:,\label{ADG}
\end{align}
where we have to make the following replacements for the decays at hand \cite{RF-psiK}:
\begin{equation}\label{replacement}
\BdtoDdDd: \, b_d e^{i\rho_d} \, = \, a e^{i\theta}\:, \qquad
\BstoDsDs: \,  b_s e^{i\rho_s} \, = \, - \epsilon a' e^{i\theta'}\:.
\end{equation}
The parameter $\eta_q$ denotes the CP eigenvalue of the final state and is given by $+1$. 
While the direct CP asymmetries ${\cal A}_{\rm CP}^{\rm dir}(\uBqtoDqDq)$ are 
caused by interference between tree and penguin contributions, the mixing-induced CP asymmetries
${\cal A}_{\rm CP}^{\rm dir}(\uBqtoDqDq)$ originate from interference between 
\Bq--\Bqb mixing and decay processes, and depend on the mixing phases 
$\phi_d$ and $\phi_s$. These quantities take the general forms
\begin{equation}\label{mix-phases}
\phi_d=2\beta + \phi_d^{\rm NP}\:, \qquad \phi_s= -2\beta_s + \phi_s^{\rm NP}\:,
\end{equation} 
where $\beta$ is the usual angle of the UT. The SM value of $\phi_s$, which 
is given by $-2\beta_s = -2\lambda^2\eta$ and hence doubly Cabibbo suppressed, can be determined
with high precision from SM fits of the UT   \cite{Charles:2015gya}:
\begin{equation}\label{phis-SM}
\phi_s^{\rm SM}=-(2.092^{+0.075}_{-0.069})^{\circ}\:.
\end{equation}
The CP-violating phases $\phi_q^{\rm NP}$ vanish in the SM and allow us to take NP 
contributions to \Bq--\Bqb mixing into account. 

It is useful to introduce ``effective mixing phases" 
\begin{equation}
\phi_{q,\Dqm\Dqp}^{\rm eff} \equiv \phi_q+ \Delta\phi_q^{\Dqm\Dqp}
\end{equation}
through the following expression \cite{FFM,DeBF-pen}:
\begin{equation}\label{phiq-eff-def}
 \frac{ {\cal A}_{\rm CP}^{\rm mix}(\uBqtoDqDq)}{\sqrt{1- 
 \left({\cal A}_{\rm CP}^{\rm dir}(\uBqtoDqDq)\right)^2}}=
 \sin(\phi_{q,\Dqm\Dqp}^{\rm eff})\:,
 \end{equation}
where the hadronic ``penguin phase shifts" $\Delta\phi_q^{\Dqm\Dqp}$ are characterised by 
\begin{equation}\label{tan-phis}
\tan  \Delta\phi_d^{\Ddm\Ddp}  =  \frac{-2 a\cos\theta\sin\gamma+a^2\sin2\gamma}{1-
2a\cos\theta\cos\gamma+a^2\cos2\gamma} =
-2a\cos\theta\sin\gamma-a^2\cos2\theta\sin 2\gamma +{\cal O}(a^3)
\end{equation}
\begin{equation}\label{tan-phid}
\tan  \Delta\phi_s^{\Dsm\Dsp}  = \frac{2 \epsilon a'\cos\theta'\sin\gamma+\epsilon^2 a'^2
\sin2\gamma}{1+2\epsilon a'\cos\theta'\cos\gamma+\epsilon^2 a'^2\cos2\gamma} 
=2\epsilon a'\cos\theta'\sin\gamma +{\cal O}(\epsilon^2 a'^2)\:.
\end{equation}

\begin{figure}[tp]
 \centering
\includegraphics[width=0.49\textwidth]{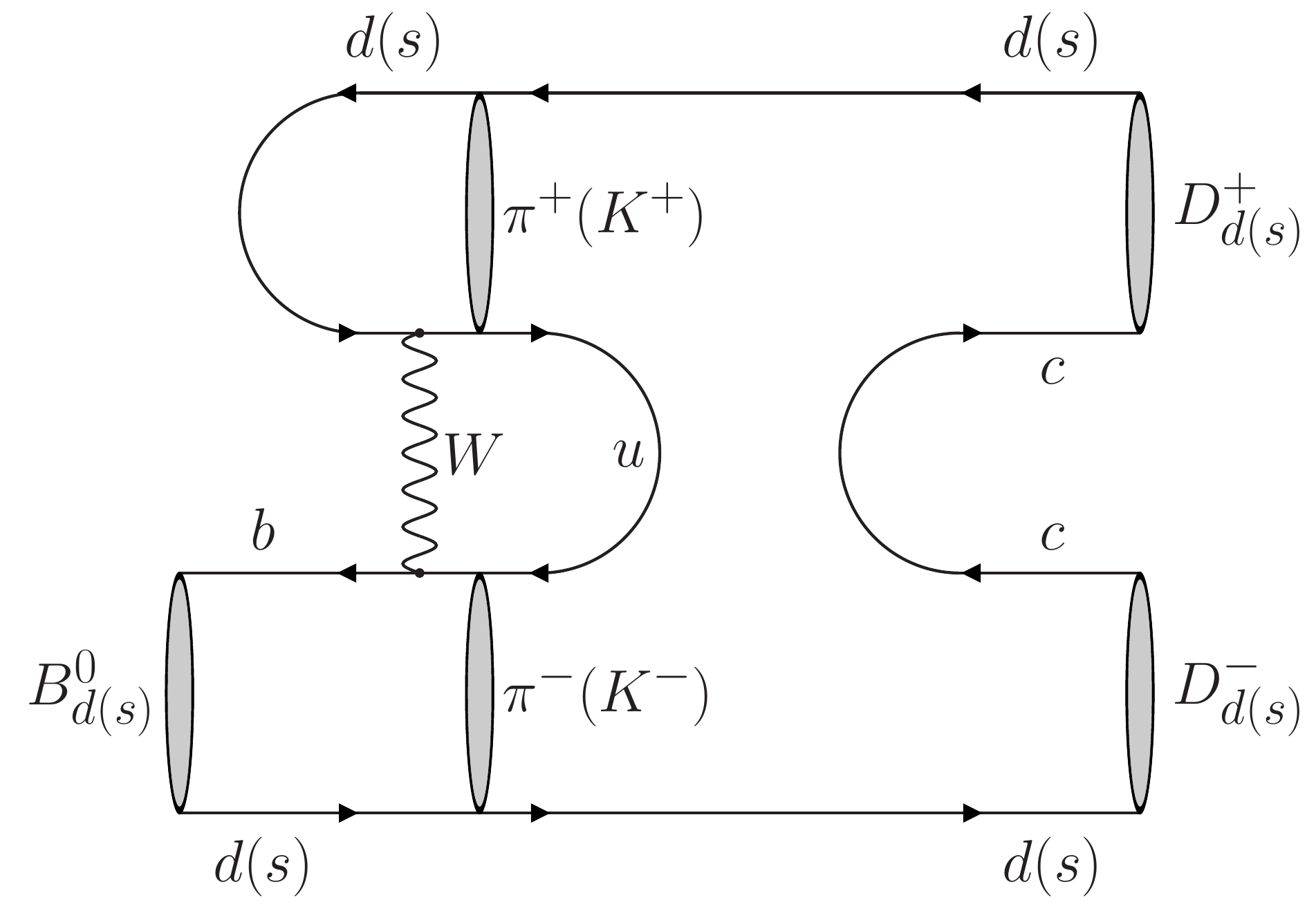}
\includegraphics[width=0.49\textwidth]{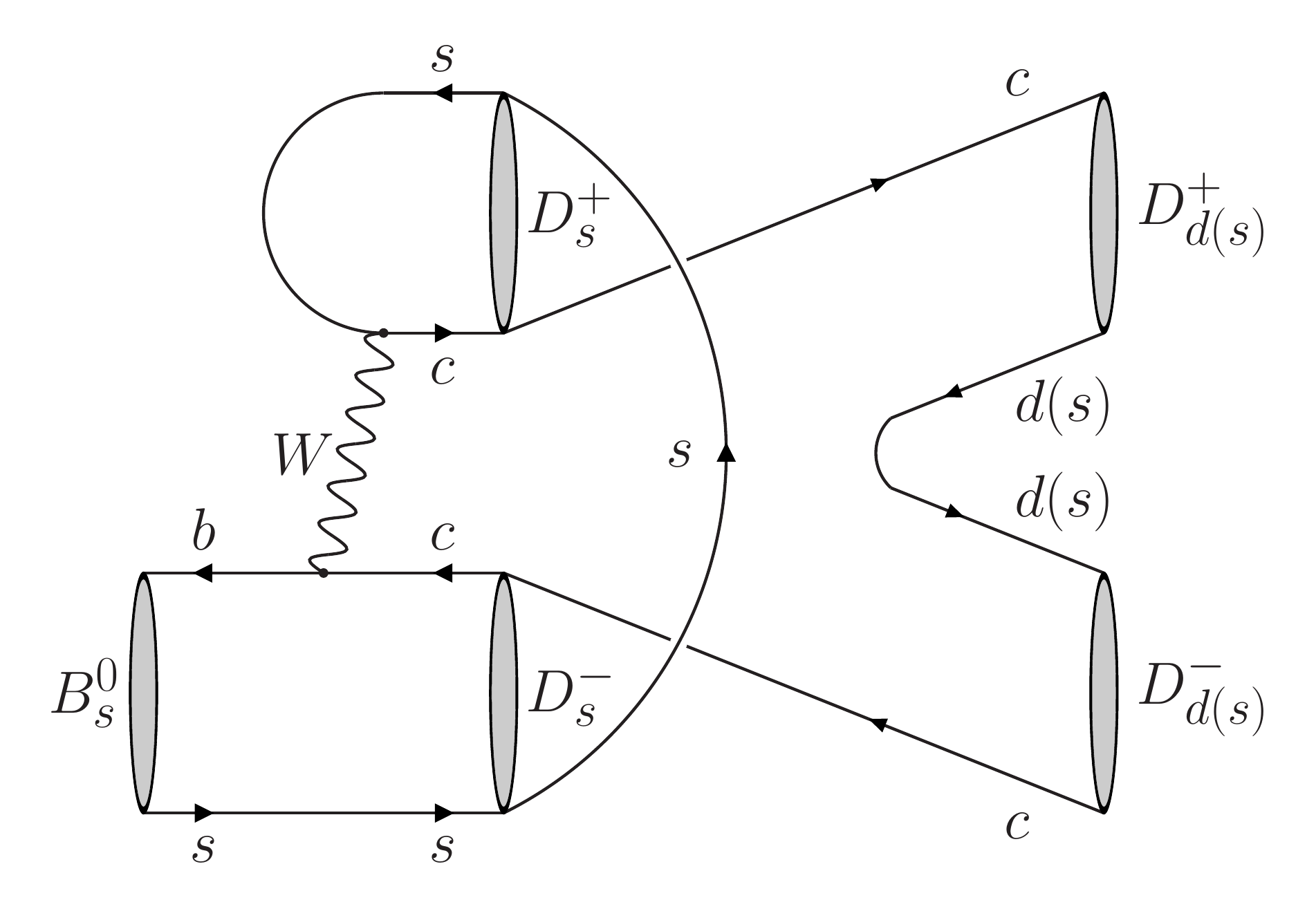}
\caption{Illustrations of the rescattering processes in Eqs.~(\ref{rescat-u}) and (\ref{rescat-c}).}\label{fig:rescat}
\end{figure}

In the limit $a=a'=0$, we simply have
\begin{equation}\label{AM-ideal-Bd}
{\cal A}_{\rm CP}^{\rm dir}(\uBdtoDdDd) |_{a=0} = 0\:, \qquad
 {\cal A}_{\rm CP}^{\rm mix}(\uBdtoDdDd) |_{a=0} = \sin \phi_d\:,
\end{equation}
\begin{equation}\label{AM-ideal-Bs}
{\cal A}_{\rm CP}^{\rm dir}(\uBstoDsDs) |_{a'=0} = 0\:, \qquad
 {\cal A}_{\rm CP}^{\rm mix}(\uBstoDsDs) |_{a'=0} = \sin \phi_s\:.
\end{equation}
The penguin parameter $ae^{i\theta}$ cannot be calculated reliably within QCD. Since this 
quantity is governed by the ratio of a penguin amplitude to a colour-allowed tree amplitude,
it is plausible to expect $a\sim 0.1\mbox{--}0.2$.  Applying the Bander--Silverman--Soni
mechanism \cite{BSS} and the formalism developed in Refs.~\cite{RF-NLO, RF-NLO-II} yields the following 
estimate \cite{RF-BDD-07}:
\begin{equation}\label{BSS-estimate}
\left.ae^{i\theta}\right|_{\rm QCD}^{\rm BSS} \sim 0.08 \times e^{i \, 205^\circ}\:.
\end{equation}
In the corresponding calculation,  form factors and decay constants cancel as the
parameter $ae^{i\theta}$ is actually defined as a ratio of hadronic amplitudes, as we 
emphasised after Eq.~(\ref{A-rel}). However, incalculable long-distance contributions, 
such as processes of the kind
\begin{equation}\label{rescat-u}
\Bd\to[\pi^-\pi^+, \rho^-\pi^+, \ldots]\to \Ddm\Ddp\:, \qquad \Bs\to[K^-K^+, K^{*-}K^+, \ldots]\to \Dsm\Dsp\:,
\end{equation}
and
\begin{equation}\label{rescat-c}
\Bd\to[\Ddm\Ddp, D_d^{*-}\Ddp, \ldots]\to \Ddm\Ddp\:, \qquad B^0_s\to[\Dsm\Dsp,D_s^{*-}\Dsp, \ldots]
\to \Dsm\Dsp\:,
\end{equation}
which can be considered as long-distance penguins with up- and charm-quark exchanges
\cite{BFM}, respectively, as illustrated in Fig.~\ref{fig:rescat}, may have an impact on $ae^{i\theta}$. 

In this paper, we discuss strategies to control these effects by means of experimental data. 
In the \BdtoDdDd case (Eq.~(\ref{AM-ideal-Bd})), the penguin effects have to be taken into account for 
the determination of $\phi_d$. In the \BstoDsDs case (Eq.~(\ref{AM-ideal-Bs})), the parameter $a'$ is 
associated with the tiny $\epsilon$ factor and is hence doubly Cabibbo-suppressed.
However, in view of the experimental precision in the LHCb upgrade era, also these effects have 
to be controlled.

\subsection{Untagged Decay Rate Information}

For the analysis of the experimental data later on, it is useful to introduce another observable, 
containing the untagged rate information.
It is defined as \cite{RF-psiK,RF-BDD-07}:
\begin{equation}\label{Eq:Hobs_Def}
H  \equiv  \frac{1}{\epsilon} \left|\frac{\mathcal{A}'}{\mathcal{A}}\right|^2
\left[\frac{m_{\uBd}}{m_{\uBs}} 
\frac{\Phi(m_{\Ds}/m_{\uBs},m_{\Ds}/m_{\uBs})}{\Phi(m_{\Dd}/m_{\uBd},m_{\Dd}/m_{\uBd})}  
 \frac{\tau_{\uBs}}{\tau_{\uBd}} \right]
\frac{\mathcal{B}\left(\uBdtoDdDd\right)}
{\mathcal{B}\left(\uBstoDsDs\right)_{\text{theo}}}\:,
\end{equation}
where
\begin{equation}
\Phi(x,y)=\sqrt{\left[1-(x+y)^2\right]\left[1-(x-y)^2\right]}
\end{equation}
is the well-known $B\to PP$ phase-space function. 
Due to the sizeable lifetime difference in the $B_s$-meson system, 
$y_s \equiv \Delta\Gamma_s/2\Gamma_s = 0.0608 \pm 0.0045$ \cite{Amhis:2014hma}, a difference arises between the 
``theoretical" branching ratio defined through the untagged decay rate at time $t=0$ 
\cite{RF-psiK} and the ``experimental" branching ratio which is extracted from the 
time-integrated untagged rate \cite{DFN}.
They can be related as \cite{BR-paper}
\begin{align}
\mathcal{B}\left(\uBstoDsDs\right)_{\text{theo}} 
& = \left[\frac{1-y_s^2}{1+\mathcal{A}_{\Delta\Gamma}(\uBstoDsDs) \, y_s }\right]
      \mathcal{B}\left(\uBstoDsDs\right) \label{BR-conv}  \\
& = (1.09 \pm 0.03) \times \mathcal{B}\left(\uBstoDsDs\right) \:, \nonumber
\end{align}
where the numerical estimate uses
\begin{equation}\label{ADG-exp}
\mathcal{A}_{\Delta\Gamma}(\uBstoDsDs) = -1.41 \pm 0.30\:,
\end{equation}
extracted from the measurement of the effective \BstoDsDs lifetime \cite{Aaij:2013bvd, RF-BDD-07}.

The observable $H$ takes the following
form in terms of the penguin parameters \cite{RF-psiK}:
\begin{equation}\label{H-exprs}
\boxed{H =  \frac{1-2\:a\cos\theta\cos\gamma+a^2}{1+2\epsilon a'\cos\theta'\cos\gamma+\epsilon^2 
a^{\prime2}}\:.}
\end{equation}
Moreover, the $U$-spin relation in Eq.~(\ref{a-theta-rel}) implies
\begin{equation}
H=-\frac{1}{\epsilon}\left[\frac{{\cal A}_{\rm CP}^{\rm dir}(\uBstoDsDs)}{{\cal A}_{\rm CP}^{\rm dir}(\uBdtoDdDd)}\right]\:.
\end{equation}
Using Eq.~(\ref{a-theta-rel}) and keeping $a$ and $\theta$ as free parameters, the
expression in Eq.~(\ref{H-exprs}) results in the following lower bound 
\cite{Fleischer:1997um, Fleischer:2004rn}:
\begin{align}\label{H-bound}
H & \geq \frac{1+\epsilon^2+2 \epsilon \cos^2\gamma - (1+\epsilon) \sqrt{1-2 \epsilon +\epsilon^2+4 
\epsilon  \cos^2\gamma}}{2\epsilon ^2\left(1-\cos^2\gamma\right)} \nonumber \\
& = \sin^2\gamma -2\epsilon\sin^2\gamma \cos^2\gamma +  \mathcal{O}(\epsilon^2)
\quad \stackrel{\gamma = 73.2^{\circ}}{\xrightarrow{\hspace{12mm}}}  \quad 0.908\:.
\end{align}
Moreover, $H$ allows us to put a lower bound on the penguin parameter $a$:
\begin{equation}\label{a-bound}
a\geq -\left(\frac{1+\epsilon H}{1-\epsilon^2 H}\right)\cos\gamma+\sqrt{\left[\left(\frac{1+\epsilon H}{1-
\epsilon^2H} \right)\cos\gamma \right]^2-\left(\frac{1-H}{1-\epsilon^2H} \right)}\:.
\end{equation}
The signs have been chosen in such a way that this expression applies to the current experimental
situation discussed in Section~\ref{ssec:global}.

\subsection{Information from Semileptonic Decays}\label{ssec:semilept}
%
%
%

\begin{figure}[tp]
\centering
\includegraphics{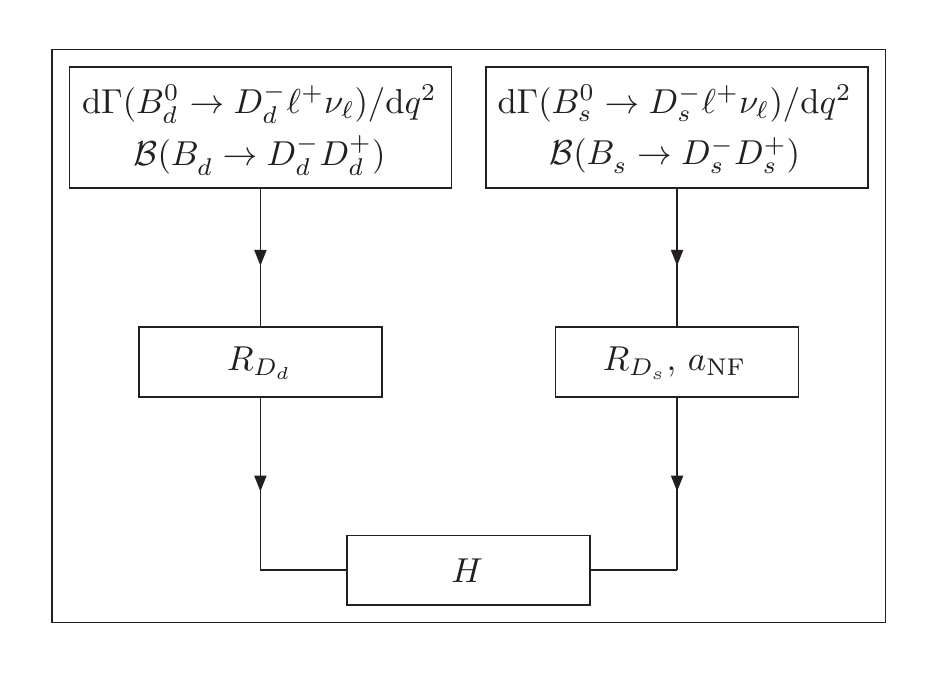}
\caption{Flow chart illustrating the new strategy to determine $H$ using data from semileptonic \BqSL decays.}
\label{Fig:Flow_Hobs_Semi}
\end{figure}

The experimental determination of $H$ through Eq.~\eqref{Eq:Hobs_Def} requires information on the amplitude
ratio $|\mathcal{A}'/\mathcal{A}|$, which is affected by $U$-spin-breaking corrections to the
relation in Eq.~(\ref{A-rel}). 
To avoid the limitations this brings, we propose a new method to determine $H$
using data from semileptonic \BqSL decays, which is illustrated by the flow chart in Fig.~\ref{Fig:Flow_Hobs_Semi}.
To this end, we introduce the ratio
\begin{align}
R_{\Dq} & \equiv \frac{\Gamma(\uBqtoDqDq)_{\text{theo}}}{[{\rm d} 
\Gamma(\BqSL)/{\rm d} q^2]|_{q^2=m_{\Dq}^2}}\label{R-def}\\
&=6\pi^2 |V_{cq}|^2  f_{\Dq}^2  X_{\Dq} \left|a_{\rm NF}^{(q)}\right|^2 \left[1-2 \,  b_q \cos \rho_q 
\cos\gamma+b_q^2 \right]\:,
\end{align}
where the parameters $b_q$ and $\rho_q$ are given in Eq.~(\ref{replacement});
$V_{cq}$ is the relevant CKM matrix element, $f_{\Dq}$ denotes the 
$\Dq$-meson decay constant defined through
\begin{equation}
\langle \Dqm(p) | \bar q \gamma_\mu \gamma_5 c|0  \rangle = -i f_{\Dq} p_\mu\:,
\end{equation}
and the factor $X_{\Dq}$ is given by 
\begin{equation}\label{Eq:XDq}
X_{\Dq}= \left[\frac{(m_{\uBq}^2-m_{\Dq}^2)^2}{m_{\uBq}^2(m_{\uBq}^2-4 \, 
m_{\Dq}^2)}\right] \left[\frac{F_0^{\uBq\Dq}(m_{\Dq}^2)}{F_1^{\uBq\Dq}(m_{\Dq}^2)} \right]^2\:,
\end{equation}
where the form factors are defined through
\begin{align}
&\langle \Dqp(k)|\bar{c}\gamma_\mu b|\Bqb(p)
\rangle\notag\\
&\quad=F_1^{\uBq\Dq}(q^2)\left[(p+k)_\mu -
\left(\frac{m_{\uBq}^2-m_{\Dq}^2}{q^2}\right)q_\mu\right]
+F_0^{\uBq\Dq}(q^2)\left(\frac{m_{\uBq}^2-m_{\Dq}^2}{q^2}\right)q_\mu\:,
\end{align}
with $q\equiv p-k$. Ratios between non-leptonic decay rates and differential semileptonic 
rates as in Eq.~(\ref{R-def}) are well-known probes for testing factorisation 
\cite{bjorken,BS,rosner,NS,BBNS,FST-fact}. 

The parameter $a_{\rm NF}^{(q)}$ measures non-factorisable effects in the
amplitudes defined through Eq.~(\ref{A-def}), which we may write as
\begin{equation}
{\cal A}_{q}\equiv \lambda^2A\,T^{(q)}_{\rm fact}a^{(q)}_{\rm NF}\:,
\end{equation}
where
\begin{equation}\label{T-fact}
T^{(q)}_{\rm fact}=-i \frac{G_{\rm F}}{\sqrt{2}} f_{\Dq} \left(m_{\uBq}^2-m_{\Dq}^2\right)
F_0^{\uBq\Dq}(m_{\Dq}^2)
\end{equation}
is the amplitude of the colour-allowed tree topology in factorisation, with $G_{\rm F}$ 
denoting Fermi’s constant. In naive factorisation, $a_{\rm NF}^{(q)}=a_1$, 
where $a_1$ represents the appropriate combination of Wilson coefficient 
functions of the current-current operators\footnote{For a detailed discussion, see Ref.~\cite{Buras}.}.
We have ${\cal A}\equiv{\cal A}_{d}$ and ${\cal A}'\equiv{\cal A}_{s}$, and
shall suppress the label $q$ in the following discussion for simplicity. Introducing the abbreviations  
\begin{equation}\label{abbrev}
P^{(ct)}\equiv P^{(c)}-P^{(t)}\:, \qquad PA^{(ct)}\equiv PA^{(c)}-PA^{(t)}\:,
\end{equation}
we obtain
\begin{equation}\label{aNF-def-0}
a_{\rm NF}=\left(1+r_P\right)\left(1+x\right)a^T_{\rm NF}\:,
\end{equation}
where 
\begin{equation}\label{rP-def}
r_P\equiv\frac{P^{(ct)}}{T}
\end{equation}
measures the importance of the penguin topologies with respect to the colour-allowed tree amplitude.
On the other hand, 
\begin{equation}\label{x-def}
x \equiv  |x|e^{i\sigma}\equiv\frac{E+PA^{(ct)}}{T+ P^{(ct)}}
\end{equation}
probes the importance of the exchange and penguin annihilation topologies. We will return to 
$x$ in Subsection~\ref{ssec:EPA}, where we determine $|x|$ and the CP-conserving strong
phase $\sigma$ from experimental data. 
The parameter $a^T_{\rm NF}$ describes the non-factorisable corrections to the ``tree'' diagram
(Eq.~(\ref{T-fact})), i.e.\ we have 
\begin{equation}\label{aNF-T}
T=T_{\rm fact} \, a^{T}_{\rm NF} \equiv T_{\rm fact} \, [1+\Delta^{T}_{\rm NF}]
\end{equation}
with $\Delta^{T}_{\rm NF}=0$ for exact factorisation.

Finally, the observable $H$ can be expressed as follows:
\begin{equation}\label{H-R}
\boxed{H=\left|\frac{V_{cs}}{V_{cd}}  \right|^2\left[\frac{R_{\Dd}}{R_{\Ds}}\right] 
\left[\frac{f_{\Ds}}{f_{\Dd}} \right]^2\left[\frac{X_{\Ds}}{X_{\Dd}}\right] 
\left|\frac{a_{\rm NF}^{(s)}}{a_{\rm NF}^{(d)}}\right|^2\:.}
\end{equation}
In comparison with Eq.~(\ref{Eq:Hobs_Def}), the advantage is that the theoretical
precision is now only limited by {\it non-factorisable} $U$-spin-breaking effects. Moreover,
as the $R_{\Dq}$ are ratios of $\uBq$ rates, the dependence on the ratio of fragmentation
functions $f_s/f_d$, which is needed for normalisation purposes \cite{FST}, drops out in this expression.

It is instructive to have a closer look at the non-factorisable $U$-spin-breaking effects
entering Eq.~(\ref{H-R}), although we will constrain them through experimental data. We obtain
the following expression:
\begin{equation}\label{aNF-rat}
\left|\frac{a_{\rm NF}^{(s)}}{a_{\rm NF}^{(d)}}\right|\equiv
\left|\frac{1+\Delta^{(s)}_{\rm NF}}{1+\Delta^{(d)}_{\rm NF}}\right|
=\Biggl|\frac{1+r_P^{(s)}}{1+r_P^{(d)}}\Biggr|
\Biggl|\frac{1+x^{(s)}}{1+x^{(d)}}\Biggr|\Biggl|\frac{1+\Delta^{T(s)}_{\rm NF}}{1+\Delta^{T(d)}_{\rm NF}}\Biggr|\:.
\end{equation}
Using heavy-meson chiral perturbation theory and the $1/N_{\rm C}$ 
expansion, non-factorisable $SU(3)$-breaking corrections to the colour-allowed tree amplitudes
of $\Bq\to \Dq\Dbar_p$ 
decays were found at the level of a few percent in Ref.~\cite{RF-BDD-93}, suggesting
small corrections from the last factor. The $U$-spin relation between 
the $\BstoDsDs$ and $\BdtoDdDd$ decays is reflected by the 
one-to-one correspondence of the $r^{(q)}_P$ and $x^{(q)}$ terms. These contributions 
enter Eq.~(\ref{aNF-rat}) only in ratios of terms with structures
$1+\Lambda^{(q)}$, where we expect the $\Lambda^{(q)}$ to be at most ${\cal O}(0.2)$. Assuming
$SU(3)$-breaking at the 30\% level for the $\Lambda^{(q)}$ terms yields a correction of only
${\cal O}(5\%)$ for the ratios, i.e. a robust situation.

Let us now exploit experimental data to probe these effects.
Using the ratio $R_{\Ds}$, we may actually determine $|a_{\rm NF}^{(s)}|$ with the help of the relation
\begin{equation}
|a_{\rm NF}^{(s)}| \equiv 1+\Delta_{\rm NF}^{(s)} 
=\left[1- \epsilon a' \cos \theta' \cos\gamma+{\cal O}\left(\epsilon^2a'^2\right) \right]
\sqrt{\frac{R_{\Ds}}{6\pi^2 |V_{cs}|^2  f_{\Ds}^2  X_{\Ds}}}\:,
\end{equation}
where $\Delta_{\rm NF}^{(s)}$ is now --- by definition --- a real parameter and the corrections 
due to the $\epsilon a'$ term are at most at the level of a few percent. Assuming
\begin{equation}
\Delta_{\rm NF}^{(d)} = \Delta_{\rm NF}^{(s)}\left[1-\xi_{SU(3)}\right]
\end{equation}
with the $SU(3)$-breaking parameter $\xi_{SU(3)}\propto m_s/\Lambda_{\rm QCD}$,
we obtain
\begin{equation}
\left|\frac{a_{\rm NF}^{(s)}}{a_{\rm NF}^{(d)}}\right|=\frac{1+\Delta_{\rm NF}^{(s)}}{1+
\Delta_{\rm NF}^{(s)}\left[1-\xi_{SU(3)}\right]}= 1+\Delta_{\rm NF}^{(s)} \, \xi_{SU(3)}
+{\cal O}\left(\Delta_{\rm NF}^{(s)2}\right)\:.
\end{equation}
Consequently, the information for the semileptonic differential rate allows us to quantify the 
non-factorisable $U$-spin-breaking corrections to the determination of $H$ (Eq.~(\ref{H-R})).

Let us now return to discuss the remaining quantities entering Eq.~(\ref{H-R}). 
The $\Dq$-meson decay constants ($q=d,s$) can be extracted from leptonic decays:
\begin{equation}
\Gamma(\Dqp\to\mu^+\nu_\mu)=\frac{G_{\rm F}^2}{8\pi}m_{\Dq} 
m_\mu^2\left[1-\left(\frac{m_\mu}{m_{\Dq}}\right)^2\right]|V_{cq}|^2f_{\Dq}^2\:.
\end{equation}
The current experimental status has been summarised in Ref.~\cite{RoSt}:
\begin{equation}\label{fD-exp}
f_{\Ds}=(257.5\pm4.6)\,{\rm MeV}\:, \quad f_{\Dd}=(204.6\pm5.0)\,{\rm MeV}\:,
\quad f_{\Ds}/f_{\Dd} = 1.258 \pm 0.038\:.
\end{equation}
A detailed overview of the status of lattice QCD calculations has been given by the FLAG 
Working Group in Ref.~\cite{Aoki:2013ldr}.

In the infinite quark-mass limit, the following consistency relation arises \cite{NeRi}:
\begin{equation}\label{Eq:FF_HQlim}
\frac{F_0^{\uBq\Dq}(q^2)}{F_1^{\uBq\Dq}(q^2)}=1-\frac{q^2}{(m_{\uBq}+m_{\Dq})^2}\:.
\end{equation}
For a discussion on QCD and $\Lambda_{\rm QCD}/m_Q$ corrections to this relation we refer the reader to
Refs.~\cite{NeRi,Neubert, BS-model}. There has recently been impressive progress in the calculation of
hadronic form factors within lattice QCD, where now the first unquenched calculations of the
$\bar B\to \bar D\ell \bar\nu_\ell$ form factors at nonzero recoil are available \cite{lat-1,lat-2}. Using
the results of Ref.~\cite{lat-2}, we obtain
\begin{equation}\label{Eq:FF_HQlat}
\frac{F_0^{\uBq\Dq}(m_{D_s}^2)}{F_1^{\uBq\Dq}(m_{D_s}^2)}=0.917\pm0.079\:,
\end{equation}
while the expression in Eq.~(\ref{Eq:FF_HQlim}) gives $0.924$, thereby indicating small
corrections. In the numerical analysis in this paper, we will use the result in Eq.~(\ref{Eq:FF_HQlat}).

Experimental data for the semileptonic decay \BsSL is not yet available.
Although it is experimentally challenging to disentangle the semileptonic \BsSL
and \BsSLstar decays, it might be feasible to distinguish them due to the
shifted invariant mass spectrum of the $\Dsp\mu^-$ combinations, and the
difference in the missing reconstructed mass, which is correlated to the
``corrected mass'' as illustrated in Ref.~\cite{LHCb-Vub}. Combined with a fit
to the angular distributions, this gives information on the different form factors.
We encourage to add this channel to the experimental agenda of the LHCb and Belle II 
experiments and perform detailed studies for the upgrade era. On the other hand, the differential 
rate of the \BdSL mode has already been measured
and will actually be used in the next section to estimate the non-factorisable effects in \Bd decays.

The lack of experimental data on semileptonic \Bs decays can be circumvented by studying the ratio
of other \Bd and \Bs decays, discussed in the next Section, and applying $SU(3)$ flavour symmetry. 
The $SU(3)$-breaking non-factorisable effects in the ratio of \Bd and \Bs decays are estimated
from the deviation from factorisation in \Bd decays.

\section{Picture Emerging from the Current Data}\label{sec:dat}
\subsection{Overview}

The main objective of this analysis is the determination of the \Bq--\Bqb mixing
phases $\phi_d$ and $\phi_s$ from the \BdtoDdDd and \BstoDsDs channels,
respectively.  High precision determinations of these phases require us to
control not only the contributions from penguin topologies, but also the impact
of additional decay topologies and non-factorisable effects.  The latter two
aspects cannot be quantified using information from the \BdtoDdDd and \BstoDsDs
decays alone.  Additional \BtoDD decays with similar dynamics to the \BdtoDdDd,
\BstoDsDs system therefore need to be studied.  An overview of the different
decay modes discussed in this section and their applications is given in
Table~\ref{tab:overview}.

\begin{table}[htp]
\begin{center}
\begin{tabular}{|l|l|ccccc|l|} 
\toprule
 Decay     & $\mathcal{A}$         &\multicolumn{5}{c|}{Topologies}&  Used for:                                                 \\
           &                       & $T$ & $P$ & $E$ & $PA$ & $A$  &                                                            \\ 
\midrule
\BdtoDdDd  &$\mathcal{A}          $& x & x & x & x  &    & determination of $a$ and $\theta$ (and $\phi_d$)                     \\
\BdtoDdDs  &$\mathcal{\tilde A}'  $& x & x &   &    &    & non-factorisable effect ${\tilde a}_{\rm NF}'$                       \\
\BdtoDsDs  &$\mathcal{A}_{EPA}    $&   &   & x & x  &    & quantify $E+PA$ contribution $\tilde{x}$                             \\
\midrule	   	       				 		  	 
\BstoDsDs  &$\mathcal{A}'         $& x & x & x & x  &    & physics goal $\phi_s$                                                \\
\BstoDsDd  &$\mathcal{\tilde A}   $& x & x &   &    &    & $SU(3)$ breaking non-fact. ${\tilde a}_{\rm NF}/{\tilde a}_{\rm NF}'$\\
\BstoDdDd  &$\mathcal{A}'_{EPA}   $&   &   & x & x  &    & quantify $E+PA$ contribution $\tilde{x}'$                            \\
\midrule	   	       				 		  	
\ButoDzDd  &$\mathcal{\tilde A}_c $& x & x &   &    & x  & quantify $A$ contribution $r_{\rm A}$ \ldots                         \\
\ButoDzDs  &$\mathcal{\tilde A}_c'$& x & x &   &    & x  & \ldots and consistency of $a_{\rm NF,c}/a_{\rm NF,c}'$               \\
\midrule	
\end{tabular}
\caption[Overview]
{Overview of the various topologies contributing to the \BtoDD decays.  
The naming convention is indicated in the second column. }\label{tab:overview}
\end{center}
\end{table}

\subsection{Preliminaries}
The direct and mixing-induced CP asymmetries of the \BdtoDdDd decay and the $H$ observable, 
when using the $U$-spin relation in Eq.~(\ref{a-theta-rel}), depend on the four 
parameters $a$, $\theta$, $\phi_d$ and $\gamma$.
In 1999, when this decay was originally suggested by one of us, 
the determination of the UT angle $\gamma$ was the main goal.
The proposed strategy therefore assumed input on $\phi_d$, determined from 
$\Bd\to J/\psi K_{\rm S}^0$ and complemented with \mbox{$\Bs\to J/\psi K_{\rm S}^0$} \cite{DeBF-pen}, 
to determine $a$, $\theta$ and $\gamma$ from 
${\cal A}_{\rm CP}^{\rm dir}(\uBdtoDdDd)$, ${\cal A}_{\rm CP}^{\rm mix}(\uBdtoDdDd)$ 
and $H$ \cite{RF-psiK}.
However, at present it is possible to extract $\gamma$ in a powerful way through pure $B\to D^{(*)}K^{(*)}$ 
tree decays.
Using current data for these channels, the CKMfitter and UTfit collaborations have obtained the following 
averages:
\begin{equation}\label{gamma-range}
\gamma  = (73.2_{-7.0}^{+6.3})^{\circ}  \quad\text{(CKMfitter \cite{Charles:2015gya})}\:, \qquad
\gamma  = (68.3 \pm 7.5)^{\circ} \quad\text{(UTfit \cite{Bevan:2014cya})\:.}
\end{equation}
For the numerical analysis in this paper, we shall use the CKMfitter result. 
By the time of the Belle II and LHCb upgrade era, much more precise 
measurements of $\gamma$ from pure tree decays will be available (see 
Section~\ref{sec:prosp}). Using $\gamma$ as an input, we may instead determine $\phi_d$ and
the penguin parameters from $H$ and the CP asymmetries of $\BdtoDdDd$
\cite{RF-BDD-07}. The penguin parameters thus determined allow us to take their 
effects into account in the determination of $\phi_s$ from the mixing-induced CP asymmetry 
${\cal A}_{\rm CP}^{\rm mix}(\uBstoDsDs)$. 

\boldmath
\subsection{Comparing \BtoDD Branching Fractions}
\unboldmath
To quantify the contributions from additional decay topologies and the impact of
non-factorisable effects in the \BdtoDdDd, \BstoDsDs system, we need to extend
the decay basis to modes with dynamics similar to the \BdtoDdDd and \BstoDsDs
decays.  If we replace the spectator quarks correspondingly, we obtain the
\BstoDsDd and \BdtoDdDs channels.  These decays are again related to each other
through the $U$-spin symmetry. However, the exchange and penguin annihilation
topologies do not have counterparts in \BstoDsDd and \BdtoDdDs, which are
characterised by the following decay amplitudes:
\begin{equation}
A\left(\BstoDsDd\right) = - \lambda \mathcal{\tilde A}
\left[1- \tilde a e^{i\tilde \theta}e^{i\gamma}\right]\
\end{equation}
\begin{equation}
A\left(\BdtoDdDs\right)= \left(1-\frac{\lambda^2}{2}\right)\mathcal{\tilde A}'
\left[1+\epsilon \tilde a'e^{i\tilde \theta'}e^{i\gamma}\right]\:,
\end{equation}
where
\begin{equation}
\mathcal{\tilde A} \equiv \lambda^2 A \left[\tilde T+ \tilde P^{(c)}-\tilde P^{(t)}\right]
\end{equation}
\begin{equation}
\tilde a e^{i\tilde \theta } \equiv R_b\left[\frac{\tilde P^{(u)}-\tilde P^{(t)}}{\tilde T+ 
\tilde P^{(c)}-\tilde P^{(t)}}\right]\:;
\end{equation}
$\mathcal{\tilde A}'$ and $ \tilde a'e^{i\tilde \theta'}$ take analogous expressions. 
If we use the $U$-spin flavour symmetry, we obtain the following relations:
\begin{equation}
\tilde a e^{i\tilde \theta } = \tilde a' e^{i\tilde \theta' }\:,\qquad \mathcal{\tilde A} =  \mathcal{\tilde A}'\:.
\end{equation}

\begin{figure}[tp]
 \centering
\includegraphics[width=0.49\textwidth]{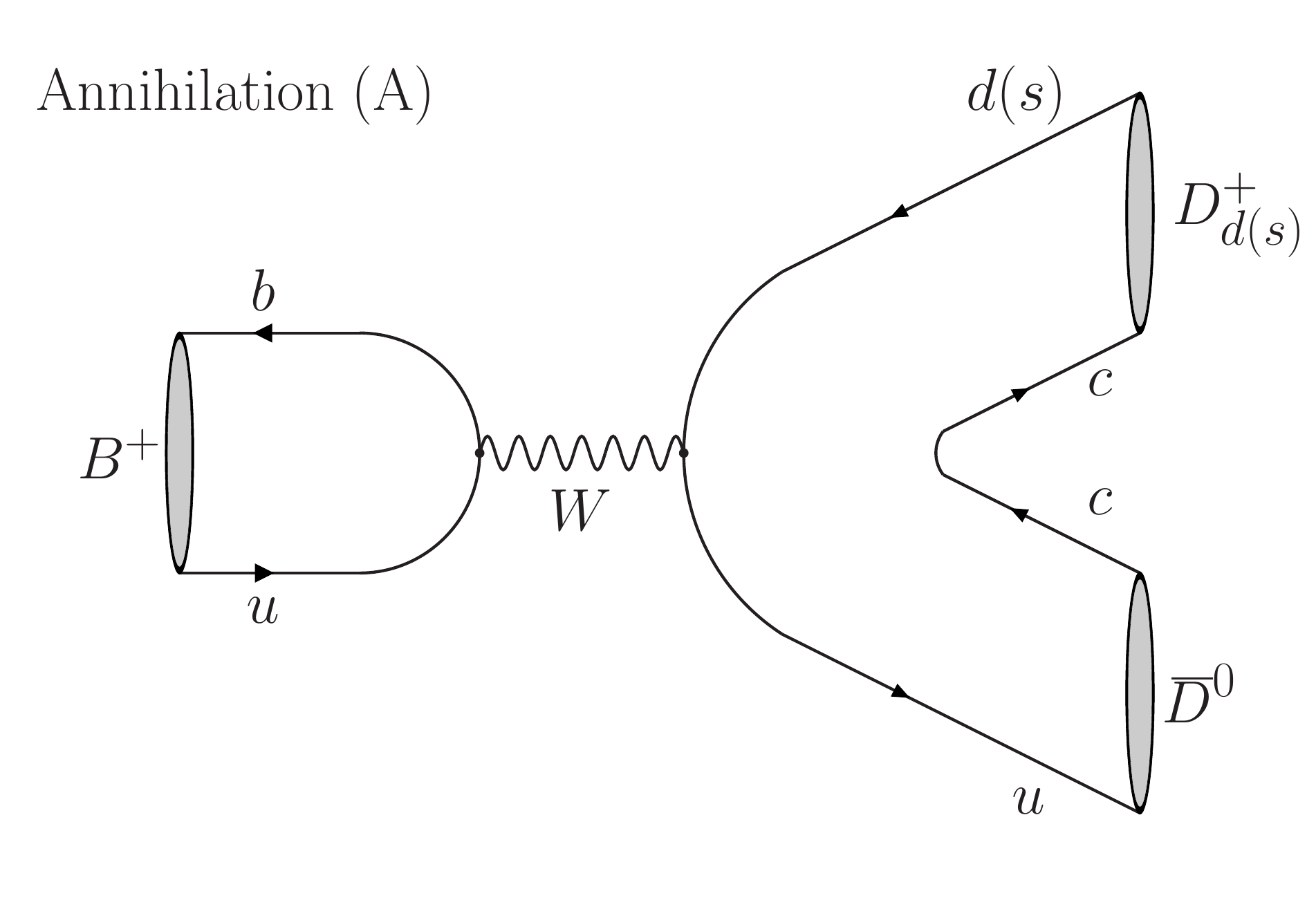}
\caption{Illustration of the annihilation topology contributing to 
$\Bu\to\Dzb D_{d(s)}^+$.}\label{fig:annihilation}
\end{figure}

Moreover, there are the charged decays \ButoDzDd and 
\ButoDzDs, which are again related to each other through the $U$-spin symmetry.
These modes also do not receive contributions from exchange and
penguin annihilation topologies. However, there are additional 
contributions from annihilation topologies, as illustrated in Fig.~\ref{fig:annihilation},
which enter with the same CKM factor as the penguin contributions with up-quark
exchanges. 

\boldmath
\subsubsection{Probing Annihilation Topologies with Charged $B$ decays}
\unboldmath

The decay amplitudes take the following forms:
\begin{equation}
A\left(\ButoDzDd\right) = - \lambda \mathcal{\tilde A}_{\rm c}
\left[1- \tilde a_{\rm c}e^{i\tilde \theta_{\rm c}}e^{i\gamma}\right]\
\end{equation}
\begin{equation}
A\left(\ButoDzDs\right)= \left(1-\frac{\lambda^2}{2}\right)\mathcal{\tilde A}'_{\rm c}
\left[1+\epsilon \tilde a'_{\rm c}e^{i\tilde \theta'_{\rm c}}e^{i\gamma}\right]\:,
\end{equation}
where
\begin{equation}
\mathcal{\tilde A}_{\rm c} \equiv \lambda^2 A \left[\tilde T_{\rm c}+ \tilde P^{(c)}_{\rm c}-
\tilde P^{(t)}_{\rm c}\right]
\end{equation}
\begin{equation}
\tilde a_{\rm c} e^{i\tilde \theta_{\rm c} } 
\equiv R_b\left[\frac{\tilde P_{\rm c}^{(u)}-\tilde P_{\rm c}^{(t)} + \tilde A_{\rm c}}{\tilde T_{\rm c}
+ \tilde P_{\rm c}^{(c)}-\tilde P_{\rm c}^{(t)}}\right]
\end{equation}
with $\tilde A_{\rm c}$ denoting the annihilation amplitude; the expressions for the primed amplitudes
are analogous. The penguin parameter satisfies 
\begin{equation}\label{ac-atilde-rel}
\tilde a_{\rm c} e^{i\tilde \theta_{\rm c} } = \tilde a e^{i\tilde \theta}\left[1+r_A  \right] 
\end{equation}
with
\begin{equation}\label{rA-def}
r_A\equiv \frac{\tilde A_{\rm c}}{\tilde P_c^{(u)}-\tilde P_c^{(t)}}\:.
\end{equation}

It is useful to introduce the following ratios:
\begin{align}\label{Eq:Xi_par}
&\Xi(\uBstoDsDd, \uButoDzDd)\notag\\
&\qquad\equiv\left[\frac{m_{\uBs}}{m_{\uBu}} 
\frac{\Phi(m_{\Du}/m_{\uBu},m_{\Dd}/m_{\uBu})}{\Phi(m_{\Ds}/m_{\uBs},m_{\Dd}/m_{\uBs})} 
 \frac{\tau_{\uBu}}{\tau_{\uBs}} \right]
\left[\frac{\mathcal{B}(\uBstoDsDd)_{\rm theo}}{\mathcal{B}(\uButoDzDd)}\right]\:,
\end{align}
\begin{align}
&\Xi(\uBdtoDdDs, \uButoDzDs)\notag\\
&\qquad\equiv\left[\frac{m_{\uBd}}{m_{\uBu}} 
\frac{\Phi(m_{\Du}/m_{\uBu},m_{\Ds}/m_{\uBu})}{\Phi(m_{\Dd}/m_{\uBd},m_{\Ds}/m_{\uBd})} 
 \frac{\tau_{\uBu}}{\tau_{\uBd}} \right]
\left[\frac{\mathcal{B}(\uBdtoDdDs)}{\mathcal{B}(\uButoDzDs)}\right]\:.
\end{align}
Using the expressions for the decay amplitudes given above yields
\begin{equation}\label{Xi-rel-1}
\Xi(\uBstoDsDd, \uButoDzDd)
=\left|\frac{\mathcal{\tilde A}}{\mathcal{\tilde A}_{\rm c}}\right|^2
\left[\frac{1-2\tilde a \cos\tilde\theta\cos\gamma+\tilde a^2}{1-
2\tilde a_{\rm c} \cos\tilde\theta_{\rm c}\cos\gamma+\tilde a_{\rm c}^2}\right]\:,
\end{equation}
\begin{equation}\label{Xi-rel-2}
\Xi(\uBdtoDdDs, \uButoDzDs)
=\left|\frac{\mathcal{\tilde A}'}{\mathcal{\tilde A}_{\rm c}'}\right|^2
\left[\frac{1+2\epsilon\tilde a' \cos\tilde\theta'\cos\gamma+\epsilon^2\tilde a'^2}{1+
2\epsilon\tilde a'_{\rm c} \cos\tilde\theta_{\rm c}'\cos\gamma+\epsilon^2\tilde a_{\rm c}^{'2}}\right]\:.
\end{equation}
If we apply the $SU(3)$ flavour symmetry (actually the $V$-spin subgroup), we obtain
\begin{equation}\label{SU3-1-rel}
\mathcal{\tilde A}_{\rm c}=\mathcal{\tilde A}\:,
\end{equation}
while the isospin symmetry of strong interactions implies
\begin{equation}
\mathcal{\tilde A}'_{\rm c}=\mathcal{\tilde A}'\:.
\end{equation}
If we neglect the annihilation contribution in Eq.~(\ref{ac-atilde-rel}) and assume the same 
penguin contributions in Eq.~(\ref{Xi-rel-1}), i.e.\ $\tilde a=\tilde a_{\rm c}$, we obtain
\begin{equation}
\Xi(\uBstoDsDd, \uButoDzDd)\approx
\left|\frac{\mathcal{\tilde A}}{\mathcal{\tilde A}_{\rm c}}\right|^2 
\quad\stackrel{V{\rm-spin}}{\xrightarrow{\hspace{12mm}}}\quad1\:.
\end{equation}
A deviation from unity of this ratio would therefore imply either the presence of non-zero 
annihilation contributions or large $SU(3)$-breaking effects through Eq.~(\ref{SU3-1-rel}).
In the case of (\ref{Xi-rel-2}), the penguin parameters are suppressed by the tiny $\epsilon$ factor
and hence play a negligible role. Consequently, the ratio
\begin{equation}
\Xi(\uBdtoDdDs, \uButoDzDs)=
\left|\frac{\mathcal{\tilde A}'}{\mathcal{\tilde A}_{\rm c}'}\right|^2 
\quad\stackrel{{\rm Isospin}}{\xrightarrow{\hspace{12mm}}}\quad1
\end{equation}
essentially relies on the strong isospin symmetry. The current experimental results compiled by the 
Particle Data Group (PDG) read as follows \cite{Agashe:2014kda}:
\begin{align}
\mathcal{B}(\uBstoDsDd)& = (3.6\pm0.8)\times10^{-4}\:,\label{BRBsDsDd} \\
\mathcal{B}(\uBdtoDdDs) & = (7.2\pm0.8)\times10^{-3}\:,\label{BRBdDDs} \\
\mathcal{B}(\uButoDzDd)& = (3.8\pm0.4)\times10^{-4}\:, \label{BRBpDDd} \\
\mathcal{B}(\uButoDzDs) & = (9.0\pm0.9)\times10^{-3}\:,\label{BRBpDDs}
\end{align}
and correspond to%
\footnote{Since \BstoDsDd is a flavour-specific final state, 
we simply have $\mathcal{A}_{\Delta\Gamma}(\uBstoDsDd) = 0$ for the conversion of 
the time-integrated branching fraction into the theoretical branching ratio.}
\begin{align}
\Xi(\uBstoDsDd, \uButoDzDd) & =  1.08 \pm 0.27\:,\\
\Xi(\uBdtoDdDs, \uButoDzDs)& = 0.89 \pm 0.13\:. \label{Xi-dir}
\end{align}
For the last decay combination, we may also employ the direct measurement of the
ratio of the relevant branching fractions \cite{LHCB-BDD-2013}, which is given by
\begin{equation}
\frac{\mathcal{B}(\uButoDzDs)}{\mathcal{B}(\uBdtoDdDs)} = 1.22 \pm 0.02 \pm 0.07\:.
\end{equation}
This leads to
\begin{equation}
\Xi(\uBdtoDdDs, \uButoDzDs) = 0.878 \pm 0.050\:,
\end{equation}
which has a significantly smaller uncertainty with respect to Eq.~(\ref{Xi-dir}) thanks to a cancellation of
uncertainties in the directly measured ratio of branching fractions. We note the deviation from one at 
the $2.4\,\sigma$ level, which is unexpected.

\subsubsection{Probing Exchange and Penguin Annihilation Topologies}\label{ssec:EPA}
The current PDG results for the CP-averaged branching ratios of the \BdtoDdDd
and \BstoDsDs decays are given as follows \cite{Agashe:2014kda}:
\begin{align}
\mathcal{B}(\uBdtoDdDd)& = (2.11\pm0.18)\times10^{-4}\:, \label{BRBdDdDd} \\
\mathcal{B}(\uBstoDsDs) & = (4.4\phantom{1}\pm0.5\phantom{1})\times10^{-3}\:. \label{BRBsDsDs}
\end{align}
In comparison with (\ref{BRBdDDs}) and (\ref{BRBpDDs}), the branching ratio in (\ref{BRBsDsDs}) 
is surprisingly small. A similar pattern --- although not as pronounced in view of the larger
uncertainties --- is observed if we compare (\ref{BRBdDdDd}) with (\ref{BRBsDsDd}) 
and (\ref{BRBpDDd}). As the \BdtoDdDd and \BstoDsDs decays
receive contributions from exchange and penguin annihilation topologies, which have no
counterparts in the \BstoDsDd, \ButoDzDd  and \BdtoDdDs, 
\ButoDzDs modes, respectively (see Table~\ref{tab:overview}), it is possible that the puzzling pattern of the
data is actually due to the presence of these exchange and penguin annihilation contributions. 

Let us first have a closer look at the ratio of the amplitudes of the \BstoDsDs
and \BdtoDdDs channels:
\begin{equation}
\frac{A(\BstoDsDs)}{A(\BdtoDdDs)}=
\left(\frac{\mathcal{A}'}{\mathcal{\tilde A}'}\right)\left[\frac{1+
\epsilon a'e^{i\theta'}e^{i\gamma}}{1+\epsilon \tilde a'e^{i\tilde \theta'}e^{i\gamma}} \right]
=\left[\frac{T'+ P^{(ct)'}}{\tilde T'+ \tilde P^{(ct)'}}+\tilde x'\right]\left[\frac{1+
\epsilon a'e^{i\theta'}e^{i\gamma}}{1+\epsilon \tilde a'e^{i\tilde \theta'}e^{i\gamma}} \right]\:,
\end{equation}
where
\begin{equation}\label{xtilde-prime-def}
\tilde x' \equiv  |\tilde x'|e^{i\tilde \sigma'}\equiv\frac{E'+PA^{(ct)'}}{\tilde T'+ \tilde P^{(ct)'}}
\end{equation}
measures, in analogy to the parameter $x$ introduced in Eq.~(\ref{x-def}),
the importance of the exchange and penguin annihilation topologies with respect 
to the dominant tree topology; we use abbreviations as in Eq.~(\ref{abbrev}).  
If we neglect the terms with the penguin parameters, which enter with the tiny $\epsilon$, and
introduce the $SU(3)$-breaking parameter
\begin{equation}\label{rhoprime-def}
\varrho'\equiv|\varrho'|e^{i\omega'}\equiv
\frac{T'+ P^{(ct)'}}{\tilde T'+ \tilde P^{(ct)'}}=
\left[\frac{T'}{\tilde T'}\right]\left[\frac{1+P^{(ct)'}/T'}{1+\tilde P^{(ct)'}/\tilde T'}\right]\:,
\end{equation}
we obtain the relation
\begin{equation}
\frac{A(\BstoDsDs)}{A(\BdtoDdDs)}=\varrho'+\tilde x'\:.
\end{equation}
The parameter $\varrho'$ is only affected by $SU(3)$-breaking effects entering at the 
spectator-quark level. Applying factorisation, where  $P^{(ct)'}/T'=\tilde P^{(ct)'}/\tilde T'$ 
(see the comment after Eq.~(\ref{BSS-estimate})), we obtain
\begin{align}
\varrho'_{\rm fact} & = 
\left[\frac{m_{\uBs}^2-m_{\Ds}^2}{m_{\uBd}^2-m_{\Dd}^2}\right]
\left[\frac{F_0^{\uBs\Ds}(m_{\Ds}^2)}{F_0^{\uBd\Dd}(m_{\Ds}^2)}\right]\\
& =\left[\frac{m_{\uBs}-m_{\Ds}}{m_{\uBd}-m_{\Dd}}\right]
\sqrt{\frac{m_{\uBs}m_{\Ds}}{m_{\uBd}m_{\Dd}}}
\left[\frac{1+w_s(m_{\Ds}^2)}{1+w_d(m_{\Ds}^2)}\right]\left[
\frac{\xi_s(w_s(m_{\Ds}^2))}{\xi_d(w_d(m_{\Ds}^2))}\right]\:.
\end{align}
Here we have taken into account the restrictions following for the corresponding $\uBq\to\Dq$ form 
factor from the heavy-quark effective theory \cite{NS}:
\begin{equation}\label{HQ-FF}
F_0^{\uBq\Dq}(q^2)=\left[\frac{m_{\uBq}+m_{\Dq}}{2\sqrt{m_{\uBq}m_{\Dq}}}\right]
\left[1-\frac{q^2}{(m_{\uBq}+m_{\Dq})^2}\right]\xi_q(w_q(q^2))\:,
\end{equation}
where $\xi_q(w_q(q^2))$ is the Isgur--Wise function with
\begin{equation}\label{wq2-def}
w_q(q^2)=\frac{m_{\uBq}^2+m_{\Dq}^2-q^2}{2m_{\uBq}m_{\Dq}}\:.
\end{equation}
Studies of the light-quark dependence of the Isgur--Wise function were performed within 
heavy-meson chiral perturbation theory, indicating an enhancement of $\xi_s/\xi_d$ at the 
level of $5\%$ \cite{HMChiPT1}. 
Applying the same formalism to $f_{\Ds}/f_{\Dd}$ leads to estimates for the value of this ratio 
of about 1.2 \cite{HMChiPT2}, which are in agreement with the experimental results in Eq.~(\ref{fD-exp}). 

Since 1992, when these calculations were pioneered, there has been a lot of progress in 
lattice QCD (for an overview of the state-of-the-art analyses, 
see Ref.~\cite{El-Khadra:2014sha}). The most recent result for the 
$SU(3)$-breaking effects in the form factors reads as follows \cite{Bailey:2012rr}:
\begin{equation}\label{Eq:FFlattice}
\left[\frac{F_0^{\uBs\Ds}(m_{\pi}^2)}{F_0^{\uBd\Dd}(m_{\pi}^2)}\right] = 1.054 \pm 0.047 \pm 0.017\:,
\end{equation}
which is in excellent agreement with the picture from heavy-meson chiral perturbation theory. 
Using this result as an input yields
\begin{equation}\label{rho-fact}
\varrho'_{\rm fact} = 1.078 \pm 0.051\:.
\end{equation}
The error quantifies only the uncertainties related to the form factors. We cannot quantify the
non-factorisable effects. However, as they enter only at the level of different spectator quarks
and as already the leading $SU(3)$-breaking effects are small, we expect a minor impact. 

The ratio\footnote{In view of the large uncertainty of Eq.~(\ref{ADG-exp}), we
use $\mathcal{A}_{\Delta\Gamma}(\uBstoDsDs) = -\cos\phi_s^{\text{SM}}$ 
for the conversion of the untagged experimental $\BstoDsDs$ branching ratio into its theoretical
counterpart.}
\begin{align}
&\Xi(\uBstoDsDs, \uBdtoDdDs)\notag\\
&\quad\equiv\left[\frac{m_{\uBs}}{m_{\uBd}} 
\frac{\Phi(m_{\Dd}/m_{\uBd},m_{\Ds}/m_{\uBd})}{\Phi(m_{\Ds}/m_{\uBs},m_{\Ds}/m_{\uBs})} 
 \frac{\tau_{\uBd}}{\tau_{\uBs}} \right]
\left[\frac{\mathcal{B}(\uBstoDsDs)_{\rm theo}}{\mathcal{B}(\uBdtoDdDs)}\right]
=0.647 \pm 0.049 \label{XiBsDsDs}
\end{align}
then takes the following form:
\begin{equation}\label{OBS-1}
\Xi(\uBstoDsDs, \uBdtoDdDs)=|\varrho'|^2+
2|\varrho'||\tilde x'|\cos(\omega'-\tilde\sigma')+|\tilde x'|^2\:,
\end{equation}
thereby fixing a circle for $\tilde x'$ in the complex plane. The numerical value in Eq.~(\ref{XiBsDsDs})
refers to a direct measurement of the corresponding ratio of branching ratios \cite{LHCB-BDD-2013}.

For the other decay combination, we obtain
\begin{equation}\label{EDA}
\frac{A(\BdtoDdDd)}{A(\BstoDsDd)}=
\left(\frac{\mathcal{A}}{\mathcal{\tilde A}}\right)\left[\frac{1-a e^{i\theta}e^{i\gamma}}{1-
\tilde a e^{i\tilde \theta}e^{i\gamma}} \right]
=\left[\frac{T+ P^{(ct)}}{\tilde T+ \tilde P^{(ct)}}+\tilde x\right]\left[\frac{1-
a e^{i\theta}e^{i\gamma}}{1- \tilde ae^{i\tilde \theta}e^{i\gamma}} \right]\:,
\end{equation}
with
\begin{equation}\label{xtilde-def}
\tilde x \equiv  |\tilde x|e^{i\tilde \sigma}\equiv\frac{E+PA^{(ct)}}{\tilde T+ \tilde P^{(ct)}}\:.
\end{equation}
In analogy to Eq.~(\ref{rhoprime-def}), we introduce a parameter
\begin{equation}
\varrho\equiv|\varrho|e^{i\omega}\equiv\frac{T+ P^{(ct)}}{\tilde T+ \tilde P^{(ct)}}=
\left[\frac{T}{\tilde T}\right]\left[\frac{1+P^{(ct)}/T}{1+\tilde P^{(ct)}/\tilde T}\right]\:,
\end{equation}
which is given in factorisation by
\begin{equation}\label{rho-rel}
\varrho_{\rm fact} = 
\left[\frac{m_{\uBd}^2-m_{\Dd}^2}{m_{\uBs}^2-m_{\Ds}^2}\right]
\left[\frac{F_0^{\uBd\Dd}(m_{\Dd}^2)}{F_0^{\uBs\Ds}(m_{\Dd}^2)}\right]=\frac{1}{\varrho_{\rm fact}'}
= 0.928 \pm 0.044\:,
\end{equation}
neglecting the tiny difference between the form-factor ratios for $q^2=m_{\Dd}^2$ and  $m_{\Ds}^2$.
As in Eq.~(\ref{rho-fact}), the error quantifies only the form factor uncertainties. 

The penguin parameters do not enter Eq.~(\ref{EDA}) with the tiny $\epsilon$. However, if we use the
$SU(3)$ relation
\begin{equation}
\frac{P^{(ut)}}{T+P^{(ct)}}=\frac{\tilde P^{(ut)}}{\tilde T+\tilde P^{(ct)}}\:,
\end{equation}
where decay constants and form factors cancel in factorisation, we get
\begin{equation}\label{a-x}
a e^{i\theta}=\tilde a e^{i\tilde \theta}
\left[\frac{1+r_{PA}}{1+x}\right]\:,
\end{equation}
with
\begin{equation}\label{rPA-def}
r_{PA}\equiv\frac{PA^{(ut)}}{P^{(ut)}}\:;
\end{equation}
the parameter $x$ was introduced in Eq.~(\ref{x-def}). Consequently, we have
\begin{equation}
\frac{1- \tilde ae^{i\tilde \theta}e^{i\gamma}}{1-a e^{i\theta}e^{i\gamma}}
=1+{\cal O}(\tilde a x)+{\cal O}(\tilde a r_{PA})\:,
\end{equation}
where the second-order terms are expected to give small corrections at the few-percent level.
Introducing the ratio
\begin{align}
& \Xi(\uBdtoDdDd, \uBstoDsDd)\notag\\
& \quad\equiv\left[\frac{m_{\uBd}}{m_{\uBs}} 
\frac{\Phi(m_{\Ds}/m_{\uBs},m_{\Dd}/m_{\uBs})}{\Phi(m_{\Dd}/m_{\uBd},m_{\Dd}/m_{\uBd})} 
 \frac{\tau_{\uBs}}{\tau_{\uBd}} \right]
\left[\frac{\mathcal{B}(\uBdtoDdDd)}{\mathcal{B}(\uBstoDsDd)_{\rm theo}}\right]
=0.59 \pm 0.14\:, \label{XiBdDdDd}
\end{align}
we obtain
\begin{equation}\label{OBS-2}
\Xi(\uBdtoDdDd, \uBstoDsDd) =
|\varrho|^2+2|\varrho||\tilde x|\cos(\omega-\tilde\sigma)+|\tilde x|^2
\end{equation}
in analogy to Eq.~(\ref{OBS-1}). 

It is interesting to consider the double ratio
\begin{align}
& \frac{\Xi(\uBstoDsDs, \uBdtoDdDs)}{\Xi(\uBdtoDdDd, \uBstoDsDd)}\notag\\
& \qquad=\frac{|\varrho'|^2+2|\varrho'||\tilde x'|
\cos(\omega'-\tilde\sigma')+|\tilde x'|^2}{|\varrho|^2+2|\varrho||\tilde x|\cos(\omega-\tilde\sigma)+
|\tilde x|^2}\approx\left|\frac{\varrho'}{\varrho}\right|^2\approx(\varrho'_{\rm fact})^4
= 1.35 \pm 0.26\:,\label{Xi-ratio}
\end{align}
where we have neglected the $|\tilde x^{(')}|$ terms and have used Eq.~(\ref{rho-rel}).
The experimental results in Eqs.~(\ref{XiBsDsDs}) and (\ref{XiBdDdDd}) give 
\begin{equation}
\frac{\Xi(\uBstoDsDs, \uBdtoDdDs)}{\Xi(\uBdtoDdDd, \uBstoDsDd)} = 1.11 \pm 0.28\:,
\end{equation}
which is in agreement with the expectation in Eq.~(\ref{Xi-ratio}). The current uncertainties are unfortunately
too large to draw any further conclusions.

\begin{figure}[tp]
 \centering
\includegraphics[width=0.49\textwidth]{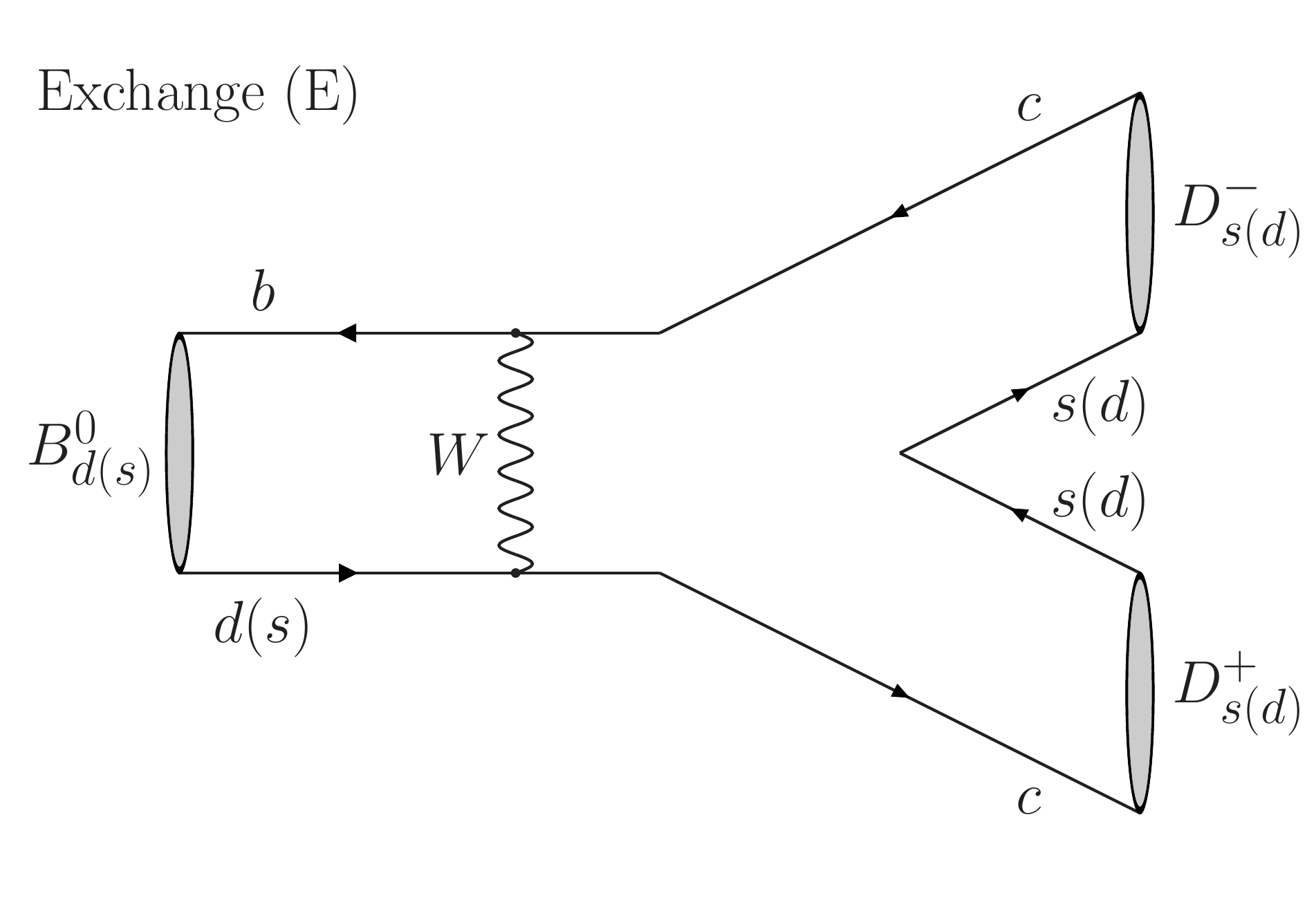}
\includegraphics[width=0.49\textwidth]{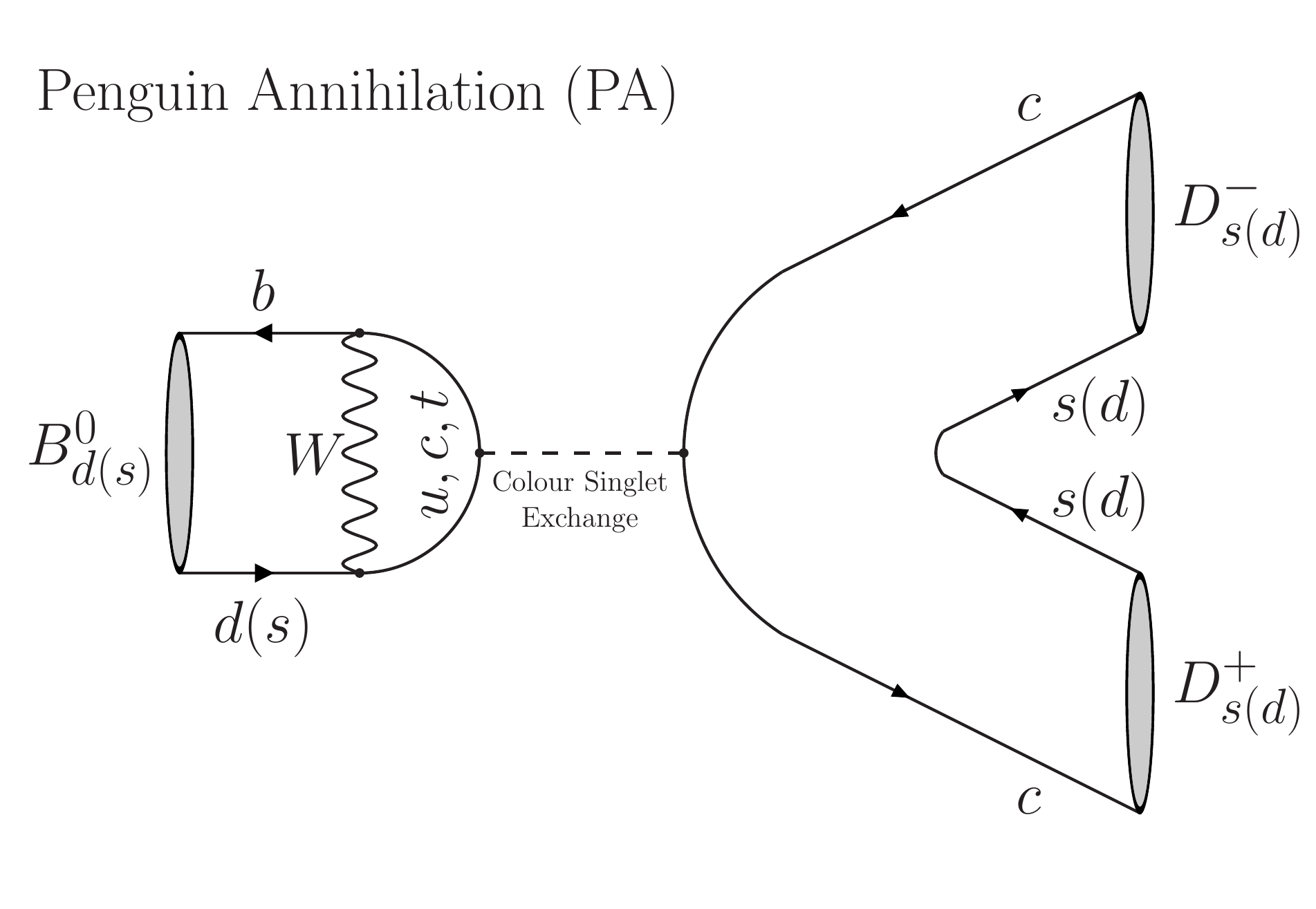}
\caption{Illustration of exchange and penguin annihilation topologies contributing to 
\BdtoDsDs  and \BstoDdDd.}\label{fig:EPA}
\end{figure}

%
%
%
\subsubsection{Probing Exchange and Penguin Annihilation Topologies Directly}
The exchange and penguin annihilation topologies can be probed in a {\it direct} way by
means of the decays \BstoDdDd and \BdtoDsDs. These modes 
receive only contributions from exchange and penguin annihilation topologies \cite{GRP,GHLR}, 
as illustrated in Fig.~\ref{fig:EPA}, and are related to each other through the 
$U$-spin symmetry. The current experimental information on the corresponding 
CP-averaged branching ratios is given as follows \cite{Agashe:2014kda}:
\begin{align}
\mathcal{B}(\uBdtoDsDs) & < 3.6 \times10^{-5}~\mbox{(90\% C.L.)}\:, \\
\mathcal{B}(\uBstoDdDd) & = (2.2\pm0.6)\times10^{-4}\:.
\end{align}
The experimental signal for the \BstoDdDd decay is in accordance 
with the picture emerging from the discussion given above. 

Let us now have a closer look at these decays. Their amplitudes can be written as
\begin{equation}
A\left(\BdtoDsDs\right) = - \lambda \mathcal{A}_{EPA}
\left[1- a_{EPA} e^{i\theta_{EPA}}e^{i\gamma}\right]\:,
\end{equation}
\begin{equation}
A\left(\BstoDdDd\right)= \left(1-\frac{\lambda^2}{2}\right)\mathcal{A}'_{EPA}
\left[1+\epsilon  a'_{EPA}e^{i \theta'_{EPA}}e^{i\gamma}\right]\:,
\end{equation}
where
\begin{equation}
\mathcal{A}_{EPA} \equiv \lambda^2 A \left[\hat{E} + \hat{PA}^{(ct)}\right]\:,
\end{equation}
\begin{equation}
\tilde a_{EPA} e^{i\tilde \theta_{EPA} } 
\equiv R_b\left[\frac{\hat{PA}^{(ut)}}{\hat{E}+ \hat{PA}^{(ct)}}\right]\:;
\end{equation}
the primed parameters are defined in an analogous way. We obtain then
\begin{equation}
\frac{A(\BstoDdDd)}{A(\BdtoDdDs)} =
\left(\frac{\mathcal{A}'_{EPA}}{\mathcal{\tilde A}'}\right)\left[\frac{1+
\epsilon a'_{EPA}e^{i\theta'_{EPA}}e^{i\gamma}}{1+
\epsilon \tilde a'e^{i\tilde \theta'}e^{i\gamma}} \right]
= \varsigma' \tilde{x}'\:,
\end{equation}
where we have neglected the terms proportional to the tiny $\epsilon$ factor and
introduced the parameter
\begin{equation}\label{E-est}
\varsigma' \equiv \frac{\hat{E}'+ \hat{PA}^{\prime(ct)}}{E'+ PA^{\prime(ct)}}\approx
\left(\frac{f_{\Dd}}{m_{\Dd}}\frac{m_{\Ds}}{f_{\Ds}}\right)^2 = 0.700 \pm 0.042\:.
\end{equation}
In the estimate of this $SU(3)$-breaking parameter, we have used that
exchange and penguin annihilation topologies, which are genuinely of non-factorisable nature, 
are expected to be suppressed by the smallness of the $B$- and $D$-meson wave functions at the 
origin, which behave as $f_B/m_B$ and $f_D/m_D$, respectively \cite{GHLR-94}. 
The $f_{B_s}/m_{B_s}$ terms cancel in Eq.~(\ref{E-est}), and the error of the numerical value describing the leading $SU(3)$-breaking effect corresponds only to the uncertainties of the $D_{d,s}$-meson 
decay constants and masses.
The ratio%
\footnote{We assume $\mathcal{A}_{\Delta\Gamma}(\uBstoDdDd) = -\cos\phi_s^{\text{SM}}$ 
for the conversion of the untagged experimental $\BstoDdDd$ branching ratio into 
the corresponding theoretical branching ratio.}
\begin{align}
& \Xi(\uBstoDdDd, \uBdtoDdDs)\notag\\
&\quad\equiv\left[\frac{m_{\uBs}}{m_{\uBd}} 
\frac{\Phi(m_{\Dd}/m_{\uBd},m_{\Ds}/m_{\uBd})}{\Phi(m_{\Dd}/m_{\uBs},m_{\Dd}/m_{\uBs})} 
 \frac{\tau_{\uBd}}{\tau_{\uBs}} \right]
\left[\frac{\mathcal{B}(\uBstoDdDd)_{\rm theo}}{\mathcal{B}(\uBdtoDdDs)}\right]
=0.031 \pm 0.009\label{XiBsDD}
\end{align}
takes then the simple form
\begin{equation}\label{OBS-3}
\Xi(\uBstoDdDd, \uBdtoDdDs) =
|\varsigma'\tilde x'|^2\:,
\end{equation}
which fixes a circle with radius $|\varsigma'\tilde x'|$ around the origin in the complex plane. 

Concerning the \BdtoDsDs decay, we have
\begin{equation}
\frac{A(\BdtoDsDs)}{A(\BstoDsDd)}=
\left(\frac{\mathcal{A}_{EPA}}{\mathcal{\tilde A}}\right)\left[\frac{1-
a_{EPA}e^{i\theta_{EPA}}e^{i\gamma}}{1 - \tilde a e^{i\tilde \theta}e^{i\gamma}} \right]
= \varsigma \tilde x\:,
\end{equation}
where we have neglected the penguin annihilation contributions on the right-hand side,
and have introduced 
\begin{equation}
\varsigma \equiv \frac{\hat{E}+ \hat{PA}^{(ct)}}{E+ PA^{(ct)}}
\approx \left(\frac{f_{\Ds}}{m_{\Ds}}\frac{m_{\Dd}}{f_{\Dd}}\right)^2
 = 1.408 \pm 0.057 \approx \frac{1}{\varsigma'}\:.
\end{equation}
As in Eq.~(\ref{E-est}), we expect that the numerical value describes the leading $SU(3)$-breaking effect
(the uncertainty corresponds only to the decay constants and masses). Non-factorisable $SU(3)$-breaking
contributions to this quantity cannot be estimated at present. 

For the comparison with the experimental data we introduce 
\begin{align}
&\Xi(\uBdtoDsDs, \uBstoDsDd)\notag\\
&\quad\equiv\left[\frac{m_{\uBd}}{m_{\uBs}} 
\frac{\Phi(m_{\Ds}/m_{\uBs},m_{\Dd}/m_{\uBs})}{\Phi(m_{\Ds}/m_{\uBd},m_{\Ds}/m_{\uBd})} 
 \frac{\tau_{\uBs}}{\tau_{\uBd}} \right]
\left[\frac{\mathcal{B}(\uBdtoDsDs)}{\mathcal{B}(\uBstoDsDd)_{\rm theo}}\right]
< 0.107~\mbox{(90\% C.L.)}\:, \label{XiBdDsDs}
\end{align}
which takes the simple form
\begin{equation}\label{OBS-4}
\Xi(\uBdtoDsDs, \uBstoDsDd) = |\varsigma\tilde x |^2\:.
\end{equation}
Also in this case it is interesting to consider the double ratio
\begin{equation}\label{Xi-xtilde-rel}
\frac{\Xi(\uBdtoDsDs, \uBstoDsDd)}{\Xi(\uBstoDdDd, \uBdtoDdDs)}
= \left|\frac{\varsigma\tilde x}{\varsigma'\tilde x'}\right|^2\:,
\end{equation}
which allows us to test the relation
\begin{equation}\label{xtilde-rel}
\tilde x'\approx \left[\left(\frac{f_{\uBs}f_{\Ds}}{f_{\uBd}f_{\Dd}}\right)\varrho_{\rm fact}'\right]\tilde x\:.
\end{equation}

\begin{figure}[tp]
 \centering
\includegraphics[width=0.49\textwidth]{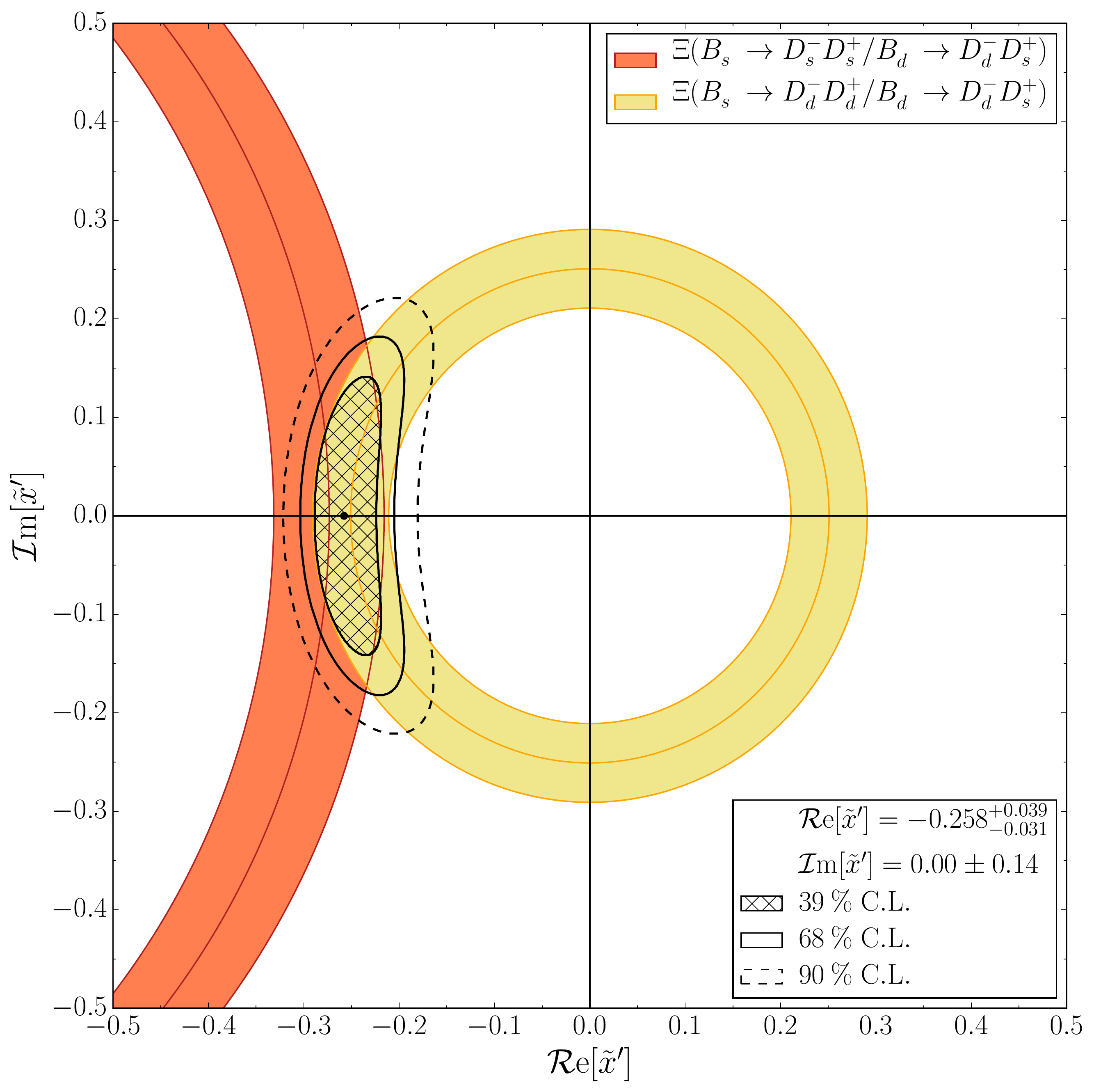}
\includegraphics[width=0.481\textwidth]{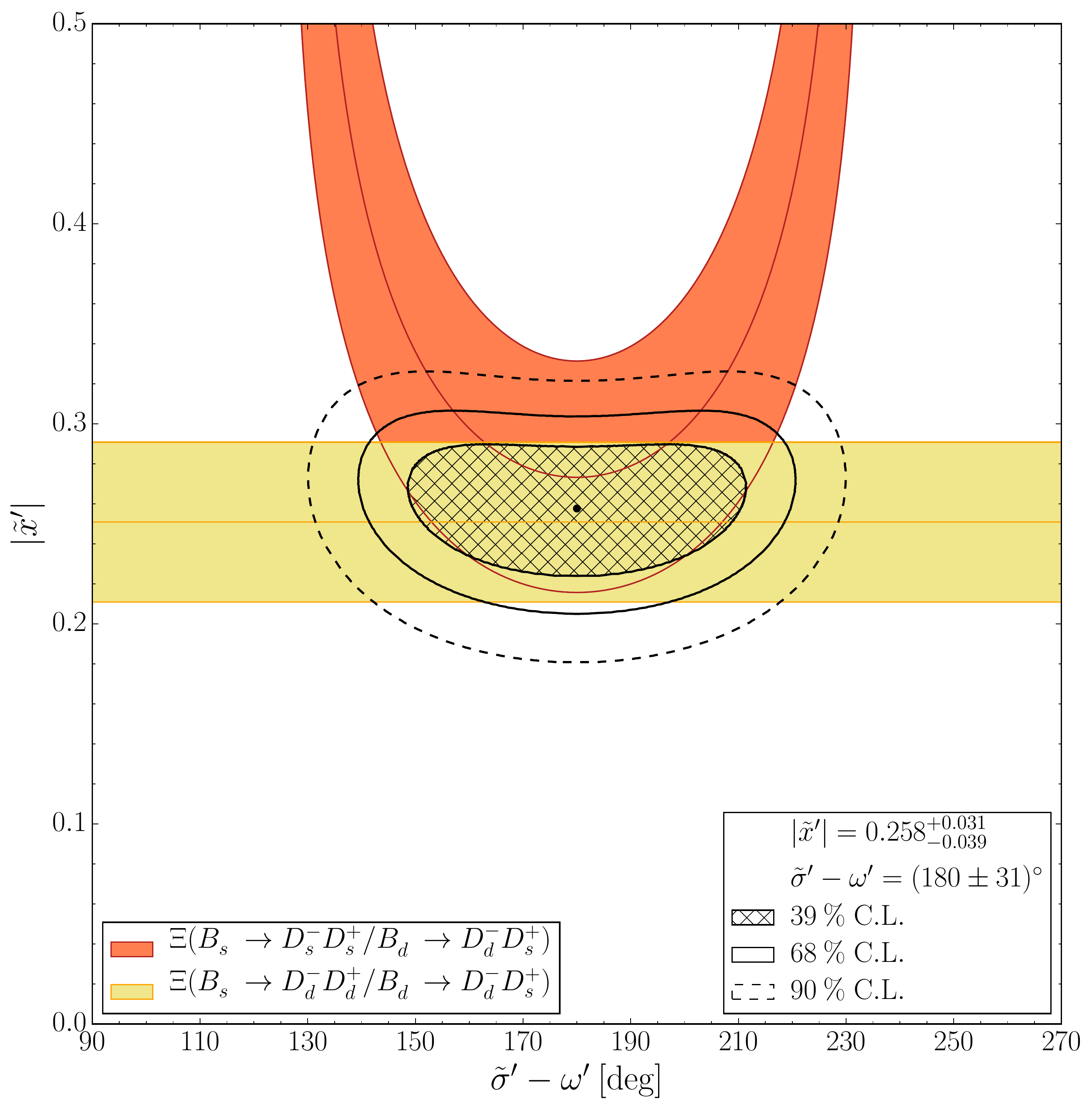}
\caption{Determination of the parameters $|\tilde x'|$ and $\tilde \sigma'$
introduced in Eq.~(\ref{xtilde-prime-def}) 
from a fit to Eqs.~(\ref{XiBsDsDs}) and (\ref{XiBsDD}), which characterise the currently 
available experimental data.
For the left plot, a value $\omega' = 0$ is assumed.}\label{fig:x-fit-p}
\end{figure}
\begin{figure}[tp]
 \centering
\includegraphics[width=0.49\textwidth]{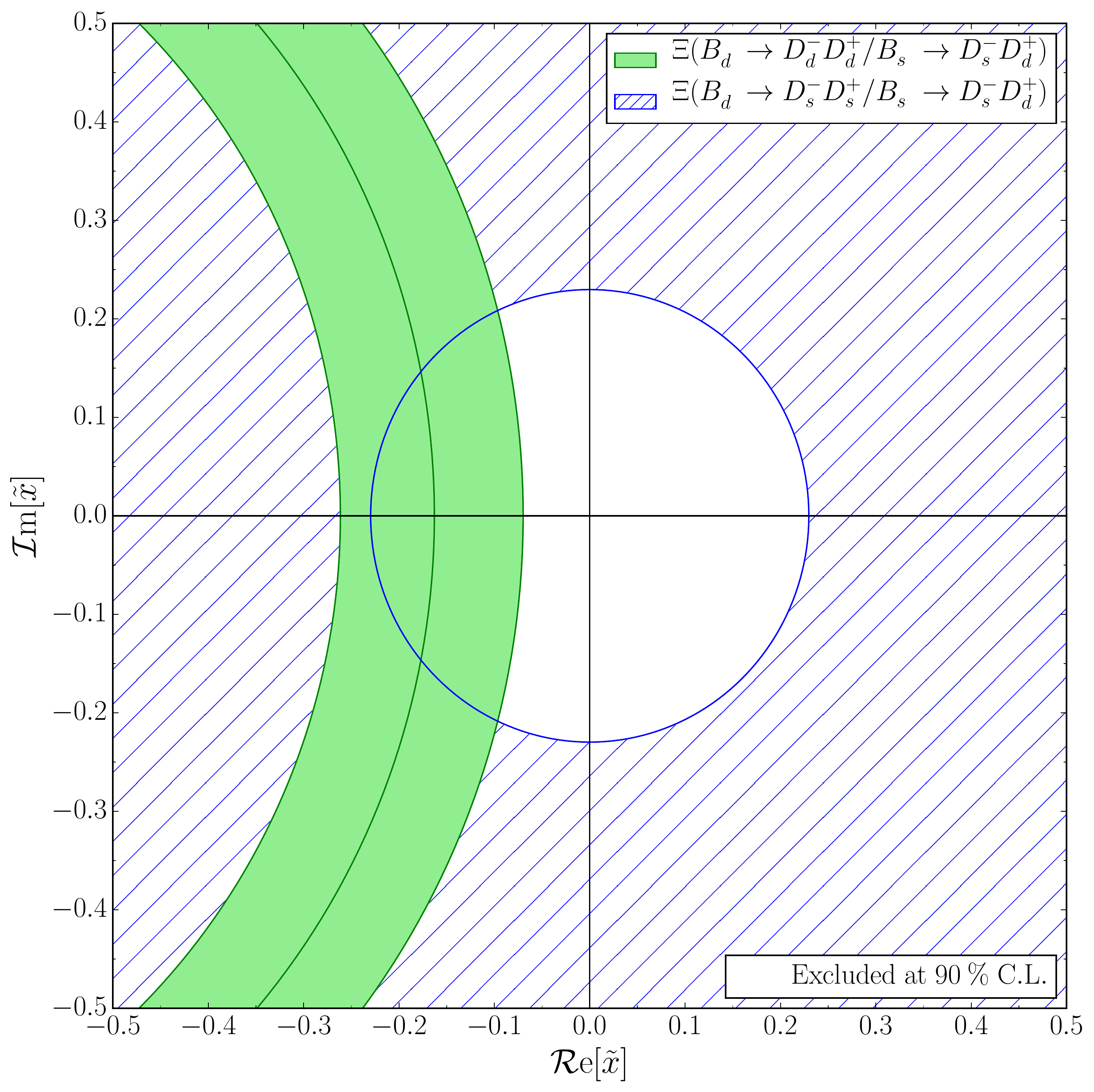}
\includegraphics[width=0.481\textwidth]{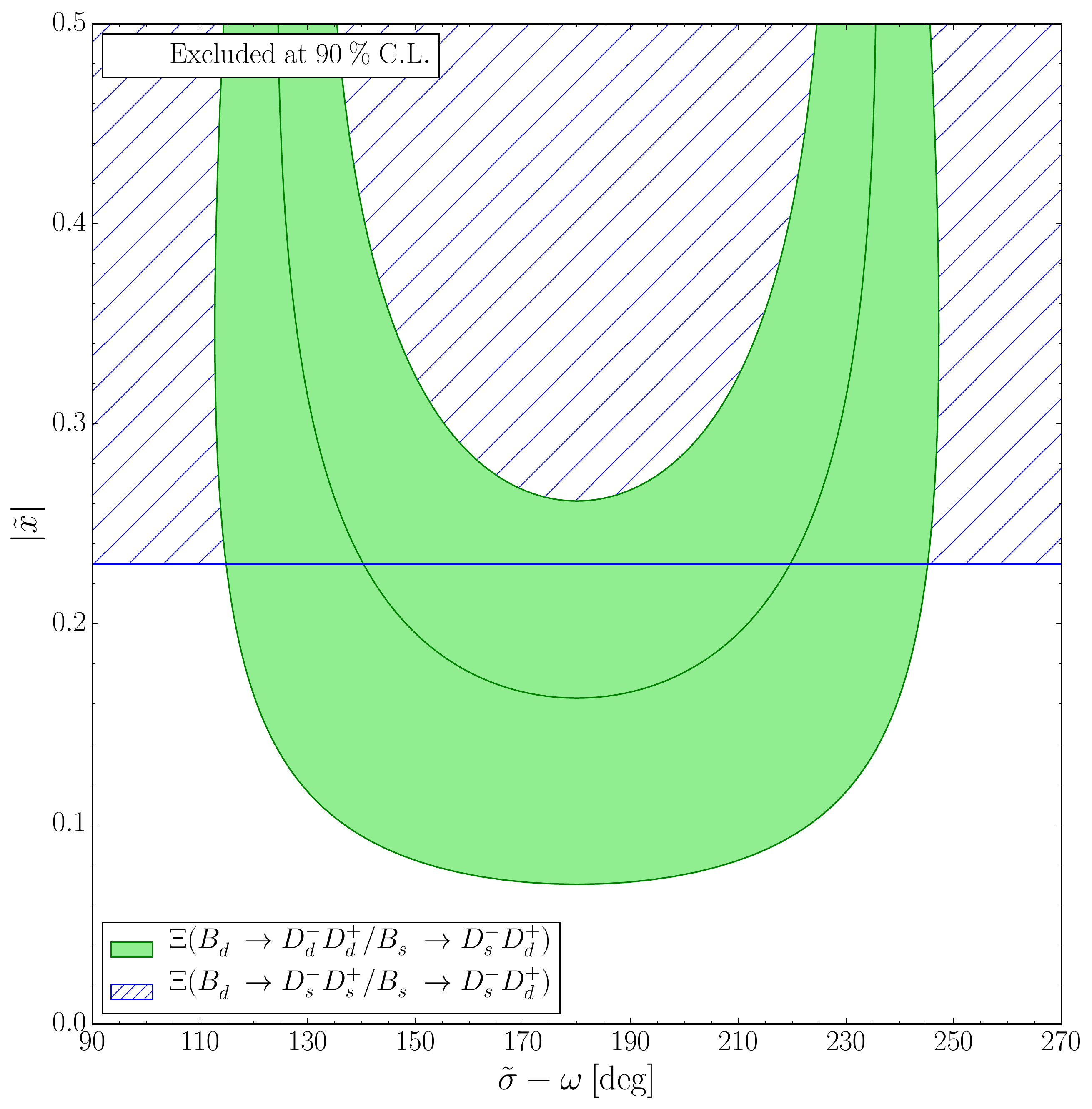}
\caption{Current experimental constraints, given in Eqs.~(\ref{XiBdDdDd}) and (\ref{XiBdDsDs}),
on the parameters $|\tilde x|$ and $\tilde \sigma$ introduced in 
Eq.~(\ref{xtilde-def}).
For the left plot, a value $\omega' = 0$ is assumed.}\label{fig:x-fit}
\end{figure}

Let us now convert the experimental results in Eqs.~(\ref{XiBsDsDs}) and (\ref{XiBsDD}) into constraints
on $|\tilde x'|$ and $\tilde \sigma'$. While the latter observable simply fixes a circle around the origin
in the complex plane, the former requires information about the $SU(3)$-breaking parameter 
$\varrho'$. We shall use the result in Eq.~(\ref{rho-fact}) for our numerical analysis as a guideline.
A fit yields 
\begin{equation}
\mathcal{R}\text{e}(\tilde x') = -0.258^{+0.039}_{-0.031}\:,\qquad \mathcal{I}\text{m}(\tilde x')= 0.0 \pm 0.14\:,
\end{equation}
or alternatively
\begin{equation}\label{tilde-xp-det}
\boxed{|\tilde x'| = 0.258^{+0.031}_{-0.039}\:,\qquad
\tilde\sigma'-\omega' = \left(180 \pm 34 \right)^{\circ}\:,}
\end{equation}
with the corresponding confidence-level contours shown in Fig.~\ref{fig:x-fit-p}. 

The exchange and penguin annihilation topologies play hence a surprisingly prominent role in the 
decays at hand, pointing towards large long-distance strong interaction effects. An example of such 
a contribution to the exchange topology is given by 
\begin{equation}
\Bs\to [\Dsm\Dsp, D_s^{*-}\Dsp, \ldots] \to \Ddm\Ddp\:,
\end{equation}
as illustrated in the right panel of Fig.~\ref{fig:rescat}. 

A similar analysis can be performed for the observables in Eqs.~(\ref{XiBdDdDd}) and
(\ref{XiBdDsDs}), which allow the determination of $|\tilde x|$ and $\tilde\sigma$. In contrast
to the determination of $|\tilde x'|$ and $\tilde\sigma'$, the penguin effects in the amplitude 
ratios do not enter with the tiny $\epsilon$ and lead to additional uncertainties. In Fig.~\ref{fig:x-fit},
we show the constraints from the current data, which are still pretty weak. Here we may have
long-distance rescattering contributions from processes of the kind
\begin{equation}
\Bd\to [\Ddm\Ddp, D_d^{*-}\Ddp, \ldots] \to \Dsm\Dsp\:.
\end{equation}
In the future, following these lines, the comparison between the values of $\tilde x'$ and $\tilde x$ 
will offer yet another test of the relation in Eq.~(\ref{xtilde-rel}), going beyond Eq.~(\ref{Xi-xtilde-rel}) 
through information on the strong phases.

\subsection{Global Analysis of the Penguin Parameters}\label{ssec:global}
\subsubsection{Information from Branching Ratios and Non-factorisable Effects}
Let us now use the currently available data to constrain the penguin parameters. Unfortunately, 
a measurement of the differential semileptonic \BsSL rate is
not available. Consequently, we may not yet apply Eq.~(\ref{H-R}) and have to follow a different 
avenue, involving larger theoretical uncertainties. In analogy to the $H$ observable 
defined in Eq.~(\ref{Eq:Hobs_Def}), we introduce the following quantities:
\begin{align}
\tilde H  & \equiv  \frac{1}{\epsilon} \left|\frac{\mathcal{\tilde A}'}{\mathcal{\tilde A}}\right|^2
\left[\frac{m_{\uBs}}{m_{\uBd}} 
\frac{\Phi(m_{\Dd}/m_{\uBd},m_{\Ds}/m_{\uBd})}{\Phi(m_{\Ds}/m_{\uBs},m_{\Dd}/m_{\uBs})}  
 \frac{\tau_{\uBd}}{\tau_{\uBs}} \right]
\frac{\mathcal{B}\left(\uBstoDsDd\right)_{\text{theo}}}
{\mathcal{B}\left(\uBdtoDdDs\right)}\:,\\
& =  \frac{1-2\:\tilde a\cos\tilde\theta\cos\gamma+\tilde a^2}{1+2\epsilon \tilde a'\cos\tilde\theta'\cos\gamma
+\epsilon^2 \tilde a^{\prime2}}\:,\label{Eq:Hobs_Def-tilde}\\
H_{\rm c}  & \equiv  \frac{1}{\epsilon} \left|\frac{\mathcal{\tilde A}_{\rm c}'}{\mathcal{\tilde A}_{\rm c}}\right|^2
\left[\frac{\Phi(m_{\Du}/m_{\uBu},m_{\Ds}/m_{\uBu})}{\Phi(m_{\Du}/m_{\uBu},m_{\Dd}/m_{\uBu})}  
  \right]
\frac{\mathcal{B}\left(\uButoDzDd\right)}
{\mathcal{B}\left(\uButoDzDs\right)}\:,\\
& =  \frac{1-2\:\tilde a_{\rm c}\cos\tilde\theta_{\rm c}\cos\gamma+\tilde a_{\rm c}^2}{1+2\epsilon 
\tilde a'_{\rm c}\cos\tilde\theta_{\rm c}'\cos\gamma+\epsilon^2 \tilde a_{\rm c}^{\prime2}}\:.\label{Eq:Hobs_Def-c}
\end{align}

If we complement $\tilde H$ with direct CP violation in $\BstoDsDd$
(and the suppressed CP asymmetry in $\BdtoDdDs$)
and $H_{\rm c}$ with direct CP violation in $\uButoDzDd$ 
(and the suppressed CP asymmetry in $\uButoDzDs$), we may determine
the penguin parameters $(\tilde a,\tilde\theta)$ and $(\tilde a_{\rm c}, \tilde\theta_{\rm c})$, respectively.
These determinations are analogous to the determination of $(a,\theta)$ from $H$ and the direct
CP asymmetry in $\BdtoDdDd$ (and the suppressed CP asymmetry in 
$\BstoDsDs$). The hence determined parameters offer insights into
the $r_{A}$ and $r_{PA}$ parameters introduced in Eqs.~(\ref{rA-def}) and (\ref{rPA-def}), respectively,
and allow for a comparison of the relative non-factorisable contributions.

In contrast to the measurements of the CP asymmetries, the extraction of the $H$, $\tilde H$ and $H_{\rm c}$ 
observables from the data requires knowledge of the following amplitude ratios:
\begin{align}
\frac{\mathcal{A}'}{\mathcal{A}} = \frac{T'+P^{(ct)'}+E'+PA^{(ct)'}}{T+P^{(ct)}+E+PA^{(ct)}}
& =\left[\frac{T'}{T}\right]\left[\frac{1+P^{(ct)'}/T'}{1+P^{(ct)}/T}\right]\left[\frac{1+x'}{1+x}\right]
\approx \left[\frac{1+x'}{1+x}\right] \frac{T'}{T}\:,\label{A-rat-1}\\
\frac{\mathcal{\tilde A}'}{\mathcal{\tilde A}} = \frac{\tilde T'+\tilde P^{(ct)'}}{\tilde T+\tilde P^{(ct)}}
& =\left[\frac{\tilde T'}{\tilde T}\right]\left[\frac{1+\tilde P^{(ct)'}/\tilde T'}{1+\tilde P^{(ct)}/\tilde T}\right]
\approx \frac{\tilde T'}{\tilde T}\:,\label{A-rat-2}\\
\frac{\mathcal{\tilde A}'_{\rm c}}{\mathcal{\tilde A}_{\rm c}} = \frac{\tilde T'_{\rm c}+
\tilde P_{\rm c}^{(ct)'}}{\tilde T+\tilde P^{(ct)}}
& =\left[\frac{\tilde T'_{\rm c}}{\tilde T_{\rm c}}\right]\left[\frac{1+\tilde P^{(ct)'}_{\rm c}/\tilde T'_{\rm c}}{1+
\tilde P^{(ct)}_{\rm c}/\tilde T_{\rm c}}\right] \approx \frac{\tilde T'_{\rm c}}{\tilde T_{\rm c}}\:.\label{A-rat-3}
\end{align}
These quantities are governed by $U$-spin-breaking effects in the ratio of the colour-allowed tree 
contributions, which we may write as
\begin{align}
\left|\frac{T'}{T}\right|=\left[\frac{m_{\uBs}^2-m_{\Ds}^2}{m_{\uBd}^2-m_{\Dd}^2}  \right]
&\left[\frac{f_{\Ds}}{f_{\Dd}}\right]\left[\frac{F_0^{\uBs\Ds}(m_{\Ds}^2)}{F_0^{\uBd\Dd}(m_{\Dd}^2)}\right]
\left[\frac{a^{T'}_{\rm NF}}{a^{T}_{\rm NF}}\right]\:,\label{Eq:TprimeT}\\
\left|\frac{\tilde T'}{\tilde T}\right|=\left[\frac{m_{\uBd}^2-m_{\Dd}^2}{m_{\uBs}^2-m_{\Ds}^2}  \right]
&\left[\frac{f_{\Ds}}{f_{\Dd}}\right]\left[\frac{F_0^{\uBd\Dd}(m_{\Ds}^2)}{F_0^{\uBs\Ds}(m_{\Dd}^2)}\right]
\left[\frac{\tilde a^{T'}_{\rm NF}}{\tilde a^{T}_{\rm NF}}\right]\:,\\
\left|\frac{\tilde T'_{\rm c}}{\tilde T_{\rm c}}\right|=
&\left[\frac{f_{\Ds}}{f_{\Dd}}\right]\left[\frac{\tilde a^{T'}_{\rm NF, c}}{\tilde a^{T}_{\rm NF, c}}\right]\:,\label{Tc-ratio}
\end{align}
where the parameters $a^{T}_{\rm NF}$ describe non-factorisable contributions affecting the
colour-allowed tree amplitude (see Eq.~(\ref{aNF-T})). If we assume that
all the $a^{T}_{\rm NF}$ parameters are equal to one another due to the $SU(3)$ flavour symmetry, 
the following relation can be derived:
\begin{equation}
\Biggl|\frac{T'}{T}\Biggr|\left|\frac{\tilde T'}{\tilde T}\right|=\left|\frac{\tilde T'_{\rm c}}{\tilde T_{\rm c}}\right|^2\:,
\end{equation}
where the decay constants and form factors cancel. In terms of branching ratios, using
Eqs.~(\ref{A-rat-1})--(\ref{A-rat-3}), this relation implies
\begin{equation}\label{G-def}
{\cal G}\equiv\frac{\mathcal{B}\left(\uButoDzDd\right)}{\mathcal{B}\left(\uButoDzDs\right)}
 \sqrt{\left[\frac{\mathcal{B}\left(\uBstoDsDs\right)}{\mathcal{B}
\left(\uBstoDsDd\right)} \right]\left[\frac{\mathcal{B}\left(\uBdtoDdDs\right)}{\mathcal{B}\left(\uBdtoDdDd\right)}\right]}\approx1\:.
\end{equation} 
The current data give
\begin{equation}
{\cal G} = 0.85 \pm 0.16\:,
\end{equation}
which is consistent with Eq.~(\ref{G-def}) within the uncertainties. 

Using data for the semileptonic \BdSL decay, the 
non-factorisable effects can be probed through
\begin{align}
&\tilde R_{\Dd}\equiv \frac{\Gamma(\BdtoDdDs)}{[{\rm d} 
\Gamma(\BdSL)/{\rm d} q^2]|_{q^2=m_{\Dq}^2}}\notag\\
&\quad=6\pi^2 |V_{cs}|^2  f_{\Ds}^2  X_{\uBd\Dd}^{\Ds} |\tilde a_{\rm NF}'|^2 \left[1+2 \, \epsilon  \tilde a'  \cos 
\tilde \theta' \cos\gamma+\epsilon^2 \tilde a'^2 \right]\:,\label{Rdtilde-deff}
\end{align}
where
\begin{equation}
 X_{\uBd\Dd}^{\Ds}= \frac{(m_{\uBd}^2-m_{\Dd}^2)^2}{\left[m_{\uBd}^2-(m_{\Dd}+m_{\Ds})^2\right]
\left[m_{\uBd}^2-(m_{\Dd}-m_{\Ds})^2\right]}
\left[ \frac{F_0^{\uBd\Dd}(m_{\Ds}^2)}{F_1^{\uBd\Dd}(m_{\Ds}^2)} \right]^2\:,
\end{equation}
and
\begin{equation}
\tilde a_{\rm NF}'=\tilde a_{\rm NF}^{T'}\left[1+\tilde r_P'  \right]\:,
\end{equation}
with
\begin{equation}
\tilde r_P'\equiv \frac{\tilde P^{(ct)'}}{\tilde T'}\:,
\end{equation}
defined in analogy to Eq.~(\ref{rP-def}). Experimentally, we find
\begin{equation}
    \tilde R_{\Dd} = (2.90 \pm 0.41)~ \text{GeV}^2\:,
\end{equation}
corresponding to a non-factorisable contribution
\begin{equation}\label{aNF-det}
\boxed{|\tilde a_{\rm NF}'| = |\tilde a_{\rm NF}^{T'}|\left|1+\tilde r_P'  \right|=0.756 \pm 0.085\:,}
\end{equation}
where Eq.~\eqref{Eq:FF_HQlat} has been used for the ratio of form factors. In this result,
the penguin effects suppressed by $\epsilon$ in Eq.~(\ref{Rdtilde-deff}) were neglected. It is plausible
to interpret the deviation from one at the $2.9\,\sigma$ level as footprints of sizeable penguin effects.
We shall discuss this parameter below.

\begin{figure}[tp]
\centering
\includegraphics{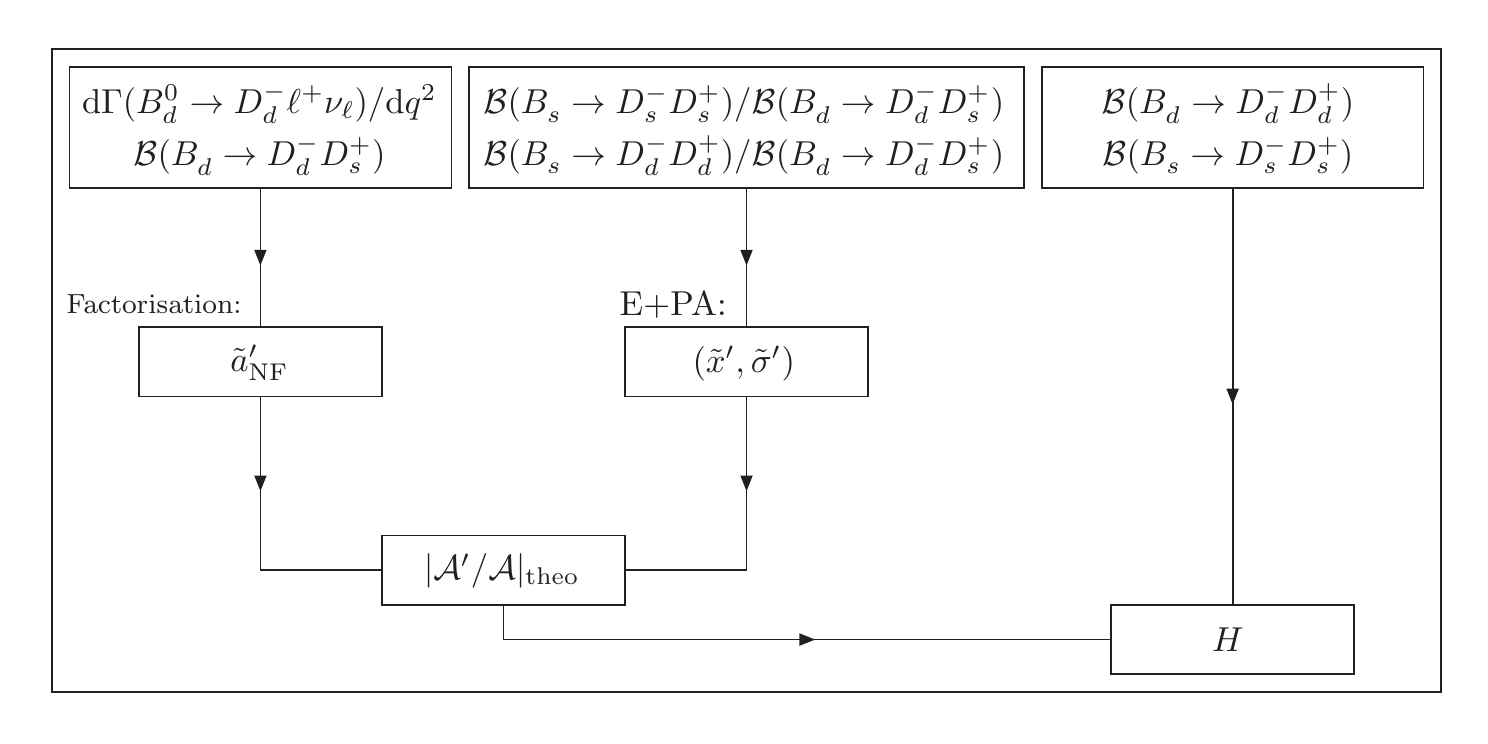}
\caption{Flow chart illustrating the \emph{classic} strategy to determine $H$ using data from \BtoDD branching ratio measurements.}
\label{Fig:Flow_Hobs_Orig}
\end{figure}

The ratios in Eqs.~(\ref{A-rat-1})--(\ref{A-rat-3}) can be written in the following forms:
\begin{align}
\left|\frac{\mathcal{A}'}{\mathcal{A}}\right| 
& =\left|\frac{1+x'}{1+x}\right|\left[\frac{a_{\rm NF}^{(0)'}}{a_{\rm NF}^{(0)}}\right]\left|\frac{T'}{T}\right|_{\rm fact}\:,\label{A-rat-1-f}\\
\left|\frac{\mathcal{\tilde A}'}{\mathcal{\tilde A}}\right| & = 
\left[\frac{\tilde a_{\rm NF}'}{\tilde a_{\rm NF}}\right]\left|\frac{\tilde T'}{\tilde T}\right|_{\rm fact}\:,\label{A-rat-2-f}\\
\left|\frac{\mathcal{\tilde A}'_{\rm c}}{\mathcal{\tilde A}_{\rm c}} \right| & = 
\left[\frac{\tilde a_{\rm NF, c}'}{\tilde a_{\rm NF, c}}\right]\left|\frac{\tilde T'_{\rm c}}{\tilde T_{\rm c}}
\right|_{\rm fact}\:,\label{A-rat-3-f}
\end{align}
which is also graphically illustrated in Fig.~\ref{Fig:Flow_Hobs_Orig}.
In the above equation, $a_{\rm NF}^{(0)}$ differs from $a_{\rm NF}$ introduced in Eq.~(\ref{aNF-def-0}) 
through the $(1+x)$ term, 
to account for the contributions from exchange and penguin annihilation topologies,
which are absent in the decays of the other two ratios.
We thus have
\begin{equation}
a_{\rm NF}=(1+x)a_{\rm NF}^{(0)}\:.
\end{equation}

As in the discussion in Subsection~\ref{ssec:semilept}, it is convenient to write
\begin{equation}\label{Eq:ANF_relations}
|\tilde a_{\rm NF}'|=1+\tilde \Delta_{\rm NF}'\:, 
\qquad |\tilde a_{\rm NF}|=1+\tilde \Delta_{\rm NF}'[1-\tilde \xi_{SU(3)}]\:,
\end{equation}
where the parameter $\tilde \xi_{SU(3)}$ describes $SU(3)$-breaking effects in the 
non-factorisable contributions, yielding
\begin{equation}
\left|\frac{\tilde a_{\rm NF}'}{\tilde a_{\rm NF}}\right|=\frac{1+\tilde \Delta_{\rm NF}'}{1+\tilde \Delta_{\rm NF}}=
1+\tilde \xi_{SU(3)}\tilde \Delta_{\rm NF}'+{\cal O}(\tilde \Delta_{\rm NF}'^2)\:.
\end{equation}
Consequently, $\tilde \Delta_{\rm NF}' = -0.244 \pm 0.085$, as determined from 
Eq.~(\ref{aNF-det}), allows us to take non-factorisable corrections to 
Eqs.~(\ref{A-rat-1-f})--(\ref{A-rat-3-f}) into account, assuming
\begin{equation}\label{NF-rel-spec}
\boxed{\left|\frac{a_{\rm NF}^{(0)'}}{a_{\rm NF}^{(0)}}\right|=
\left|\frac{\tilde a_{\rm NF}'}{\tilde a_{\rm NF}}\right|=\left|\frac{\tilde a_{\rm NF, c}'}{\tilde a_{\rm NF, c}}\right|\:.}
\end{equation}
In these relations, $SU(3)$-breaking effects enter only through different spectator quarks and are
expected to be small.

In order to determine the $SU(3)$-breaking effects in the ratio of the colour-allowed 
tree amplitudes $|T'/T|$, we use again Eq.~(\ref{HQ-FF}) to derive the following 
expression \cite{RF-psiK}:
\begin{equation}
\left|\frac{T'}{T}\right|_{\rm fact} =\left[\frac{m_{\uBs}^2-m_{\Ds}^2}{m_{\uBd}^2-m_{\Dd}^2}\right]
\left[\frac{f_{\Ds}}{f_{\Dd}}\right]\left[\frac{F_0^{\uBs\Ds}(m_{\Ds}^2)}{F_0^{\uBd\Dd}(m_{\Dd}^2)}\right]
= \rho'_{\text{fact}}\left[\frac{f_{\Ds}}{f_{\Dd}}\right] = 1.356 \pm 0.076\:.
\end{equation}
For the calculation of the numerical value, we have used Eq.~\eqref{Eq:FFlattice} and the values of the 
decay constants in Eq.~\eqref{fD-exp}.
In analogy, we get
\begin{equation}\label{Eq:Tratio_tilde}
\left|\frac{\tilde T'}{\tilde T}\right|_{\rm fact}=\left[\frac{m_{\uBd}^2-m_{\Dd}^2}{m_{\uBs}^2-m_{\Ds}^2}\right]
\left[\frac{f_{\Ds}}{f_{\Dd}}\right]\left[\frac{F_0^{\uBd\Dd}(m_{\Ds}^2)}{F_0^{\uBs\Ds}(m_{\Dd}^2)}\right]
= \rho_{\text{fact}}\left[\frac{f_{\Ds}}{f_{\Dd}}\right] = 1.167 \pm 0.066\:.
\end{equation}
In the case of the charged decays, a particularly 
simple situation arises in factorisation as the form factors cancel:
\begin{equation}
\left|\frac{\tilde T'_{\rm c}}{\tilde T_{\rm c}}\right|_{\rm fact}=\frac{f_{\Ds}}{f_{\Dd}}\:,
\end{equation}
as is also evident from Eq.~(\ref{Tc-ratio}). 

Using the amplitude ratio in Eq.~(\ref{A-rat-1-f}), we can now determine the observable $H$ 
(see Eq.~\eqref{Eq:Hobs_Def}). For this, we also have to quantify the uncertainty from the $SU(3)$-breaking 
corrections to the term involving the $x$ and $x'$ parameters. 
The analysis discussed in Subsection~\ref{ssec:EPA} allows us 
to accomplish this task. Using the results for $\tilde x'$ and $\tilde\sigma'$ in Eq.~(\ref{tilde-xp-det}) 
and the relations
\begin{equation}
x'=\frac{\tilde x'}{\varrho'}\:, \qquad x\approx
\left(\frac{m_{\uBs}m_{\Ds}}{m_{\uBd}m_{\Dd}} \right)\left(\frac{f_{\uBd}f_{\Dd}}{f_{\uBs}f_{\Ds}} \right)\tilde x'\:,
\end{equation}
yields
\begin{equation}
\left|\frac{1+x'}{1+x}\right| = 0.930 \pm 0.020\:.
\end{equation}

\begin{figure}[tp]
 \centering
\includegraphics[height=0.4\textwidth]{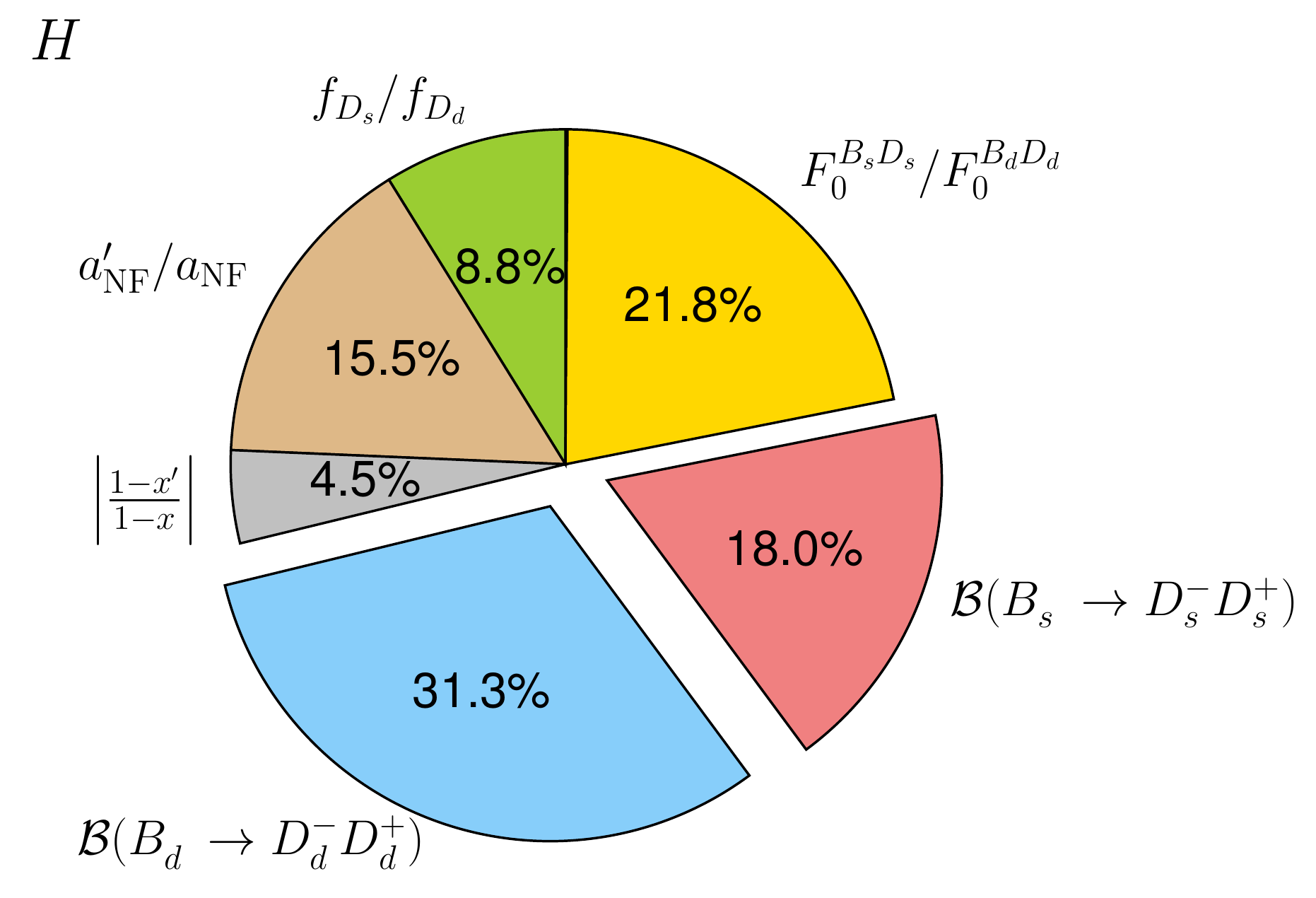}
\\~\\
\includegraphics[height=0.4\textwidth]{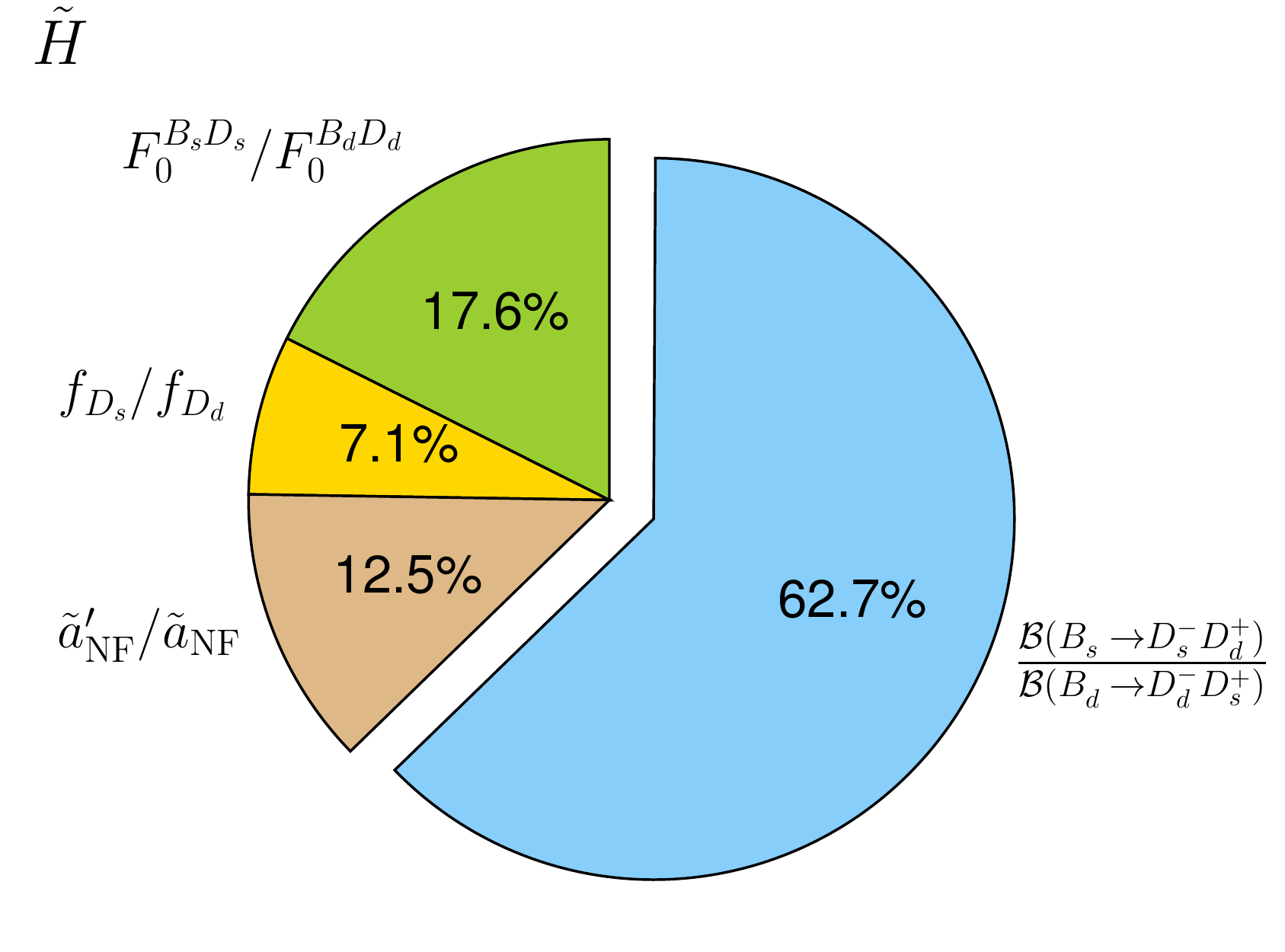}
\\~\\
\includegraphics[height=0.4\textwidth]{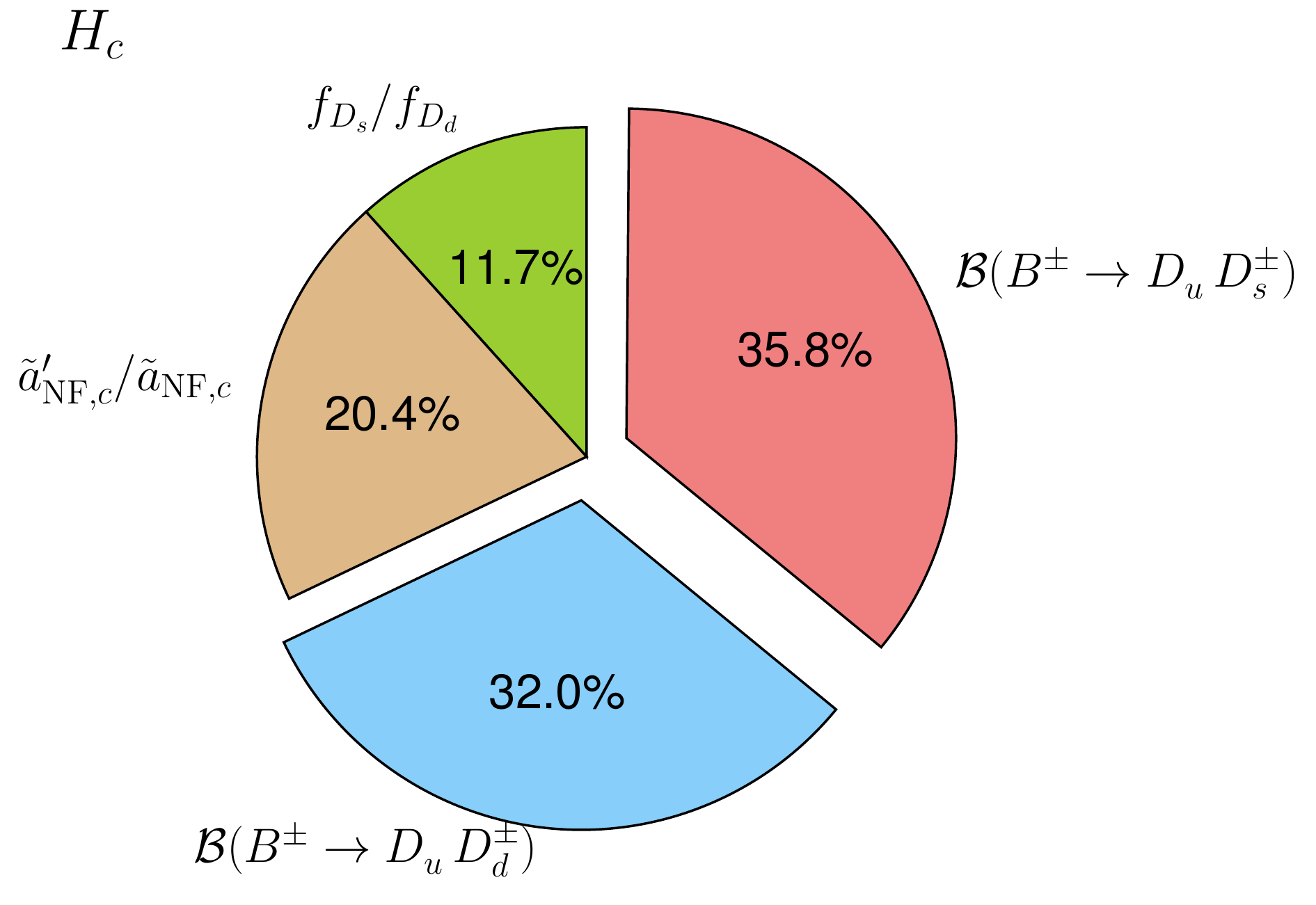}
\caption{Pie charts illustrating the uncertainty budget of the $H$ observables.}
\label{Fig:ErrorBudget_Hobs}
\end{figure}

Finally, we obtain
\begin{align}
\left|\frac{\mathcal{A}'}{\mathcal{A}}\right| & = 1.261 \pm 0.091\:,\label{Acal-rat-theo}\\
\left|\frac{\mathcal{\tilde A}'}{\mathcal{\tilde A}}\right| & = 1.167 \pm 0.081\:,\\
\left|\frac{\mathcal{\tilde A}'_{\rm c}}{\mathcal{\tilde A}_{\rm c}} \right| & = 1.258 \pm 0.063\:,
\end{align}
and correspondingly
\begin{align}
H & = 1.30 \pm 0.26\:,     \label{H-value}\\
\tilde{H} & = 1.28 \pm 0.29\label{Htilde-value}\:,\\
H_{\rm c} & = 1.18 \pm 0.21\label{Hc-value}\:.
\end{align}
In Fig.~\ref{Fig:ErrorBudget_Hobs} we show the corresponding uncertainty budgets. Should the
relations in Eq.~(\ref{NF-rel-spec}) actually receive corrections, they would affect this error budget. 
In the future, implementing the strategy proposed in Section~\ref{ssec:semilept}, this 
approximation/assumption is no longer needed.

The result on $H$ allows us to put first constraints on the penguin parameters $a$
with the help of the lower bound in Eq.~(\ref{a-bound}):
\begin{equation}\label{a-bounds}
\boxed{a\geq 0.052 \:,}
\end{equation}
where we have used the lower 
value of $H$ at one standard deviation.

\subsubsection{Information from CP Asymmetries}
\begin{figure}[tp]
 \centering
\includegraphics[width=0.49\textwidth]{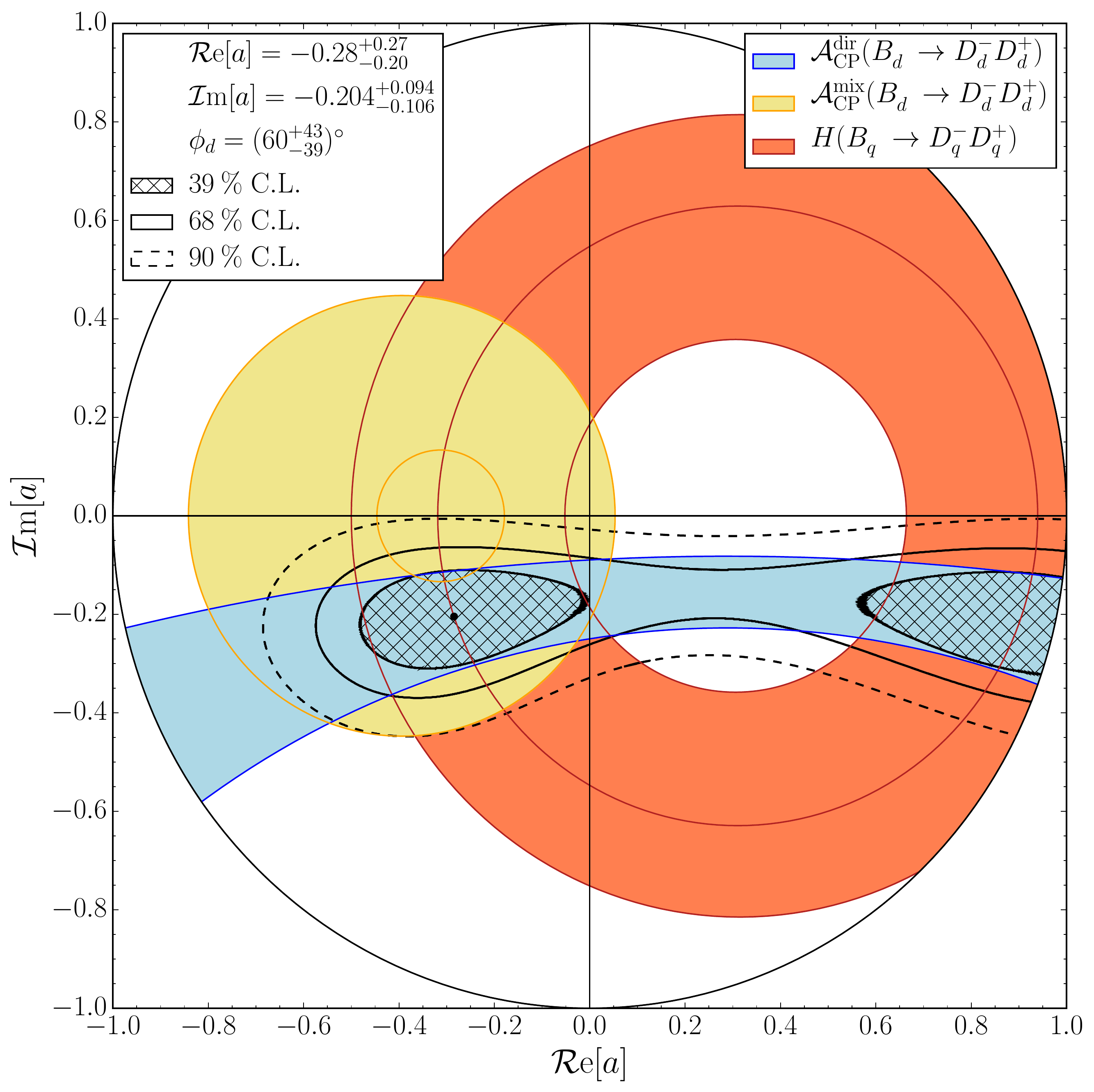}
\includegraphics[width=0.481\textwidth]{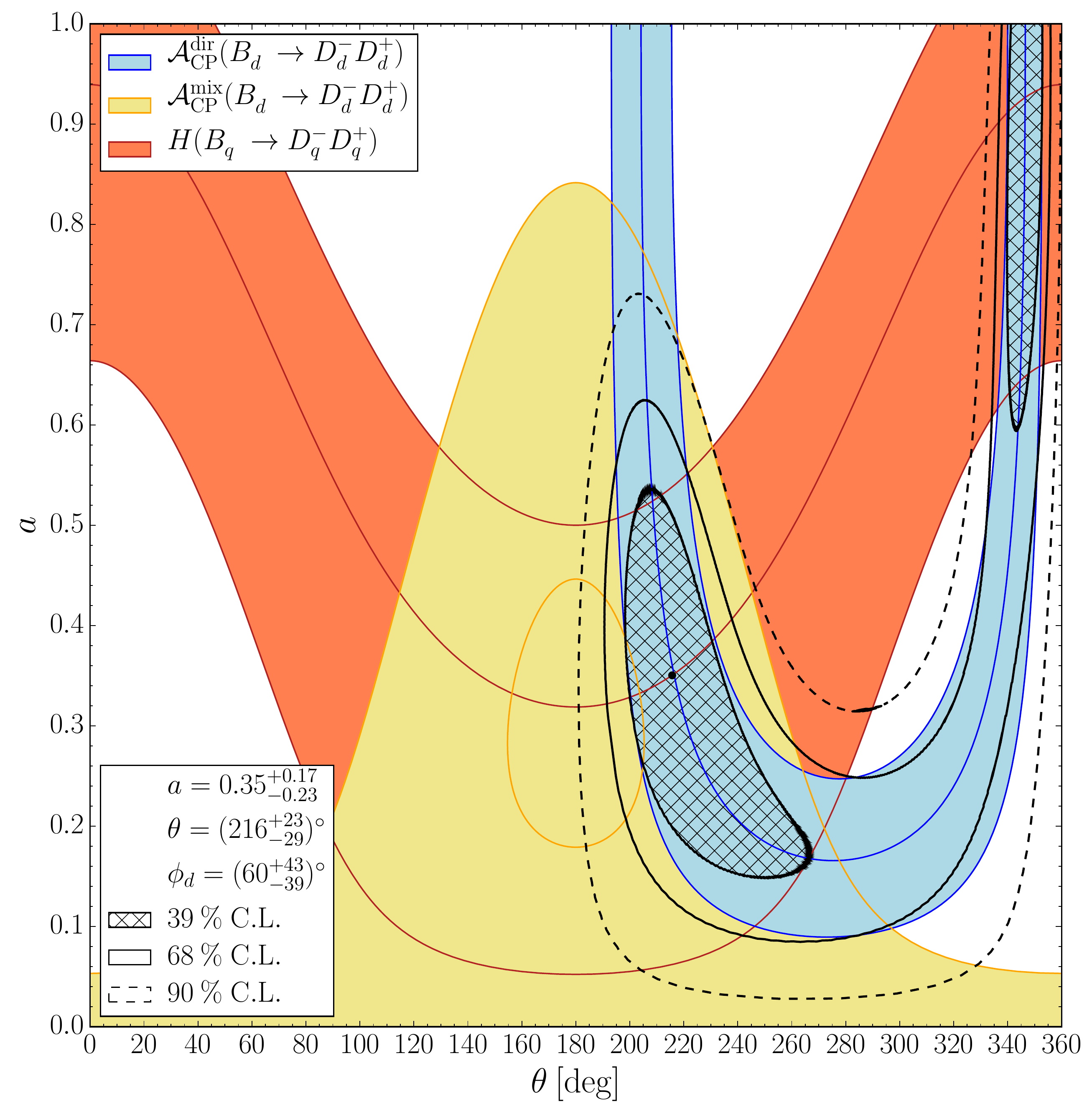}
\caption{Illustration of the determination of the penguin parameters $a$ and $\theta$ from a $\chi^2$ 
fit to the CP asymmetries of the decay \BdtoDdDd and the observable $H$.}
\label{fig:a-fit-neut}
\end{figure}

Let us now add experimental information on CP violation to our analysis. 
Concerning the decay $\BdtoDdDd$, the current status of the measurement 
of the direct and mixing-induced CP asymmetries is given as follows:
\begin{equation}\label{ACP-dir-BDD-ex}
{\cal A}_{\rm CP}^{\rm dir}(\uBdtoDdDd)=\left\{
\begin{array}{ll}
-0.07\pm0.23\pm0.03 & \mbox{(BaBar \cite{BaBar-BDD})}\\
-0.43\pm0.16\pm0.05 & \mbox{(Belle \cite{Belle-BDD})}\:,
\end{array}\right.
\end{equation}
\begin{equation}\label{ACP-mix-BDD-ex}
{\cal A}_{\rm CP}^{\rm mix}(\uBdtoDdDd)=\left\{
\begin{array}{ll}
+0.63\pm0.36\pm0.05 & \mbox{(BaBar \cite{BaBar-BDD})}\\
+1.06^{+0.14}_{-0.21}\pm0.08 & \mbox{(Belle \cite{Belle-BDD})}\:.
\end{array}\right.
\end{equation}
The measurements by the BaBar and Belle collaborations are not in good agreement with one another, 
in particular for the mixing-induced CP asymmetry. HFAG gives the following averages \cite{Amhis:2014hma}:
\begin{equation}\label{CP-HFAG}
{\cal A}_{\rm CP}^{\rm dir}(\uBdtoDdDd)=-0.31 \pm 0.14\:, \qquad
{\cal A}_{\rm CP}^{\rm mix}(\uBdtoDdDd)=0.98 \pm 0.17\:,
\end{equation}
which have to be taken with great care. It is nevertheless interesting to use these results as input
for the strategy discussed above.  A $\chi^2$ fit to Eq.~(\ref{CP-HFAG})  and the value of 
$H$ in Eq.~(\ref{H-value}) yields 
$\chi^2_{\text{min}} = 0.028$ for 4 degrees of freedom $(a,\theta, \phi_d,\gamma)$, and results 
in the solution
\begin{equation}
\mathcal{R}\text{e}[a] = -0.29^{+0.27}_{-0.20} \:,\qquad \mathcal{I}\text{m}[a] = -0.204^{+0.094}_{-0.105}\:,
\end{equation}
corresponding to
\begin{equation}\label{Eq:Penguin_chi2fit}
\boxed{a = 0.35^{+0.19}_{-0.20} \:,\qquad \theta = \left(215^{+51}_{-17}\right)^{\circ}\:,}
\end{equation}
and
\begin{equation}\label{Eq:phid_chi2fit}
\boxed{\phi_d = \left(60^{+43}_{-39}\right)^{\circ}\:.}
\end{equation}
Following Refs.~\cite{FFJM,DeBF-pen}, we illustrate the various constraints entering the fit 
through contour bands of the individual observables in Fig.~\ref{fig:a-fit-neut}. For the 
${\cal A}_{\rm CP}^{\rm mix}(\uBdtoDdDd)$ range, we have used the value of 
$\phi_d$ in Eq.~\eqref{Eq:phid_chi2fit}.
The penguin parameters in Eq.~\eqref{Eq:Penguin_chi2fit} result in the 
penguin phase shift
\begin{equation}
\Delta\phi_d^{\Ddm\Ddp} = \left(30^{+23}_{-32}\right)^{\circ}\:.
\end{equation}
For the fit, we use the expressions for the CP asymmetries in Eqs.~(\ref{AD}) and (\ref{AM}), and
the expression for $H$ in Eq.~(\ref{H-exprs}) with the $U$-spin relation in Eq.~(\ref{a-theta-rel}). 
Since $a'$ enters Eq.~(\ref{H-exprs}) with the tiny $\epsilon$, $U$-spin-breaking corrections to
Eq.~(\ref{a-theta-rel}) have a very minor impact.

Although the uncertainties are large, the results  from the fit may indicate significant penguin contributions.
Should this actually be the case, long-distance effects, such as those illustrated in Fig.~\ref{fig:rescat}, 
would be at work in the penguin sector of the $\BqtoDqDq$ decays. It is interesting
to have a closer look at the result of non-factorisable contributions to \uBdtoDdDs in Eq.~(\ref{aNF-det}).
It corresponds to
\begin{equation}\label{aNF-det-1}
\left|1+\tilde r_P'  \right|=(0.750 \pm 0.055)/|\tilde a_{\rm NF}^{T'}|\:.
\end{equation}
Assuming that $|\tilde a_{\rm NF}^{T'}|= 1$, i.e.\ that the colour-allowed tree topologies have negligible non-factorisable contributions, and that all parameters are real, this results in
\begin{equation}
\tilde r_P' = -0.250 \pm 0.055 \:.
\end{equation}
Note that the uncertainty on $\tilde r_P'$ only reflects the uncertainty on the input quantity \eqref{aNF-det}, but does not take into account further theoretical uncertainties associated with the made approximation.
Further assuming $\tilde P^{(ut)'} = \tilde P^{(ct)'}$, which leads to the approximation $ae^{i\theta} \approx R_b\, \tilde r_P'/(1+\tilde r_P')$, we obtain
\begin{equation}
a = 0.130 \pm 0.039\:, \qquad \theta \sim 180^\circ\:,
\end{equation}
which is in the ballpark of the theoretical estimate in Eq.~\eqref{BSS-estimate}.
Consequently, it is still premature to draw definite conclusions on anomalously enhanced penguin contributions at this point and future analyses are required to shed light on this issue. 

Large penguins would have important consequences for the determination of the 
\Bs--\Bsb mixing phase $\phi_s$ from the mixing-induced CP violation 
of the \BstoDsDs decay.  The LHCb collaboration has recently presented 
the first analysis of CP violation in this channel \cite{LHCb-BsDsDs}, yielding
\begin{equation}\label{Eq:phis_eff_BsDsDs}
\phi_{s,\Dsm\Dsp}^{\text{eff}} = \left(1.1 \pm 9.7 \pm 1.1\right)^{\circ}\:,\qquad
\left|\lambda_{\Dsm\Dsp}\right| = 0.91 \pm 0.18 \pm 0.02\:,
\end{equation}
which can be converted into
\begin{equation}
{\cal A}_{\rm CP}^{\rm dir}(\uBstoDsDs)= 0.09 \pm 0.20\:, \qquad
{\cal A}_{\rm CP}^{\rm mix}(\uBstoDsDs)=0.02 \pm 0.17\:.
\end{equation}
These results are in good agreement with the predictions based on a global $SU(3)$ analysis of the \BtoDD system \cite{Jung:2014jfa}.
Moreover, we obtain
\begin{equation}
\mathcal{A}_{\Delta\Gamma}(\uBstoDsDs) = -0.995 \pm 0.019\:,
\end{equation}
which should be compared with Eq.~(\ref{ADG-exp}) corresponding to the direct 
measurement of the effective lifetime $\tau_{\Dsm\Dsp}^{\rm eff}$.

If we generalise the $U$-spin relation in (\ref{a-theta-rel}) as
\begin{equation}\label{a-theta-rel-U-break}
a'= \xi a\:, \qquad \theta'=\theta +\delta\:,
\end{equation}
with the assumed $U$-spin-breaking parameters $\xi=1.00\pm0.20$ and $\delta=(0\pm20)^\circ$ 
(which are of similar size as the corresponding parameters in Ref.~\cite{DeBF-pen}) and use the 
expression in Eq.~(\ref{tan-phid}), the penguin parameters in Eq.~(\ref{Eq:Penguin_chi2fit}) determined 
from the fit can be converted into
\begin{equation}\label{Eq:Deltaphis_DD}
\Delta\phi_s^{\Dsm\Dsp} = -\left(1.7^{+1.6}_{-1.2}\:(\text{stat})^{+0.3}_{-0.7}\:(U\text{-spin})\right)^{\circ}\:.
\end{equation}
Finally, we can extract $\phi_s$ form the effective mixing phase in  
Eq.~\eqref{Eq:phis_eff_BsDsDs}, which yields
\begin{equation}
\phi_s = -\left(0.6^{+9.8}_{-9.9}\:(\text{stat})^{+0.3}_{-0.7}\:(U\text{-spin})\right)^{\circ}\:.
\end{equation}
Despite the suppression through the parameter $\epsilon$, penguins may have a significant 
impact on the extraction of $\phi_s$ and have to be taken into account. This will be particularly 
relevant for the LHCb upgrade era. In this new round of precision, we will also get valuable insights
into the validity of the $U$-spin symmetry, parameterised through Eq.~(\ref{a-theta-rel-U-break}). 

Unfortunately,  there is no measurement of CP violation in  \BstoDsDd available, 
which would be very interesting, in particular in view of the situation for \BdtoDdDd.
Consequently, we may not yet determine $\tilde a$ and $\tilde \theta$ from the data. 

\begin{figure}[tp]
 \centering
\includegraphics[width=0.49\textwidth]{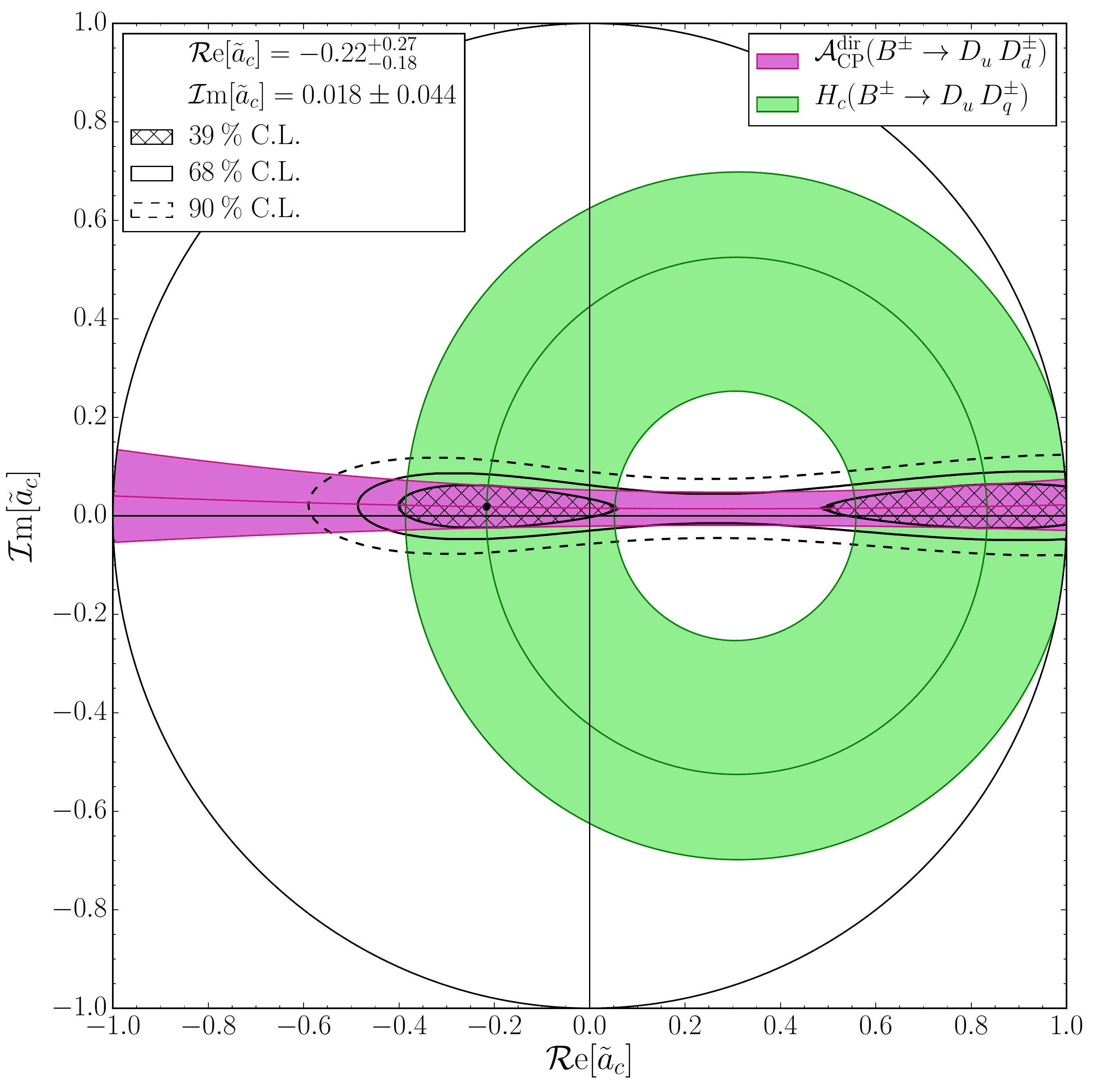}
\includegraphics[width=0.481\textwidth]{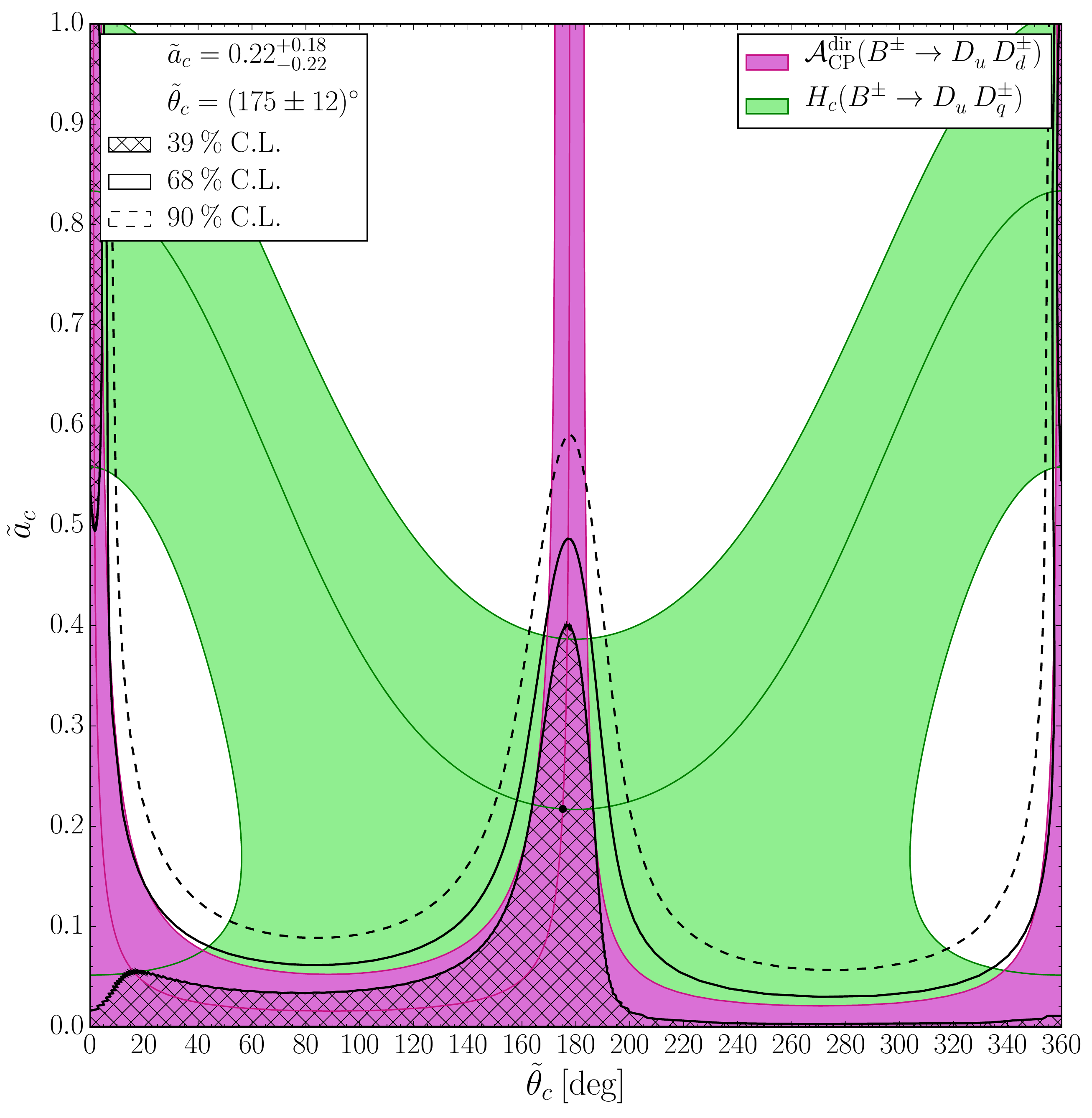}
\caption{Illustration of the determination of the penguin parameters $\tilde a_{\rm c}$ and 
$\tilde\theta_{\rm c}$ from a $\chi^2$ fit to the direct CP asymmetries of the decay 
\ButoDzDd and the observable $H_{\rm c}$.}
\label{fig:a-fit-c}
\end{figure}

However, for the charged \ButoDzDd decay, the PDG gives
\begin{equation}\label{ACP-dir-c}
{\cal A}_{\rm CP}^{\rm dir}(\uButoDzDd)=+0.03 \pm 0.07\:.
\end{equation}
If we complement this measurement with the value of $H_{\rm c}$ in Eq.~(\ref{Hc-value}), we may
perform a fit to the charged decays. It has three degrees of freedom $(a,\theta,\gamma)$, and
results in the solution
\begin{equation}
\mathcal{R}\text{e}[\tilde a_{\rm c}] = -0.22^{+0.27}_{-0.18} \:,\qquad \mathcal{I}\text{m}[\tilde a_{\rm c}] 
= 0.018 \pm 0.044\:,
\end{equation}
which corresponds to
\begin{equation}\label{Eq:Penguin_chi2fit_II}
\tilde a_{\rm c} = 0.22^{+0.18}_{-0.22} \:,\qquad \tilde \theta_{\rm c} = \left(175 \pm 12\right)^{\circ}\:.
\end{equation}
In Fig.~\ref{fig:a-fit-c}, we illustrate the corresponding situation, which complements Fig.~\ref{fig:a-fit-neut}.

It is interesting to compare the penguin parameter $ae^{i\theta}$ with its charged decay counterpart
$\tilde a_{\rm c} e^{\tilde \theta_{\rm c}}$. We obtain
\begin{equation}\label{rPAA}
r_{PA}^A\equiv\left[\frac{1}{1+x}\right]\left[\frac{\tilde a_{\rm c} e^{\tilde \theta_{\rm c}}}{ae^{i\theta}}\right]=
\frac{1+r_A}{1+r_{PA}}\:,
\end{equation}
where we have used Eqs.~(\ref{ac-atilde-rel}) and (\ref{a-x}).
The precision that can be obtained with the current data does not yet allow us to draw any conclusions regarding $r_{PA}^A$.
However, in the future it will be interesting 
to monitor this quantity as the experimental precision improves. Moreover, a measurement of the direct 
CP violation in the \BstoDsDd channel will allow us to determine $\tilde a e^{i\tilde \theta}$ 
from the information from $\tilde H$. The comparison with $\tilde a_{\rm c} e^{\tilde \theta_{\rm c}}$ 
will yield the $r_A$ parameter from Eq.~(\ref{ac-atilde-rel}), so that Eq.~(\ref{rPAA}) 
will then allow the determination of $r_{PA}$. Consequently, following these lines, we may reveal the 
impact of the annihilation and penguin annihilation topologies in the decays at hand.

\section{Prospects for the LHCb Upgrade and Belle II Era}\label{sec:prosp}

Let us conclude the discussion on the \BtoDD decays by exploring the potential
of these decay modes in the Belle II era and at the LHCb upgrade.  We do this
using several scenarios, examined in Section \ref{Sec:Benchmark}, that reflect
the different possibilities still allowed by the current data.  The inputs used
in these scenarios are discussed first.  Section \ref{Sec:Exp_Input} gives the
experimental prospects for the relevant CP and branching ratio information of
the \BtoDD decays, while Section \ref{Sec:Exchange_Prospects} deals with the
future constraints on the additional decay topologies.

\subsection{Extrapolating from Current Results}\label{Sec:Exp_Input}

The $B$-factories have pioneered the study of \BtoDD decays, including the
discoveries of numerous \BtoDD decay modes~\cite{Belle-observation}, the
measurements of branching fractions~\cite{Belle-CPV-BDD-2008,Zupanc-Belle-2007},
and the analyses of CP asymmetries~\cite{Belle-BDD,Belle-CPV-BDD-2008,
BaBar-BDD}.  The LHCb collaboration subsequently continued the study of \BtoDD
decays, notably focusing on the analysis of \Bs
decays~\cite{LHCB-BDD-2013,Aaij:2013bvd,LHCb-BsDsDs}, which are
abundantly produced at the LHC. Based on the successful performance of LHCb during
run I of the LHC, an estimate can be made of its performance with the data samples that are 
expected to be collected after the upgrade of the LHCb detector. 
For these extrapolations, an integrated luminosity of 5 fb$^{-1}$ in run II, from 2015 until 2018, is
assumed. In addition, the $B$ production cross section will increase
at a centre-of-mass energy of 13~TeV compared to 8~TeV by about 60\%.  For the
upgrade scenario, an integrated luminosity of 50 fb$^{-1}$ is assumed with
increased trigger efficiency, leading to about a three times larger data sample
per fb$^{-1}$ compared to the $B$ yield per fb$^{-1}$ at run I.
Similarly, a prognosis can be made for
measurements at Belle II, which is expected to start taking data in 2018.
Here we assume that 50 times more data will be collected than currently is available (1 ab$^{-1}$) .

\begin{table}[!b]
\begin{center}
\begin{tabular}{|l|r|c|l|}
\toprule
Observable & \multicolumn{1}{c|}{Current Measurement} & Upgrade & Experiment\\
\midrule
  $\mathcal{A}^{\textrm{dir}}_{CP}(\uBdtoDdDd) $                              & $-0.43 \pm 0.16 \pm 0.05 $ \cite{Belle-BDD}         & $\pm 0.05$ & Belle\\
  $\mathcal{A}^{\textrm{mix}}_{CP}(\uBdtoDdDd) $                              & $1.06 ~^{+0.14}_{-0.21} \pm 0.08 $  \cite{Belle-BDD}& $\pm 0.08$ & Belle \\
  $\mathcal{A}^{\textrm{dir}}_{CP}(\uBu\to\Du^{\vphantom{\pm}}D_{(s)}^{\pm})$ & $0.00 \pm 0.08 \pm 0.02 $~\cite{Belle-CPV-BDD-2008}& $\pm 0.02$ & Belle\\
  $\phi_s^{\rm eff}(\uBstoDsDs) $                                             & $(1 \pm 10 \pm 1)^\circ $~\cite{LHCb-BsDsDs}       & $\pm 2^\circ$ & LHCb \\
\bottomrule
\end{tabular}
\caption[Experimental prospects]
{Experimental prospects for the currently available CP asymmetry measurements of \BtoDD decays from Belle and LHCb.
}
\label{tab:Aprospects}
\end{center}
\end{table}

The expectations for the CP asymmetry parameters are listed in Table \ref{tab:Aprospects}.
The extrapolations are done for the currently available measurements only; no attempt is made 
to forecast the precision on yet-to-be-performed analyses.
For example, the LHCb collaboration will also determine the CP
asymmetries of the \BdtoDdDd decay, but it remains to be seen what the
accuracy will be in comparison with possible Belle II results.

The expectations for the branching fractions are listed in Table \ref{tab:BRprospects}.
The ``current measurement'' column reflects the best available knowledge at this moment,
which in some cases could have been more precise if the ratio of branching
fractions were determined directly, rather than dividing the individually measured branching
fractions. In the extrapolations it is assumed that the ratios of branching
fractions are determined. Moreover, it is assumed that the systematic uncertainties due
to $f_s/f_d$ (4.7\%), due to the $D$-meson branching fractions (3.9\%, 2.1\% and
1.3\% for $\Dsp$, $\Ddp$ and $\Dz$ mesons, respectively) and due to the
different \Bs lifetimes (2.9\% (1.5\%) for a CP (flavour) eigenstate),
remain the same.
We assume that the total experimental systematic uncertainty will
decrease from 5.0\% to 4.0\%.  In some ratios, the uncertainty on the $D$
branching fractions cancels with their contribution to the $f_s/f_d$
uncertainty. This is taken into account where appropriate.
In our upgrade era scenario, systematic uncertainties will be the limiting factor on
the ratio of branching fractions. Therefore, we would like to encourage research
into $f_s/f_d$, $\mathcal{B}(\Dsp \rightarrow K^+ K^-
\pi^+)/\mathcal{B}(\Ddp \rightarrow K^- \pi^+ \pi^+)$ and $B$ lifetime differences. If
these three factors could be reduced to a level of about 2\%, then that would
lead to a systematic uncertainty of (5--6)\% for all of these decays, assuming an
experimental uncertainty of 4\%. 
Finally, the prospects for improvements on external input parameters are listed in Table~\ref{tab:Extprospects}.

\begin{table}[tp]
\begin{center}
\resizebox{\textwidth}{!}{
\begin{tabular}{|c|c|r|rr|}
\toprule
Obs & Decay Ratio & \multicolumn{1}{c|}{Current Measurement} & \multicolumn{2}{c|}{LHCb Uncertainty} \\
    &             & of ratio of BRs                          & 2011 & Upgrade\\
\midrule
$H$         & $\BdtoDdDd/\BstoDsDs$
            &  $0.048 \pm 0.007 $~\cite{Agashe:2014kda}                    & 14\%(12\%)        & 8\%   \\
$\tilde H$  &  $\BstoDsDd/\BdtoDdDs$
            & $0.050\pm 0.008\pm 0.004$~\cite{LHCB-BDD-2013}               & 18\%              & 7\%   \\
$H_c$       &  $\ButoDzDd/\ButoDzDs$ 
            & $0.042 \pm 0.006$~\cite{Agashe:2014kda}                      & 15\%(7\%)         & 6\%   \\
\midrule
$\Xi$ & $\BstoDsDs/\BdtoDdDs $   
      & $0.56\pm 0.03\pm 0.04$~\cite{LHCB-BDD-2013}                        &  9\%              & 7\%   \\
$\Xi$ & $\BdtoDdDd/\BstoDsDd$     
      & $0.59 \pm 0.14  $~\cite{Agashe:2014kda}                            &  24\%(20\%)       & 6\%   \\
\midrule
$\Xi$ & $\BstoDdDd/\BdtoDdDs$       
      & $0.031 \pm 0.009 $~\cite{Agashe:2014kda}                           &  24\%(20\%)       & 11\%  \\
$\Xi$ & $\BdtoDsDs/\BstoDsDd$     
      & Not observed                                                       &                   &        \\
\bottomrule
\end{tabular}
} 
\caption[Experimental prospects]
{Experimental prospects for ratios of branching fractions. The second and fourth
ratios are obtained from direct determinations of the ratios of branching
fractions, whereas the others are calculated from individual branching
fractions. The value in brackets indicates the possible uncertainty if this ratio
were determined directly.  Note that for the calculation of the $H$ observables,
additional uncertainties due to $|\mathcal{A}'/\mathcal{A}|$ arise.}\label{tab:BRprospects}
\end{center}
\end{table}

\begin{table}[tp]
\begin{center}
\begin{tabular}{|c|r|c|}
\toprule
Observable& \multicolumn{1}{c|}{Current Measurement} & Upgrade\\
\midrule 
$f_{\Dd}$ & $(204.6 \pm 5.7 \pm 2.0) \,{\rm MeV}$~\cite{Huang-BESIII-2012} & $\pm3.0\,{\rm MeV}$~\cite{BESIII-2008} \\  
$f_{\Ds}$ & $(255.5 \pm 4.2 \pm 5.1) \,{\rm MeV}$~\cite{Zupanc-Belle-2013} & $\pm3.6\,{\rm MeV}^*$                  \\
$\gamma$  &  $(68.3 \pm 7.5        ) ^\circ     $~\cite{Bevan:2014cya}     & $\pm0.9^\circ$~\cite{LHCb-implications}\\
\bottomrule
\end{tabular}
\caption[Experimental prospects]
{Experimental prospects for external input. A $^*$ indicates an average of the 
extrapolated measurement with the current PDG average. }\label{tab:Extprospects}
\end{center}
\end{table}

%
%
%
\subsection{Exchange and Penguin Annihilation Contributions}\label{Sec:Exchange_Prospects}

\begin{figure}[tp]
 \centering
\includegraphics[width=0.49\textwidth]{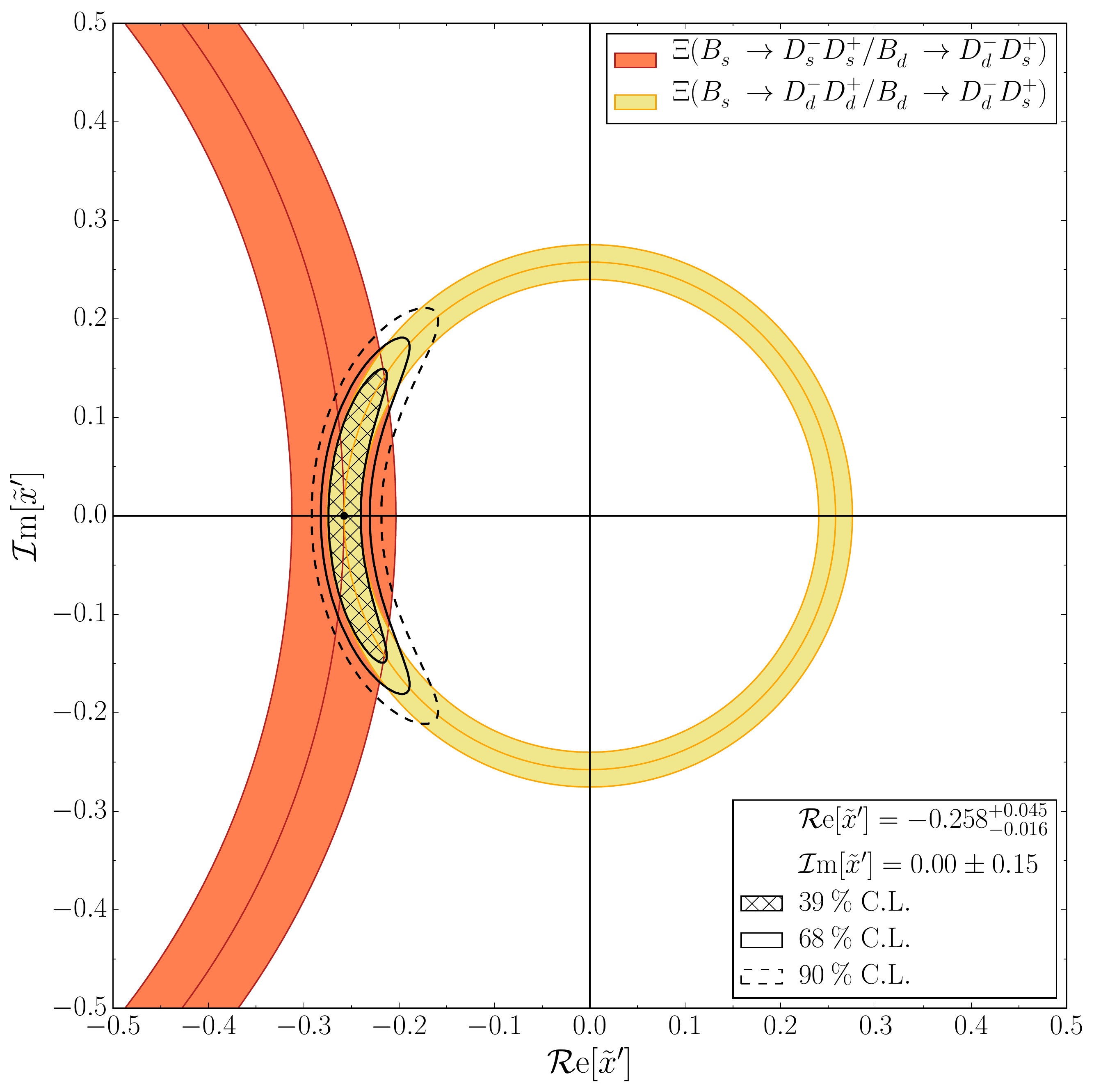}
\includegraphics[width=0.481\textwidth]{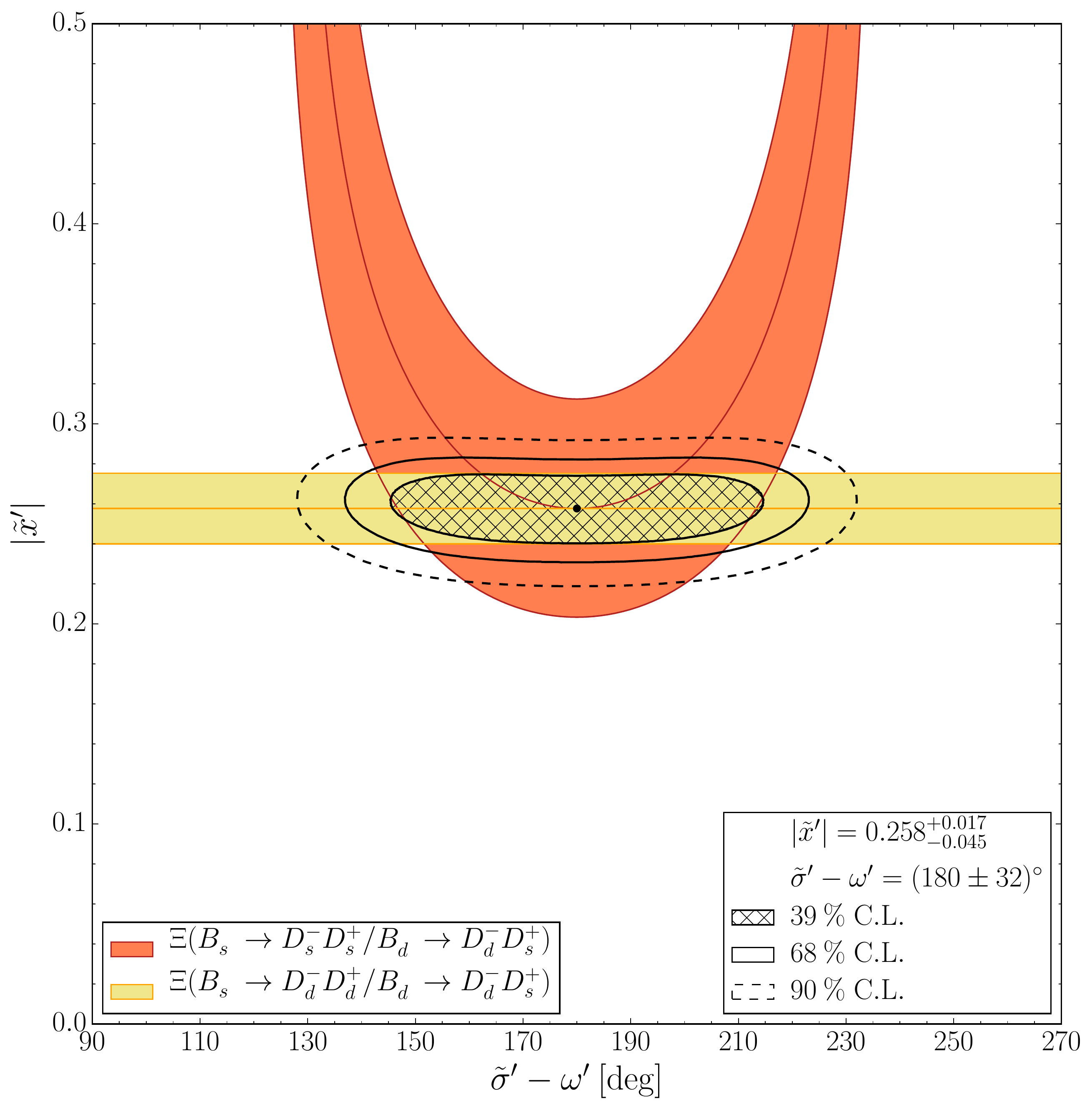}
\caption{Illustration of the determination of exchange and penguin-annihilation 
contributions through the parameter $x$ (Eq.~(\ref{x-def}))
in the Belle II/LHCb upgrade era. The precision should be compared to Fig.~\ref{fig:x-fit-p}.}\label{fig:x-fit-upgrade}
\end{figure}

For the construction of the $H$ observable based on the ratio of hadronic
amplitudes in Eq.~\eqref{A-rat-1-f}, the contributions from exchange and penguin
annihilation topologies, represented by the parameters $\tilde x'$ and
$\tilde\sigma'$, need to be quantified.  With future, improved measurements of
the \BtoDD branching ratios it is expected that the picture emerging from the
current data, discussed in Section \ref{ssec:EPA}, can be sharpened further.  We
therefore explore the precision that can be achieved towards the end of the
Belle II era and of the LHCb upgrade.  Based on the best fit solution obtained
from the current data, i.e.\ Eq.~\eqref{tilde-xp-det}, and the prospects in
Table \ref{tab:BRprospects}, we start from the following input measurements:
\begin{align}
\Xi(\uBstoDsDs, \uBdtoDdDs) & = 0.673 \pm 0.043\:,\\
\Xi(\uBstoDdDd, \uBdtoDdDs) & = 0.033 \pm 0.004\:.
\end{align}
A fit yields
\begin{equation}
{\rm Re}[\tilde x'] = -0.258^{+0.045}_{-0.016}\:, \qquad
{\rm Im}[\tilde x'] = 0.00 \pm 0.15\:,
\end{equation}
corresponding to
\begin{equation}
\tilde x' = 0.258^{+0.017}_{-0.045}\:, \qquad
\tilde\sigma' -\omega'= \left(180 \pm 15\right)^{\circ}\:.
\end{equation}
The associated confidence-level contours are shown in Fig.~\ref{fig:x-fit-upgrade}.
Note that at first sight these uncertainties do not seem to have improved significantly
with respect to the present experimental situation. However, this is merely caused by
the shape of the confidence contour in Fig.~\ref{fig:x-fit-upgrade}.

%
%
%
%
%
\subsection{Future Scenarios}\label{Sec:Benchmark}

To achieve the smallest theoretical uncertainty on the $H$ observable, it should
be constructed from the semileptonic decay information, see Eq.~\eqref{H-R}, as
explained in Section \ref{ssec:semilept}.  This method is preferred over the
direct ratio of hadronic branching fractions in Eq.~\eqref{Eq:Hobs_Def}, as it
does not rely on form factor information, and is not experimentally limited by
$f_s/f_d$.  However, as the necessary information on
$\text{d}\Gamma/\text{d}q^2(\BsSL)$ is currently not yet available, we also do
not have any estimates for the precision that can be achieved at LHCb or Belle
II.  For the following discussion, we will thus, like for the fits to the
current data, rely on the original definition using the ratio of hadronic
branching fractions, Eq.~\eqref{Eq:Hobs_Def}.
\newline

Using the currently available data, we have illustrated in Section
\ref{ssec:global} that it is possible to simultaneously determine the penguin
contributions and the \Bd--\Bdb mixing phase $\phi_d$ using the CP and branching
ratio information of the \BdtoDdDd decay.  However, as the result of $\phi_d$ in
Eq.~\eqref{Eq:phid_chi2fit} shows, the precision on $\phi_d$ is very limited.
Also for the Belle II and LHCb upgrade era it will be challenging to reach
precisions below the (10--$20)^{\circ}$ level.  A high-precision determination
of $\phi_d$ using the \BdtoDdDd decay is only possible if the direct CP
asymmetry and the $H$ observable together are sufficient to unambiguously pin
down the penguin parameters $a$ and $\theta$.  In such a situation, the phase
$\phi_d$ can then be determined from the mixing-induced CP asymmetry.  As
Fig.~\ref{fig:a-fit-neut} illustrates, $a$ and $\theta$ cannot precisely be
determined.  The situation would arise either for very large values of $a$,
which looks unrealistic, or if the $H$ observable can be determined with a
precision of well below the 5\% level, which in view of the prospects in Table
\ref{tab:BRprospects} is unlikely.  The determination of $\phi_d$ from the
\BdtoDdDd decay is therefore not competitive with the results from the $B_d^0\to
J/\psi K_{\rm S}^0$ decay.  It is thus advantageous to use external input on
$\phi_d$ for the analysis of the \BtoDD decays.  This extra information breaks
the ambiguity that is still present in the confidence-level contours shown in
Fig.~\ref{fig:a-fit-neut}, and can therefore improve the precision on $a$ and
$\theta$, and thus also on $\Delta\phi_s$.  For the future benchmark scenarios
we therefore only focus on the high precision determination of $\phi_s$.
\newline

\begin{figure}[tp]
\centering
\includegraphics{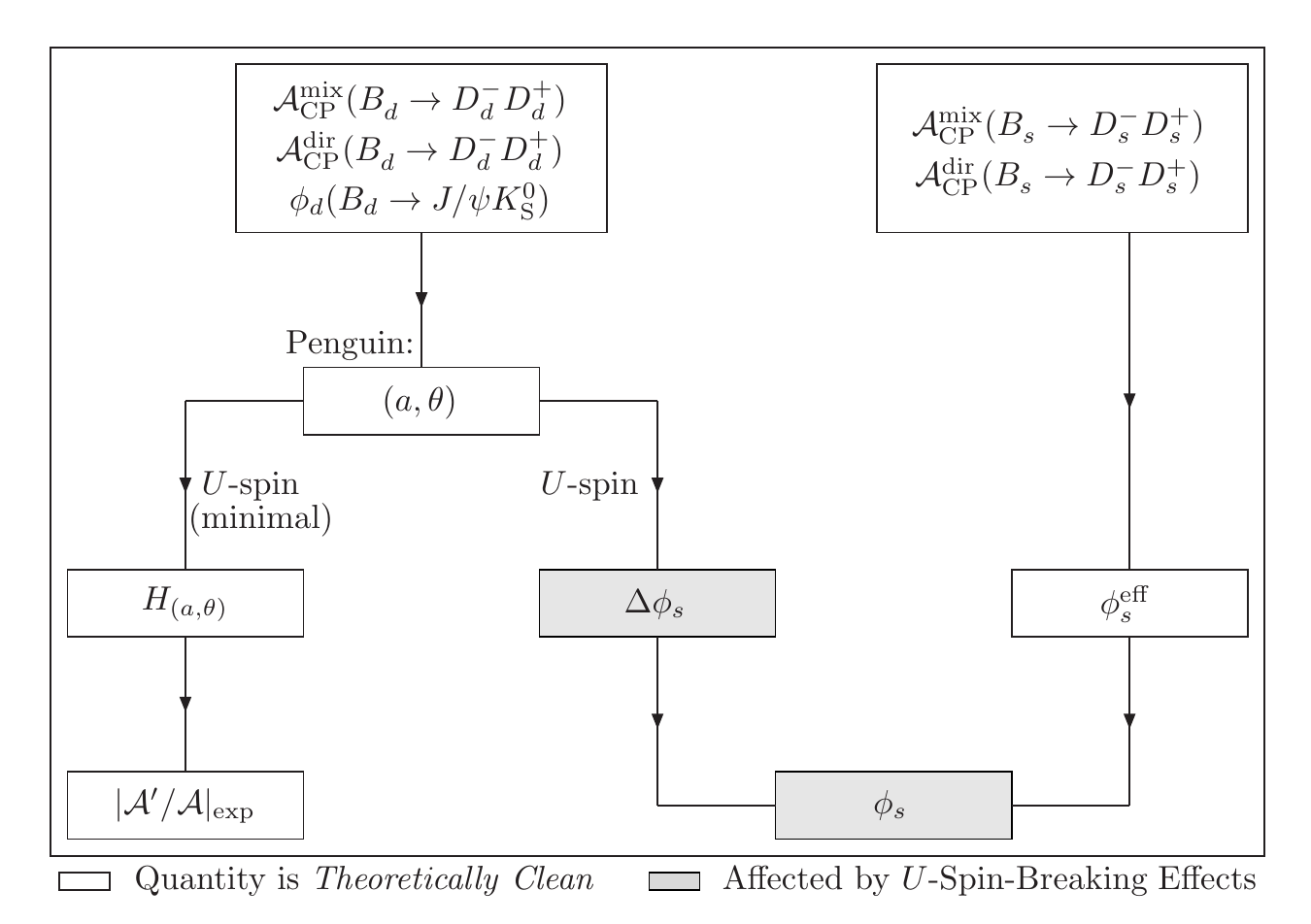}
\caption{Flow chart illustrating the favourable strategy to control
$\Delta\phi_s$, which only requires information on the \BdtoDdDd CP asymmetries.}
\label{Fig:Flow_withoutH}
\end{figure}

Using external input for $\phi_d$ in principle makes one of the three
observables used by the fit (${\cal A}_{\rm CP}^{\rm dir}$, ${\cal A}_{\rm
CP}^{\rm mix}$ or $H$) superfluous.  As the $H$ observable receives corrections
from possible $U$-spin-breaking effects through $|\mathcal{A}'/\mathcal{A}|$,
which result in large theoretical uncertainties, it is preferred to determine
the penguin parameters using information on the CP asymmetries only, omitting
$H$.  Such a determination is \emph{theoretically clean}.  This will ultimately
lead to the highest precision on $a$, $\theta$ and $\Delta\phi_s$.  In this
situation, the branching ratio information can instead be used to gain insight
into the hadronic physics of the \BdtoDdDd, \BstoDsDs system.  We can then
follow the opposite path where the fit results for $a$ and $\theta$ can be used
to determine the $H$ observable with the help of Eq.~\eqref{H-exprs}, labelled
$H_{(a,\theta)}$ below.  Since $a'$ enters there with the tiny $\epsilon$, the
$U$-spin-breaking corrections affecting the $U$-spin relation in
Eq.~\eqref{a-theta-rel-U-break} have a very minor impact.  Since we now know the
value of $H$, the relation \eqref{Eq:Hobs_Def} can be inverted to instead
determine the ratio of hadronic amplitudes
\begin{equation}
\left|\frac{{\cal A}'}{{\cal A}}\right|=\sqrt{\epsilon H_{(a,\theta)}
\left[\frac{m_{\uBs}}{m_{\uBd}}
\frac{\Phi(m_{\Ddm}/m_{\uBd},m_{\Ddp}/m_{\uBd})}{\Phi(m_{\Dsm}/m_{\uBs},m_{\Dsp}/m_{\uBs})}
 \frac{\tau_{\Bd}}{\tau_{\Bs}} \right]
\frac{\mathcal{B}\left(\uBstoDsDs\right)_{\text{theo}}}{\mathcal{B}\left(\uBdtoDdDd\right)}}
\end{equation}
from the measured ratio of branching ratios.
This experimental measurement of the ratio of hadronic amplitudes can be compared with the theoretical result in Eq.~\eqref{Acal-rat-theo}.
This favourable strategy is illustrated by the flow chart in Fig.~\ref{Fig:Flow_withoutH}.
\newline

\begin{figure}[tp]
\centering
\includegraphics{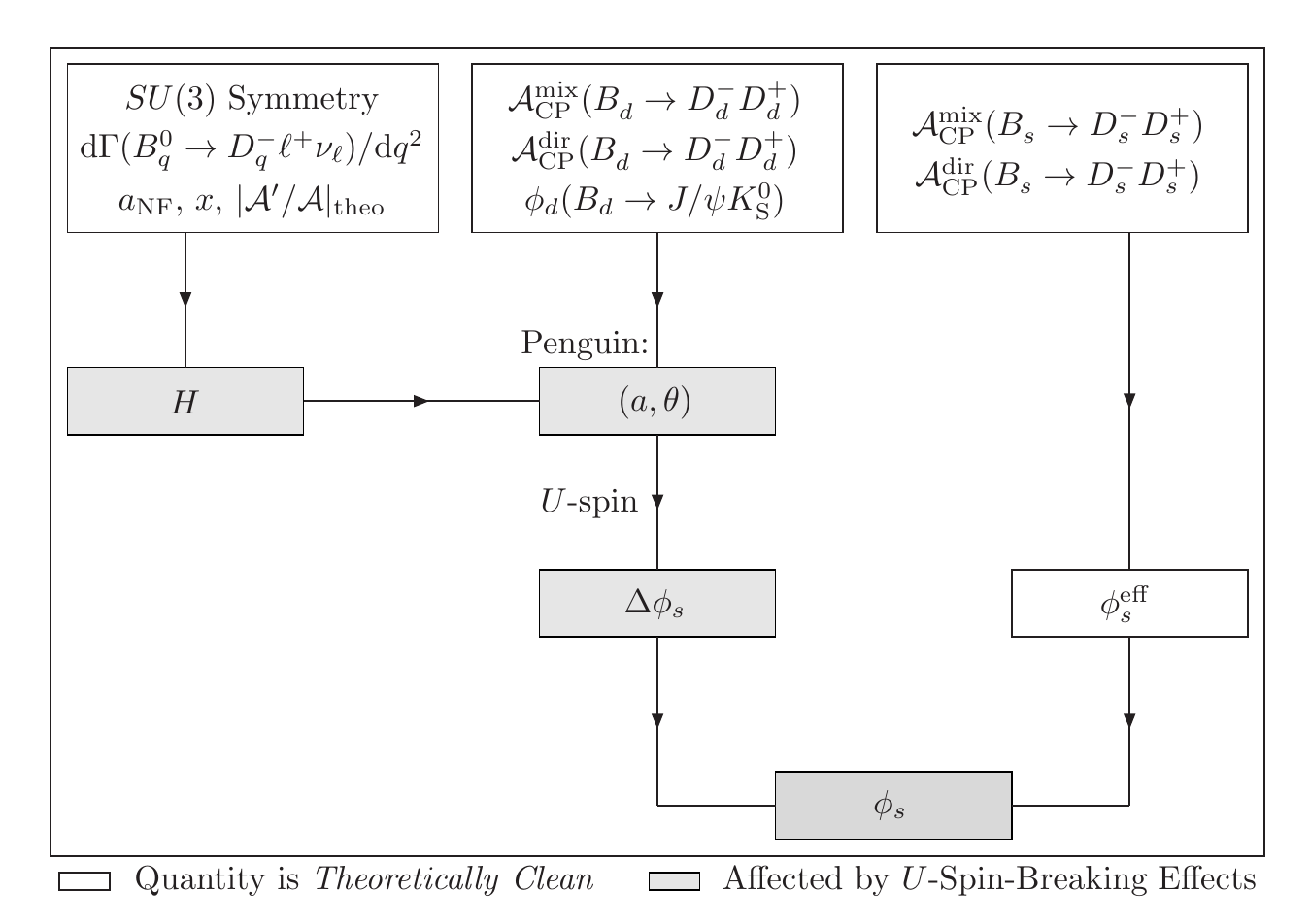}
\caption{Flow chart illustrating the less favourable strategy to control
$\Delta\phi_s$, which requires information on the $H$ observable in addition to
the \BdtoDdDd CP asymmetries.}
\label{Fig:Flow_withH}
\end{figure}

However, the ideal scenario described above cannot always be realised.  When the
value of the mixing-induced CP asymmetry is compatible with $1$ (at the
$1\sigma$ level), its power to constrain the penguin parameters $a$ and $\theta$
is limited.  This can best be illustrated using the contour plots, like
Fig.~\ref{fig:a-fit-neut} or Fig.~\ref{Fig:Scan_Fits} below.  In this situation,
the annular constraint originating from the mixing-induced CP asymmetry becomes
a closed disk, leading to a large overlap region with the direct CP asymmetry
constraint.  Consequently, it is not possible to conclusively pin down $a$ and
$\theta$ in such a situation.  Additional information is thus needed to improve
the picture, and reach our target of matching the foreseen experimental
precision on $\phi_s$ with an equally precise determination of $\Delta\phi_s$.
In this situation, the $H$ observable forms an essential ingredient in the fit,
and it can therefore not be used to experimentally constrain the ratio of
hadronic amplitudes.  This less favourable strategy is illustrated by the flow
chart in Fig.~\ref{Fig:Flow_withH}.
\newline

\begin{figure}[tp]
 \centering
\includegraphics[width=0.49\textwidth]{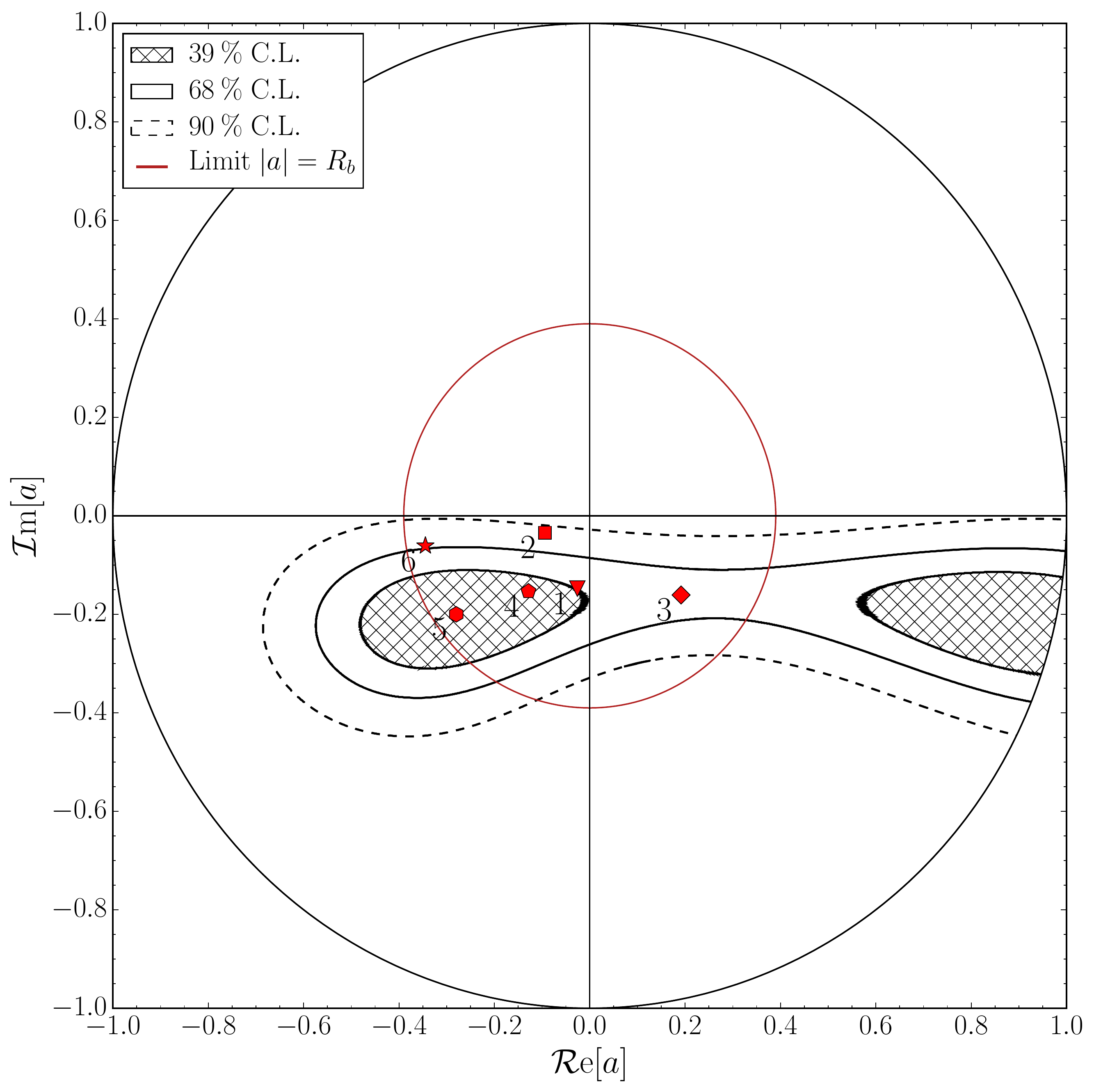}
\hfill
\includegraphics[width=0.481\textwidth]{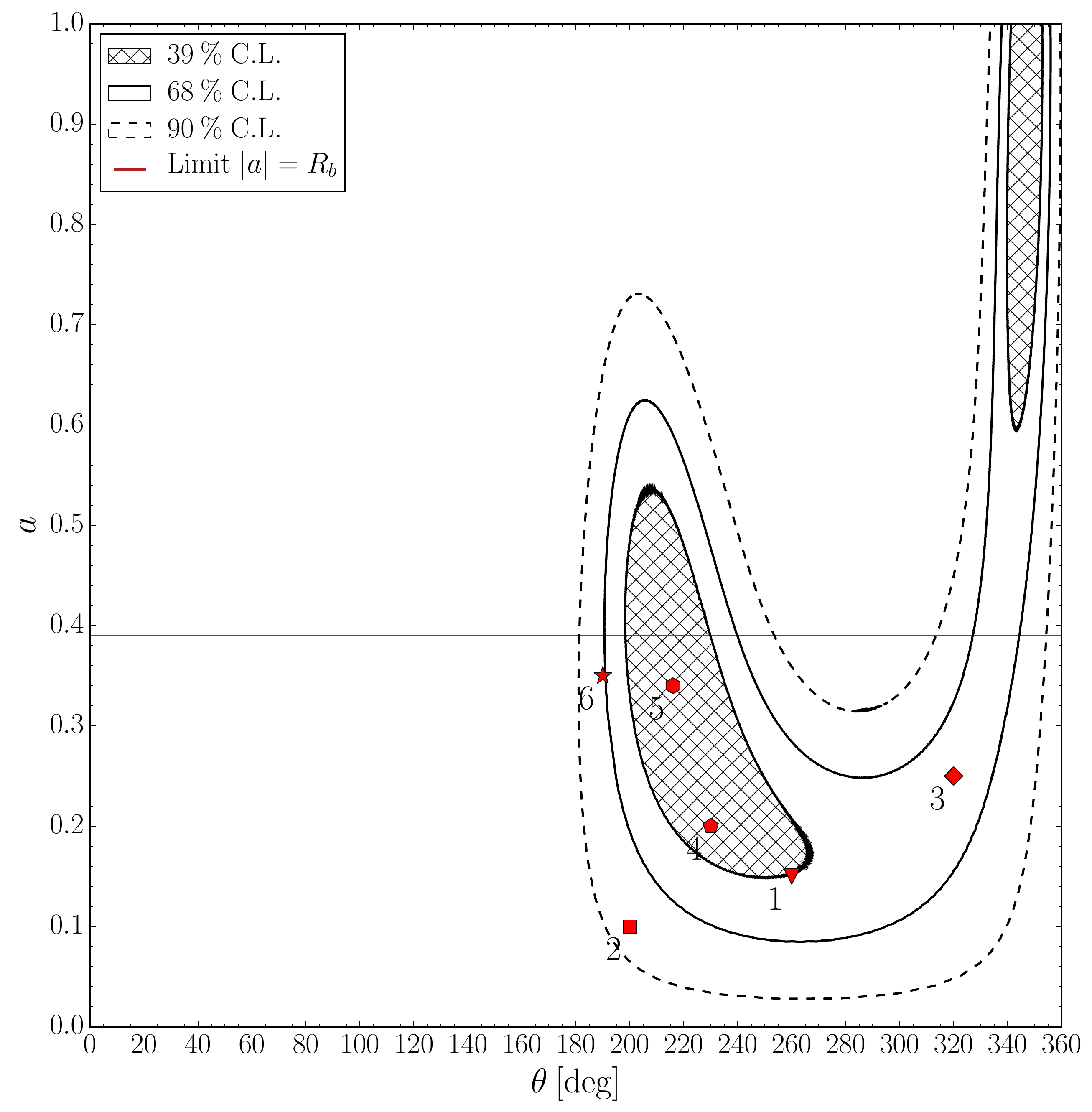}
\caption{Distribution of six scenarios in the
$\mathcal{R}\text{e}[a]$--$\mathcal{I}\text{m}[a]$ plane [Left] and in the
$\theta$--$a$ plane [Right].  Superimposed are the confidence-level contours
from the fit to the current data.  The red circle and horizontal line represent
the natural upper limit $a = R_b$.}
\label{Fig:Scan_input_points}
\end{figure}

\begin{figure}[p]
 \centering
\includegraphics[width=0.49\textwidth]{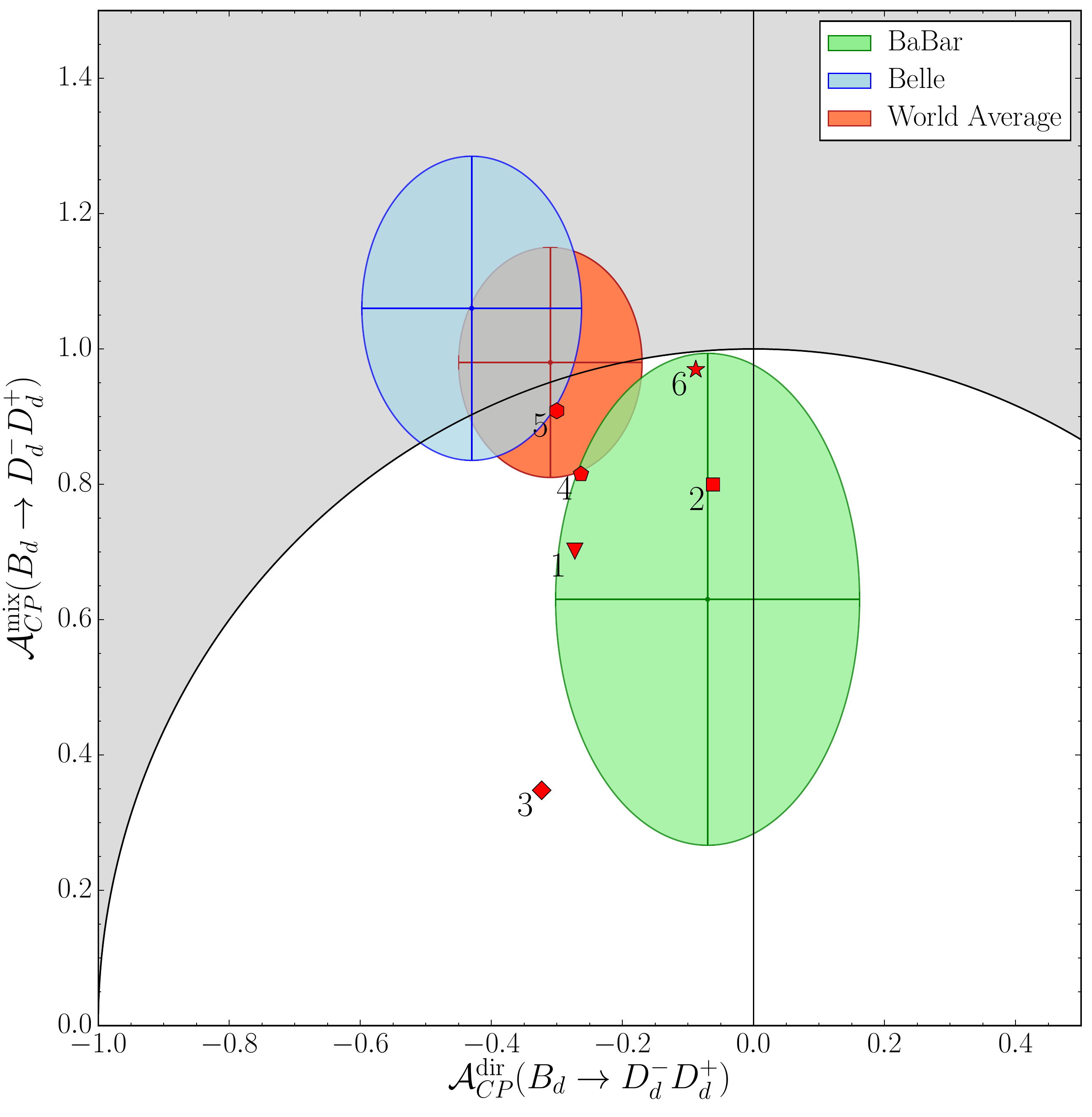}
\caption{Distribution of the six scenarios in the ${\cal A}_{\rm CP}^{\rm dir}$--${\cal A}_{\rm CP}^{\rm mix}$ 
plane, which can be compared to the current BaBar [Green] and Belle [Blue] measurement and the world average [Red].}
\label{Fig:Scan_CPobs}
\end{figure}

Given the current experimental situation, either of the two situations sketched
above can still be realised, depending on the future world average for ${\cal
A}_{\rm CP}^{\rm mix}(\BdtoDdDd)$.  To demonstrate the variety of situations in
which we may ultimately end up, we made an overview of six different scenarios,
covering both situations.  Scenarios 1--3 represent the favourable situation in
which the $H$ observable can be omitted from the fit, in which we can determine
$a$ and $\theta$ in a theoretical clean way, and get experimental access to the
ratio of hadronic amplitudes.  Scenarios 4--6, on the other hand, fall in the
second category and do require information on $H$ to conclusively pin down $a$
and $\theta$.  All six scenarios are chosen to be compatible with the current
experimental situation, with scenario 5 representing the current best fit point,
and have $a< R_b$, which is suggested by Eq.~\eqref{a-pen}.  Although it is
mathematically possible for $a$ to be larger than $R_b$, it would imply that the
penguin topologies are larger than the tree contribution, which seems very
unlikely.  The condition $a=R_b$ thus serves as a naturally upper limit for the
size of the penguin contributions.  The different scenarios we consider, and the
resulting input values of the three observables (${\cal A}_{\rm CP}^{\rm dir}$,
${\cal A}_{\rm CP}^{\rm mix}$ and $H$) are listed in Table \ref{Tab:Scan_Input}.
\begin{table}[tp]
\center
\begin{tabular}{|r|c|c|c|c|c|}
\toprule
No. & $a$ & $\theta$ [deg] & ${\cal A}_{\rm CP}^{\rm dir}$ & ${\cal A}_{\rm CP}^{\rm mix}$ & $H$ \\
\midrule
1 & $0.15$ & $260$ & $-0.27$ & $0.70$ & $1.04$ \\
2 & $0.10$ & $200$ & $-0.06$ & $0.80$ & $1.07$ \\
3 & $0.25$ & $320$ & $-0.32$ & $0.35$ & $0.95$ \\
4 & $0.20$ & $230$ & $-0.26$ & $0.82$ & $1.12$ \\
5 & $0.34$ & $216$ & $-0.30$ & $0.91$ & $1.29$ \\
6 & $0.35$ & $190$ & $-0.09$ & $0.97$ & $1.34$ \\
\bottomrule
\end{tabular}
\caption{Penguin parameters and observables corresponding to the six different scenarios.}
\label{Tab:Scan_Input}
\end{table}
The choice of input points can be compared with the current fit solution for $a$
and $\theta$ in Fig.~\ref{Fig:Scan_input_points}, and with the current
measurements of the \BdtoDdDd CP asymmetry parameters in
Fig.~\ref{Fig:Scan_CPobs}.

For each of the six scenarios, the individual constraints coming from ${\cal
A}_{\rm CP}^{\rm dir}$, ${\cal A}_{\rm CP}^{\rm mix}$ and $H$ are illustrated in
Fig.~\ref{Fig:Scan_Fits}.  The three left-most plots represent the favourable
situation, while the three right-most plots fall in the second category.  For
the three right-most plots the constraint from the mixing-induced CP asymmetry
is more disk-like as the central value of ${\cal A}_{\rm CP}^{\rm mix}$ is
closer to one.  As a consequence, the overlap with the direct CP asymmetry,
which forms a narrow band in all cases, is too large to pin down $a$ and
$\theta$.  In this respect, scenario 4 should be seen as a limiting case;
information of $H$ is not strictly necessary if one is only interested in the 1
sigma results.

For each of the six scenarios we also performed a $\chi^2$ fit, similar to the
one described in Section \ref{ssec:global}, but including $\phi_d$ as a Gaussian
constraint.  The fit results for $a$ and $\theta$, and the associated values for
the shifts $\Delta\phi_d$ and $\Delta\phi_s$ are listed in Table
\ref{Tab:Scan_Fit}.  $U$-spin-breaking effects, parametrised by
Eq.~\eqref{a-theta-rel-U-break}, have been included in the results for
$\Delta\phi_s$.  The associated confidence-level contours are shown in
Fig.~\ref{Fig:Scan_Fits}.  In all cases we succeed in our goal of matching the
foreseen experimental precision on $\phi_s$, see Table \ref{tab:Aprospects},
with an equally precise determination of $\Delta\phi_s$.  This is also the case
for $\phi_d^{\rm eff}$ and $\Delta\phi_d$, as illustrated in Table
\ref{Tab:Scan_Eff}.  For the first three scenarios, which do not include the $H$
observable in the fit, the resulting solution for $H_{(a,\theta)}$ and the
values for the ratio of hadronic amplitudes are listed in Table
\ref{Tab:Scan_FF}.  The resulting uncertainties are about a factor two smaller
that the current theoretical uncertainties derived within the factorisation
framework, and of comparable size to the experimental precision that can be
obtained on the ratio of hadronic amplitudes describing the $B_d^0\to J/\psi
K_{\rm S}^0$ and $B_s^0\to J/\psi K_{\rm S}^0$ decays \cite{DeBF-pen}.
Consequently, the experimental determination of $|\mathcal{A}'/\mathcal{A}|$ is
yet another interesting topic for Belle II and the LHCb upgrade. It will provide
valuable insights into possible non-factorisable $U$-spin-breaking effects and
the hadronisation dynamics of the \BtoDD decays.

\begin{table}[p]
\center
\begin{tabular}{|r|l|c|c|r|r|}
\toprule
No. & $H$ & $a$ & $\theta$ [deg] & $\Delta\phi_d$ [deg] & \multicolumn{1}{c|}{$\Delta\phi_s$ [deg]} \\
\midrule
1 & No & $0.150_{-0.029}^{+0.032}$ & $260.0_{-21.8}^{+25.8}$ & $3.6_{-6.3}^{+7.5}$ & $-0.15_{-0.34}^{+0.40}(\text{stat})_{-0.30}^{+0.32}(U\text{-spin})$ \\
2 & No & $0.100_{-0.063}^{+0.079}$ & $200.0_{-19.5}^{+22.1}$ & $10.1_{-7.0}^{+8.9}$ & $-0.55_{-0.39}^{+0.49}(\text{stat})_{-0.08}^{+0.20}(U\text{-spin})$ \\
3 & No & $0.250_{-0.037}^{+0.036}$ & $320.0_{-7.8}^{+7.6}$ & $-21.6_{-6.1}^{+5.8}$ & $1.12_{-0.25}^{+0.24}(\text{stat})_{-0.40}^{+0.40}(U\text{-spin})$ \\
4 & Yes & $0.200_{-0.052}^{+0.063}$ & $230.0_{-16.7}^{+20.8}$ & $14.4_{-7.7}^{+9.8}$ & $-0.75_{-0.42}^{+0.53}(\text{stat})_{-0.33}^{+0.39}(U\text{-spin})$ \\
5 & Yes & $0.340_{-0.087}^{+0.106}$ & $216.0_{-11.5}^{+14.0}$ & $29.1_{-12.1}^{+14.7}$ & $-1.62_{-0.62}^{+0.75}(\text{stat})_{-0.48}^{+0.62}(U\text{-spin})$ \\
6 & Yes & $0.350_{-0.115}^{+0.130}$ & $190.0_{-6.4}^{+6.9}$ & $33.6_{-14.0}^{+15.8}$ & $-2.04_{-0.69}^{+0.78}(\text{stat})_{-0.32}^{+0.56}(U\text{-spin})$ \\
\bottomrule
\end{tabular}
\caption{Fit results for the different scenarios.}
\label{Tab:Scan_Fit}
\end{table}
\begin{table}[p]
\center
\begin{tabular}{|r|r|r|}
\toprule
No. & \multicolumn{1}{c|}{$\phi_d^{\rm eff}$ [deg]} & \multicolumn{1}{c|}{$\Delta\phi_d$ [deg]} \\
\midrule
1 & $46.8 \pm 6.9$ & $3.6_{-6.3}^{+7.5}$\\
2 & $53.3 \pm 7.5$ & $10.1_{-7.0}^{+8.9}$\\
3 & $21.6 \pm 5.1$ & $-21.6_{-6.1}^{+5.8}$\\
4 & $57.7 \pm 8.8$ & $14.4_{-7.7}^{+9.8}$\\
5 & $72.3 \pm 15.7$ & $29.1_{-12.1}^{+14.7}$\\
6 & $76.8 \pm 19.6$ & $33.6_{-14.0}^{+15.8}$\\
\bottomrule
\end{tabular}
\caption{Comparison between the effective mixing phase $\phi_d^{\rm eff}$ and the shift $\Delta\phi_d$.}
\label{Tab:Scan_Eff}
\end{table}
\begin{table}[p]
\center
\begin{tabular}{|r|c|c|}
\toprule
No. & $H_{(a,\theta)}$ & $|\mathcal{A}'/\mathcal{A}|$ \\
\midrule
1 & $1.038 \pm 0.039 (a,\theta) \pm 0.002 (\xi,\delta)$ & $1.163 \pm 0.048$\\
2 & $1.067 \pm 0.053 (a,\theta) \pm 0.001 (\xi,\delta)$ & $1.179 \pm 0.053$\\
3 & $0.946 \pm 0.014 (a,\theta) \pm 0.002 (\xi,\delta)$ & $1.110 \pm 0.042$\\
\bottomrule
\end{tabular}
\caption{Constraints on the ratio of hadronic amplitudes for those scenarios where the $H$ observable is not needed.}
\label{Tab:Scan_FF}
\end{table}

\begin{figure}[p]
\center
    \begin{picture}(440,650)(0,0)
      \put(  0,440){\includegraphics[width=0.49\textwidth]{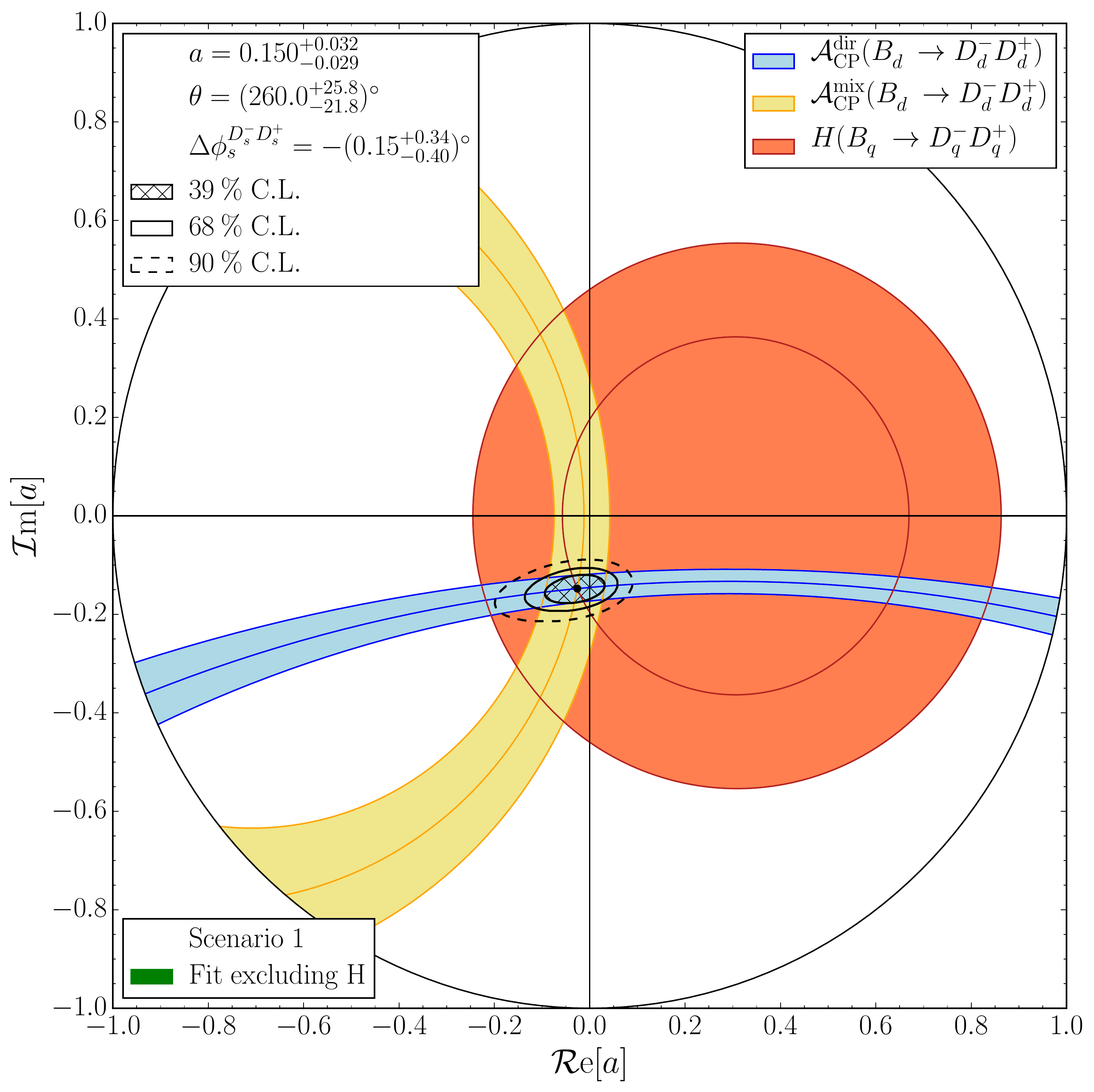}}
      \put(220,440){\includegraphics[width=0.49\textwidth]{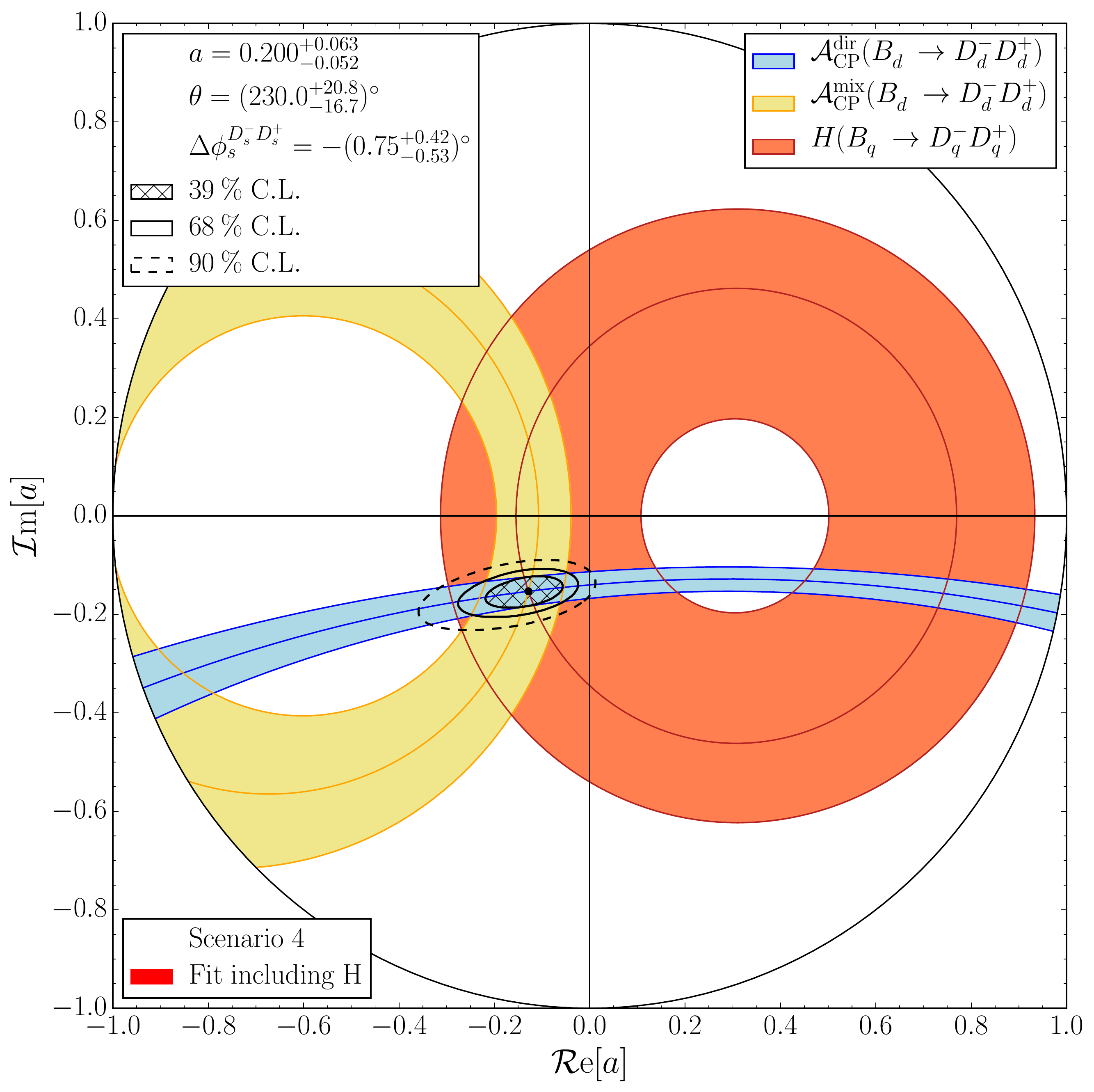}}
      \put(  0,220){\includegraphics[width=0.49\textwidth]{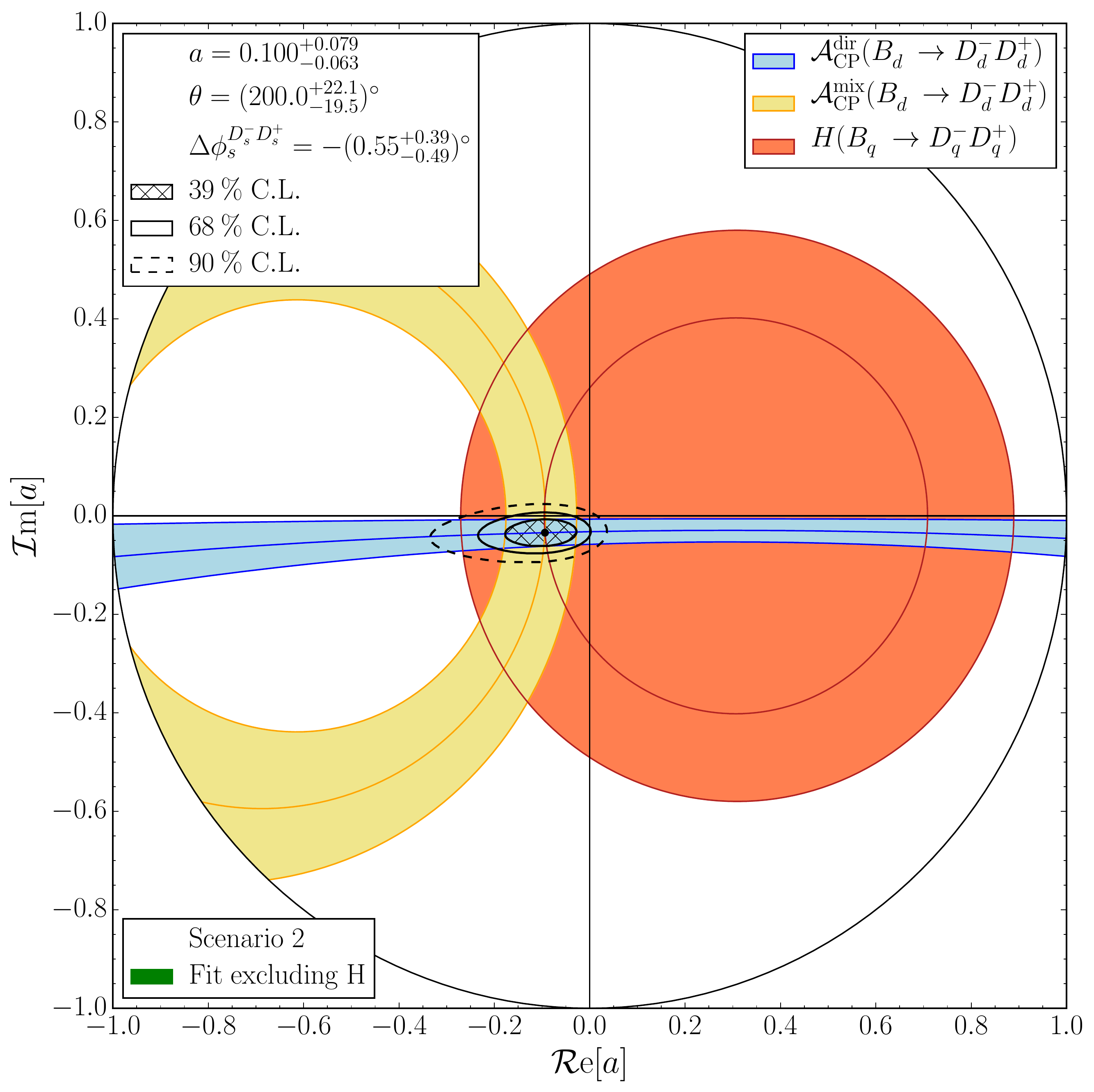}}
      \put(220,220){\includegraphics[width=0.49\textwidth]{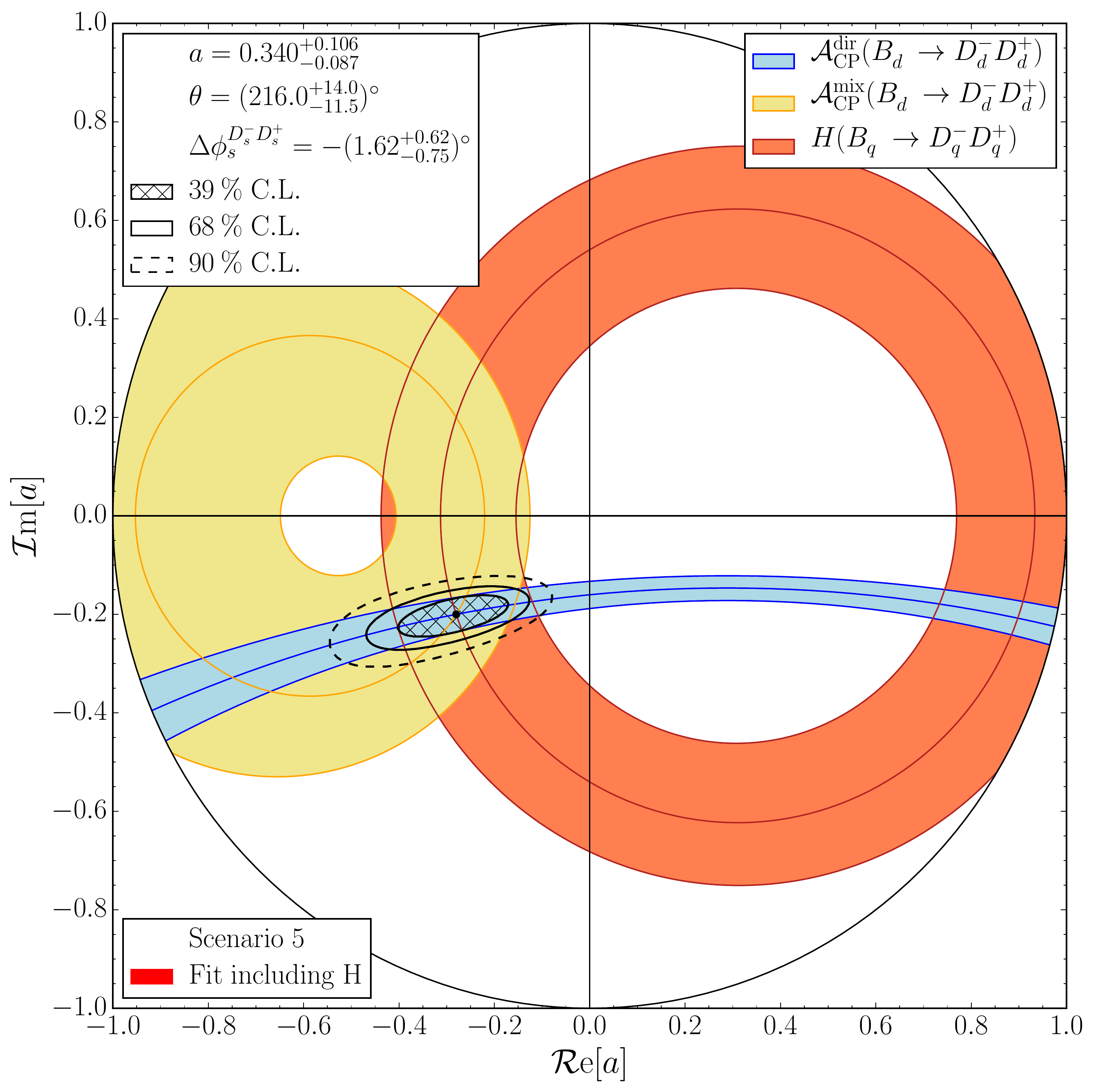}}
      \put(  0,  0){\includegraphics[width=0.49\textwidth]{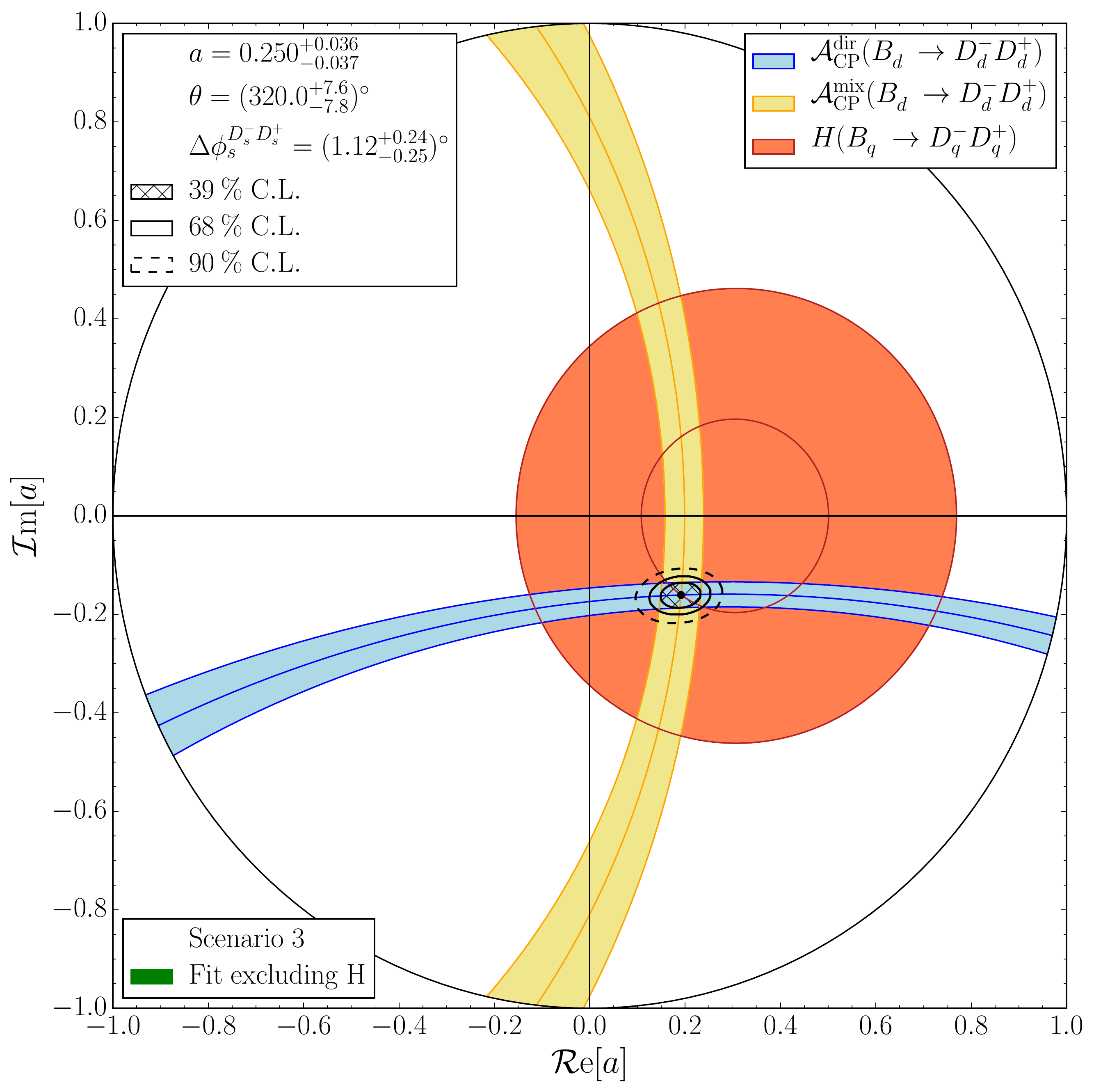}}
      \put(220,  0){\includegraphics[width=0.49\textwidth]{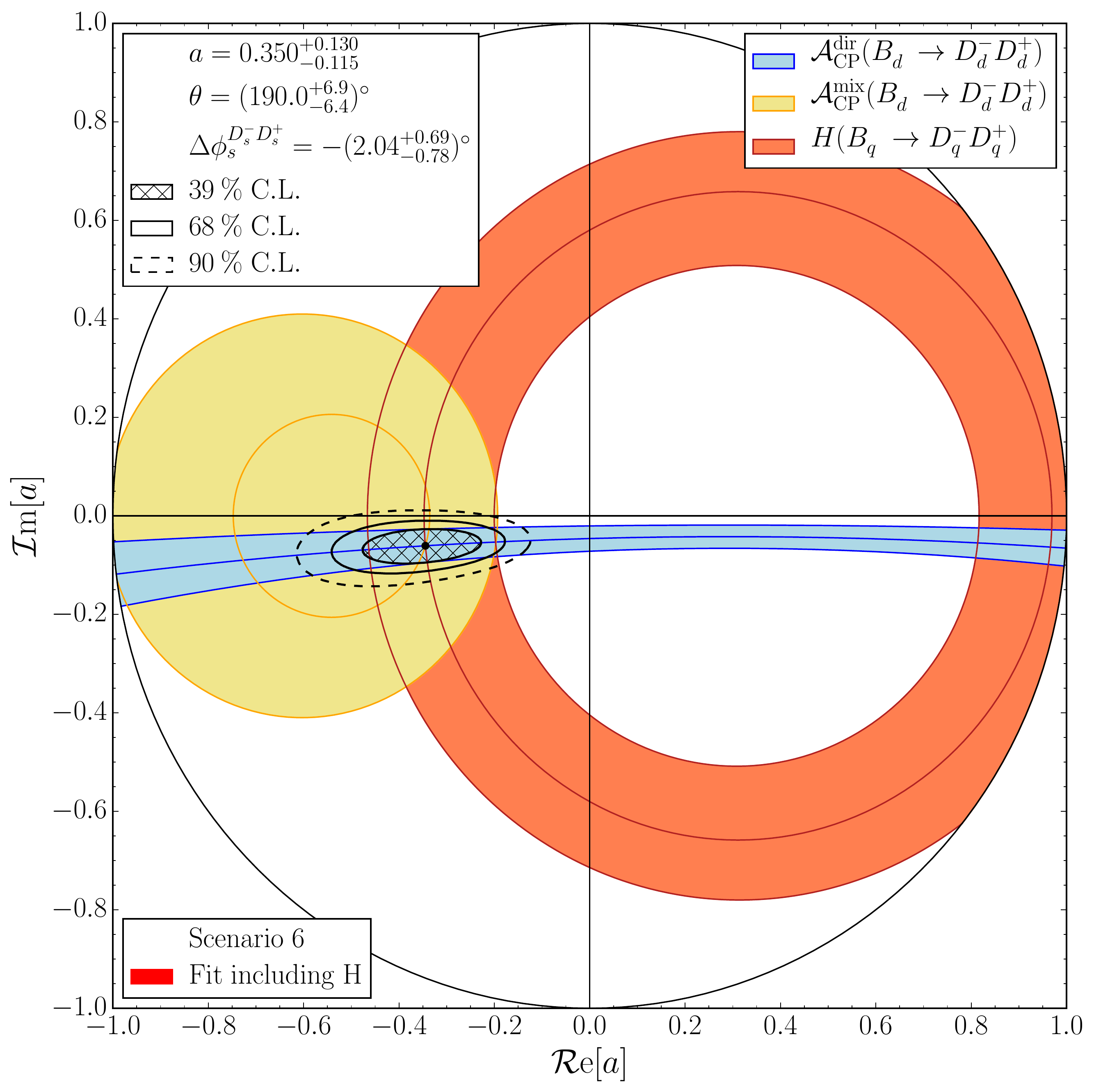}}
      \put( 20,665){Scenarios 1--3:}
      \put(240,665){Scenarios 4--6:}
    \end{picture}
\caption{Illustration of the determination of the penguin parameters $a$ and $\theta$ for the scenarios introduced above.
The three left scenarios have favourable values for  ${\cal A}_{\rm CP}^{\rm dir}$--${\cal A}_{\rm CP}^{\rm mix}$,
while scenarios 5 and 6 at the right bottom would require knowledge from $H$.}
\label{Fig:Scan_Fits}
\end{figure}

\clearpage

\section{Conclusions}\label{sec:concl}
In this paper, we have presented a detailed study of the system of \BtoDD decays, exploring the 
picture emerging from the currently available data and proposing new strategies for the era of Belle II
and the LHCb upgrade. We find that patterns in the current branching ratio measurements 
can be accommodated through sizeable contributions from exchange and penguin annihilation 
topologies, which play a more prominent role than naively expected. This feature suggests that 
long-distance effects of strong interactions are at work, which cannot be understood within 
perturbation theory. 

Using data for the differential semileptonic $\BdSL$ rate, we have
determined non-factorisable contributions to the \BdtoDdDs decay, finding effects at the 25\% level, 
including certain penguin contributions. We have pointed out that a future measurement of the 
semileptonic \BsSL channel would allow an optimal determination of the observable $H$ from
the \BdtoDdDd and \BstoDsDs branching ratios, with uncertainties given only by non-factorisable 
$U$-spin-breaking corrections, which can be further quantified through the comparison of 
the differential \BsSL rate with the \BstoDsDs rate. Experimental studies of the \BsSL modes 
are encouraged. 

In view of our new insights into the prominent role of the exchange and penguin annihilation
topologies, 
the penguin contributions may also be more important than naively expected. The current experimental 
situation for CP violation in the  \BdtoDdDd channel is not satisfactory, with measurements of the 
CP-violating asymmetries by the BaBar and Belle collaborations that are not in good agreement with 
one another. 
Measurements of CP-violation in the \BstoDsDd channel might help to clarify the situation.
In the case of  the \BstoDsDs mode, the LHCb collaboration has presented a first 
analysis of CP violation, with large experimental uncertainties. In the future, the experimental errors
can be significantly reduced. 
Using only information from branching ratios, we find the lower bound $a\geq0.052$ for the
penguin effects from the \BdtoDdDd  and  \BstoDsDs branching ratios. Adding the current 
measurements of CP violation in \BdtoDdDd to the analysis, we obtain 
\begin{displaymath}
a = 0.35^{+0.19}_{-0.20} \:,\qquad \theta = \left(215^{+51}_{-17}\right)^{\circ}\:, \quad 
\phi_d = \left(60^{+43}_{-39}\right)^{\circ}
\end{displaymath}
from a $\chi^2$ fit to the data. These results indicate potentially sizeable penguin effects, although
the
large uncertainties do not allow us to draw further conclusions.

Since the determination of $\phi_d$ from \BdtoDdDd will not be competitive with the $\Bd\to J/\psi 
K_{\rm S}^0$ analysis and the control of penguins though $\Bs\to J/\psi K_{\rm S}^0$, we advocate to 
use $\phi_d$ as an input from the latter analysis for the determination of the penguin parameters  
from \BdtoDdDd and relating them to their counterparts in  \BstoDsDs with the help of the 
$U$-spin symmetry of strong interactions. Following these lines, it will be possible to control the
penguin
effects in the determination of $\phi_s$ from the CP violation in the \BstoDsDs channel. 

We find that the implementation of this strategy depends strongly on the values of the measured 
CP asymmetries, as we illustrated through a variety of future scenarios which are consistent with the 
current experimental situation, giving us a guideline for the LHCb upgrade era. We distinguish between 
two kinds of scenarios: in the first, the direct and mixing-induced CP asymmetries of the \BdtoDdDd
channel
are sufficient to determine $a$ and $\theta$ in a theoretically clean way, allowing us to determine
the ratio $|{\cal A}'|/{\cal A}|$ from the observable $H$, providing valuable insights into
non-factorisable
$U$-spin-breaking effects. In the second --- less favourable --- class of scenarios, information both
from 
$H$ and the CP asymmetries is needed to determine the penguin parameters. We have 
demonstrated that the resulting theoretical uncertainty for the penguin shift $\Delta\phi_s$ 
of the \BstoDsDs channel will be smaller than the experimental uncertainty for 
$\phi_s^{\rm eff}(\uBstoDsDs)$ in both classes of scenarios. Similar analyses can be performed
for $\Bd\to D_d^{*-}D_d^{*+}$ and $\Bs\to D_s^{*-}D_s^{*+}$ decays, where time-dependent 
measurements of the angular distribution of the decay products of the two vector mesons are
required \cite{DF,DDF}.

Analyses of \BtoDD decays in the era of Belle II and the LHCb upgrade will offer interesting new
insights both into the physics of strong interactions and into CP violation. We look forward to 
confronting the strategies discussed in this paper with future data!

\subsubsection*{Acknowledgements}
We would like to thank Greg Ciezarek for very interesting discussions on the experimental prospects
of measuring form factors with semileptonic $\Bs$ decays,
and Patrick Koppenburg for carefully reading the manuscript.

\vspace*{2.5truecm}

\appendix
\section*{Appendix: Notation}\label{appendix}
\addcontentsline{toc}{section}{Appendix: Notation}
In this Appendix, we give an overview of the notation, parameters and observables
used in our analysis of the \BtoDD system.

\vspace*{1.5truecm}

\begin{table}[htp]
\begin{center}
\small
\begin{tabular}{|l|r|c|l|} 
\toprule
Var.       & Eq.                & Amplitude ratio                                & Description                                   \\
\midrule	
$a$        &(\ref{a-pen})       &$(P^{(ut)}+PA^{(ut)})/(T$+$E$$+P^{(ct)}$+$PA^{(ct)})$ & Penguin contribution wrt.\ total  ampl.\       \\
$x$        &(\ref{x-def})       &$(E+PA^{(ct)})(T+ P^{(ct)})$                    & Exchange and penguin annihil.\ contr.          \\ 
$r_P$      &(\ref{rP-def})      &$P^{(ct)}/T$                                    & Penguin contribution wrt. tree                \\
$r_A$      &(\ref{rA-def})      &$A/P^{(ut)}$                                    & Annihilation contr.\ (in charged $B$'s)        \\
$r_{PA}$   & \eqref{rPA-def}    & $PA^{(ut)}/P^{(ut)}$                           & Penguin-annihilation contribution             \\
$r_{PA}^{A}$ & \eqref{rPAA}     & $(1+r_A)/(1+r_{PA})$                           & Comparison between $A$ and $PA$               \\
$|a_{NF}|^2$ &(\ref{aNF-def-0}) &$\sim \Gamma(B\to DD')/{\rm d} \Gamma(B\to D\ell\nu)/{\rm d} q^2$ & Non-factorisable effects    \\
\midrule
$\rho$       & \eqref{rhoprime-def}       &                                      & $SU(3)$-breaking in $T$+$P$ contributions     \\
$\varsigma$  & \eqref{E-est}              &                                      & $SU(3)$-breaking in $E$+$PA$ contributions    \\
$\xi,\delta$ & \eqref{a-theta-rel-U-break}&                                      & $SU(3)$-breaking in $a$ and $\theta$          \\
\bottomrule	
\end{tabular}
\caption[Overview]
{Overview of the various amplitude ratios and theoretical quantities.}\label{tab:variables-A}
\end{center}
\end{table}

\begin{table}[htp]
\begin{center}
\small
\begin{tabular}{|l|l|lllll|l|} 
\toprule
 Decay     & Amplitude             &\multicolumn{5}{c|}{Topologies} & Variables                \\
           &                       &$T$&$P$&$E$&$PA$&$A$            &                          \\ 
\midrule
\BdtoDdDd  &$\mathcal{A}          $& x & x & x & x  &               & $a, x , r_P$             \\
\BstoDsDs  &$\mathcal{A}'         $& x & x & x & x  &               & $a',x', r'_P$            \\
\midrule	   	       	    
\BstoDsDd  &$\mathcal{\tilde A}   $& x & x &   &    &               & $\tilde a, \tilde x$     \\
\BdtoDdDs  &$\mathcal{\tilde A}'  $& x & x &   &    &               & $\tilde a', \tilde x'$   \\
\midrule	   	       	    
\BdtoDsDs  &$\mathcal{A}_{EPA}    $&   &   & x & x  &               & $x$                      \\
\BstoDdDd  &$\mathcal{A}'_{EPA}   $&   &   & x & x  &               & $x'$                     \\
\midrule	   	       	    
\ButoDzDd  &$\mathcal{\tilde A}_c $& x & x &   &    & x             & $\tilde a_c, r_A$        \\
\ButoDzDs  &$\mathcal{\tilde A}_c'$& x & x &   &    & x             & $\tilde a_c', r'_A$      \\
\bottomrule	
\end{tabular}
\caption[Overview]
{Overview of the variables relevant to the various decays.}\label{tab:variables-AA}
\end{center}
\end{table}

\begin{table}[tp]
\begin{center}
\small
\begin{tabular}{|l|l|l|l|l|}
\toprule
Decay Ratio            & Observable  & Eq.                      & Value                 & Used for     \\
\midrule 
$B\to DD'/B\to D\ell\nu$& $R_D$      &(\ref{R-def},\ref{Rdtilde-deff})&                 & $|a_{NF}|$   \\
\midrule 
$\BdtoDdDd/\BstoDsDs$  & $H$         & (\ref{Eq:Hobs_Def})      & (\ref{H-value})       &  $a$         \\
$\BstoDsDd/\BdtoDdDs$  & $\tilde H$  & (\ref{Eq:Hobs_Def-tilde})& (\ref{Htilde-value})  &  $\tilde a$  \\
$\ButoDzDd/\ButoDzDs$  & $H_c$       & (\ref{Eq:Hobs_Def-c})    & (\ref{Hc-value})      &  $a_c$       \\
\midrule			            
$\BstoDsDs/\BdtoDdDs$  & $\Xi$       & (\ref{OBS-1})            &  (\ref{XiBsDsDs})     &  $\tilde x'$ \\
$\BdtoDdDd/\BstoDsDd$  & $\Xi$       & (\ref{OBS-2})            &  (\ref{XiBdDdDd})     &  $\tilde x$  \\	  
\midrule			            
$\BstoDdDd/\BdtoDdDs$  & $\Xi$       & (\ref{OBS-3})            &  (\ref{XiBsDD})       &  $\tilde x'$ \\
$\BdtoDsDs/\BstoDsDd$  & $\Xi$       & (\ref{OBS-4})            &  (\ref{XiBdDsDs})     &  $\tilde x$  \\	  
\bottomrule			     
\end{tabular}			     
\caption[]{List of decay ratios and the corresponding observables and variables.}
\end{center}
\end{table}

\clearpage

\begin{thebibliography}{99}
%
%
%
\bibitem{Agashe:2014kda}
  K.~A.~Olive {\it et al.}  [Particle Data Group],
  Chin.\ Phys.\ C {\bf 38} (2014) 090001.

\bibitem{Amhis:2014hma}
  Y.~Amhis {\it et al.}  [Heavy Flavor Averaging Group],
  \href{http://arxiv.org/abs/1412.7515}{arXiv:1412.7515 [hep-ex]}.
  for updates, see  \href{http://www.slac.stanford.edu/xorg/hfag/}{\tt http://www.slac.stanford.edu/xorg/hfag/}.

\bibitem{LHCb-implications}R.~Aaij {\it et al.}  [LHCb Collaboration],
  Eur.\ Phys.\ J.\ C {\bf 73} (2013) 4,  2373
   \href{http://arxiv.org/abs/1208.3355}{[arXiv:1208.3355 [hep-ex]]}.

\bibitem{Abe:2010gxa} 
  T.~Abe {\it et al.}  [Belle II Collaboration],
  \href{http://arxiv.org/abs/1011.0352}{arXiv:1011.0352 [physics.ins-det]}.

\bibitem{RF-psiK}R.~Fleischer,
  Eur.\ Phys.\ J.\ C {\bf 10} (1999) 299
  \href{http://arxiv.org/abs/hep-ph/9903455}{[arXiv:hep-ph/9903455]}.

\bibitem{RF-BDD-07}R.~Fleischer,
  Eur.\ Phys.\ J.\ C {\bf 51} (2007) 849
  \href{http://arxiv.org/abs/0705.4421}{[arXiv:0705.4421 [hep-ph]]}.

\bibitem{GRP}M.~Gronau, J.~L.~Rosner and D.~Pirjol,
  Phys.\ Rev.\ D {\bf 78} (2008) 033011
  \href{http://arxiv.org/abs/0805.4601}{[arXiv:0805.4601 [hep-ph]]}.

\bibitem{RF-ang}R.~Fleischer,
  Phys.\ Rev.\ D {\bf 60} (1999) 073008
  \href{http://arxiv.org/abs/hep-ph/9903540}{[arXiv:hep-ph/9903540]}.
 
\bibitem{RF-B99}R.~Fleischer,
  Nucl.\ Instrum.\ Meth.\ A {\bf 446} (2000) 1
  \href{http://arxiv.org/abs/hep-ph/9908340}{[arXiv:hep-ph/9908340]}.
 
\bibitem{CPS}M.~Ciuchini, M.~Pierini and L.~Silvestrini,
  Phys.\ Rev.\ Lett.\  {\bf 95} (2005) 221804
  \href{http://arxiv.org/abs/hep-ph/0507290}{[arXiv:hep-ph/0507290]};
  \href{http://arxiv.org/abs/1102.0392}{arXiv:1102.0392 [hep-ph]}.
  
\bibitem{FFJM} S.~Faller, R.~Fleischer, M.~Jung, and T.~Mannel,
  Phys.\ Rev.\ D {\bf 79} (2009) 014030
  \href{http://arxiv.org/abs/0809.0842}{[arXiv:0809.0842 [hep-ph]]}.
 
\bibitem{FFM}S.~Faller, R.~Fleischer and T.~Mannel,
  Phys.\ Rev.\ D {\bf 79} (2009) 014005
  \href{http://arxiv.org/abs/0810.4248}{[arXiv:0810.4248 [hep-ph]]}.
  
\bibitem{GR-09}M.~Gronau and J.~L.~Rosner,
  Phys.\ Lett.\ B {\bf 672} (2009) 349
  \href{http://arxiv.org/abs/0812.4796}{[arXiv:0812.4796 [hep-ph]]}.
  
\bibitem{MJ}M.~Jung,
  Phys.\ Rev.\ D {\bf 86} (2012) 053008
  \href{http://arxiv.org/abs/1206.2050}{[arXiv:1206.2050 [hep-ph]]}.
  
\bibitem{LWX}X.~Liu, W.~Wang and Y.~Xie,
  Phys.\ Rev.\ D {\bf 89} (2014) 094010
  \href{http://arxiv.org/abs/1309.0313}{[arXiv:1309.0313 [hep-ph]]}.

\bibitem{DeBF-pen}K.~De Bruyn and R.~Fleischer,
  JHEP {\bf 1503} (2015) 145
  \href{http://arxiv.org/abs/1412.6834}{[arXiv:1412.6834 [hep-ph]]}.

\bibitem{FNW}P.~Frings, U.~Nierste and M.~Wiebusch,
  \href{http://arxiv.org/abs/1503.00859}{arXiv:1503.00859 [hep-ph]}.

\bibitem{Okubo}S.~Okubo,
  Phys.\ Lett.\  {\bf 5} (1963) 165.

\bibitem{Zweig}G.~Zweig,
   CERN-TH-412, NP-8419 (1964).
 
\bibitem{Iizuka}J.~Iizuka,
  Prog.\ Theor.\ Phys.\ Suppl.\  {\bf 37} (1966) 21.

\bibitem{RF-pen}R.~Fleischer,
  Int.\ J.\ Mod.\ Phys.\ A {\bf 12} (1997) 2459
  \href{http://arxiv.org/abs/hep-ph/9612446}{[arXiv:hep-ph/9612446]}.

\bibitem{FM-NP}R.~Fleischer and T.~Mannel,
  Phys.\ Lett.\ B {\bf 506} (2001) 311
  \href{http://arxiv.org/abs/hep-ph/0101276}{[arXiv:hep-ph/0101276]}.

\bibitem{BFRS}A.~J.~Buras, R.~Fleischer, S.~Recksiegel and F.~Schwab,
  Phys.\ Rev.\ Lett.\  {\bf 92} (2004) 101804
  \href{http://arxiv.org/abs/hep-ph/0312259}{[arXiv:hep-ph/0312259]}.
  
\bibitem{BFRS_II}A.~J.~Buras, R.~Fleischer, S.~Recksiegel and F.~Schwab,
  Nucl.\ Phys.\ B {\bf 697} (2004) 133
  \href{http://arxiv.org/abs/hep-ph/0402112}{[arXiv:hep-ph/0402112]}.

\bibitem{FJPZ}R.~Fleischer, S.~J\"ager, D.~Pirjol and J.~Zupan,
  Phys.\ Rev.\ D {\bf 78} (2008) 111501
  \href{http://arxiv.org/abs/0806.2900}{[arXiv:0806.2900 [hep-ph]]}.

\bibitem{Zp}V.~Barger, L.~Everett, J.~Jiang, P.~Langacker, T.~Liu and C.~Wagner,
  Phys.\ Rev.\ D {\bf 80} (2009) 055008
  \href{http://arxiv.org/abs/0902.4507}{[arXiv:0902.4507 [hep-ph]]}.

\bibitem{Zp_II}V.~Barger, L.~L.~Everett, J.~Jiang, P.~Langacker, T.~Liu and C.~E.~M.~Wagner,
  JHEP {\bf 0912} (2009) 048
  \href{http://arxiv.org/abs/0906.3745}{[arXiv:0906.3745 [hep-ph]]}.
  
\bibitem{cab}N.~Cabibbo,
  Phys.\ Rev.\ Lett.\  {\bf 10} (1963) 531.

\bibitem{KM}M.~Kobayashi and T.~Maskawa,
  Prog.\ Theor.\ Phys.\  {\bf 49} (1973) 652.

\bibitem{GHLR}M.~Gronau, O.~F.~Hernandez, D.~London and J.~L.~Rosner,
  Phys.\ Rev.\ D {\bf 52} (1995) 6356
  \href{http://arxiv.org/abs/hep-ph/9504326}{[arXiv:hep-ph/9504326]}.

\bibitem{wolf}L.~Wolfenstein,
  Phys.\ Rev.\ Lett.\  {\bf 51} (1983) 1945.

\bibitem{Charles:2015gya} 
  J.~Charles, O.~Deschamps, S.~Descotes-Genon, H.~Lacker, A.~Menzel, S.~Monteil, V.~Niess and J.~Ocariz {\it et al.},
  \href{http://arxiv.org/abs/1501.05013}{arXiv:1501.05013 [hep-ph]};
  for updates, see  \href{http://ckmfitter.in2p3.fr}{\tt http://ckmfitter.in2p3.fr}.

\bibitem{RF-rev}R.~Fleischer,
  Phys.\ Rept.\  {\bf 370} (2002) 537
  \href{http://arxiv.org/abs/hep-ph/0207108}{[arXiv:hep-ph/0207108]}.

\bibitem{BSS}M.~Bander, D.~Silverman and A.~Soni,
  Phys.\ Rev.\ Lett.\  {\bf 43} (1979) 242.

\bibitem{RF-NLO}R.~Fleischer,
  Z.\ Phys.\ C {\bf 58} (1993) 483.

\bibitem{RF-NLO-II}R.~Fleischer,
  Z.\ Phys.\ C {\bf 62} (1994) 81.

\bibitem{BFM}A.~J.~Buras, R.~Fleischer and T.~Mannel,
  Nucl.\ Phys.\ B {\bf 533} (1998) 3
  \href{http://arxiv.org/abs/hep-ph/9711262}{[arXiv:hep-ph/9711262]}.

\bibitem{DFN}I.~Dunietz, R.~Fleischer and U.~Nierste,
  Phys.\ Rev.\ D {\bf 63} (2001) 114015
  \href{http://arxiv.org/abs/hep-ph/0012219}{[arXiv:hep-ph/0012219]}.

\bibitem{BR-paper}K.~De Bruyn, R.~Fleischer, R.~Knegjens, P.~Koppenburg, M.~Merk and N.~Tuning,
  Phys.\ Rev.\ D {\bf 86} (2012) 014027
  \href{http://arxiv.org/abs/1204.1735}{[arXiv:1204.1735 [hep-ph]]}.

\bibitem{Aaij:2013bvd} 
  R.~Aaij {\it et al.}  [LHCb Collaboration],
  Phys.\ Rev.\ Lett.\  {\bf 112} (2014) 11, 111802
  \href{http://arxiv.org/abs/1312.1217}{[arXiv:1312.1217 [hep-ex]]}.

\bibitem{Fleischer:1997um}
  R.~Fleischer and T.~Mannel,
  Phys.\ Rev.\ D {\bf 57} (1998) 2752
  \href{http://arxiv.org/abs/hep-ph/9704423}{[arXiv:hep-ph/9704423]}.

\bibitem{Fleischer:2004rn}
  R.~Fleischer and S.~Recksiegel,
  Phys.\ Rev.\ D {\bf 71} (2005) 051501
  \href{http://arxiv.org/abs/hep-ph/0409137}{[arXiv:hep-ph/0409137]}.

\bibitem{bjorken}J.~D.~Bjorken,
  Nucl.\ Phys.\ Proc.\ Suppl.\  {\bf 11} (1989) 325.
  
\bibitem{BS}D.~Bortoletto and S.~Stone,
  Phys.\ Rev.\ Lett.\  {\bf 65} (1990) 2951.
  
\bibitem{rosner}J.~L.~Rosner,
  Phys.\ Rev.\ D {\bf 42} (1990) 3732.

\bibitem{NS}M.~Neubert and B.~Stech,
  Adv.\ Ser.\ Direct.\ High Energy Phys.\  {\bf 15} (1998) 294
  \href{http://arxiv.org/abs/hep-ph/9705292}{[arXiv:hep-ph/9705292]}.

\bibitem{BBNS}M.~Beneke, G.~Buchalla, M.~Neubert and C.~T.~Sachrajda,
  Nucl.\ Phys.\ B {\bf 591} (2000) 313
  \href{http://arxiv.org/abs/hep-ph/0006124}{[arXiv:hep-ph/0006124]}.

\bibitem{FST-fact}R.~Fleischer, N.~Serra and N.~Tuning,
  Phys.\ Rev.\ D {\bf 83} (2011) 014017
  \href{http://arxiv.org/abs/1012.2784}{[arXiv:1012.2784 [hep-ph]]}.
      
\bibitem{Buras}A.~J.~Buras,
  Nucl.\ Phys.\ B {\bf 434} (1995) 606
  \href{http://arxiv.org/abs/hep-ph/9409309}{[arXiv:hep-ph/9409309]}.
  
\bibitem{FST}R.~Fleischer, N.~Serra and N.~Tuning,
  Phys.\ Rev.\ D {\bf 82} (2010) 034038
  \href{http://arxiv.org/abs/1004.3982}{[arXiv:1004.3982 [hep-ph]]}.
 
\bibitem{RF-BDD-93}R.~Fleischer,
  Nucl.\ Phys.\ B {\bf 412} (1994) 201.
  
\bibitem{RoSt}J.~L.~Rosner and S.~Stone,
  \href{http://arxiv.org/abs/1309.1924}{arXiv:1309.1924 [hep-ex]}.
    
\bibitem{Aoki:2013ldr}
  S.~Aoki {\it et al.} [Flavour Lattice Averaging Group], 
  Eur.\ Phys.\ J.\ C {\bf 74} (2014) 9,  2890
  \href{http://arxiv.org/abs/1310.8555}{[arXiv:1310.8555 [hep-lat]]}.
   
\bibitem{NeRi}M.~Neubert and V.~Rieckert,
  Nucl.\ Phys.\ B {\bf 382} (1992) 97.
  
\bibitem{Neubert}M.~Neubert,
  Nucl.\ Phys.\ B {\bf 371} (1992) 149.
  
\bibitem{BS-model}A.~Abd El-Hady, A.~Datta, K.~S.~Gupta and J.~P.~Vary,
  Phys.\ Rev.\ D {\bf 55} (1997) 6780
  \href{http://arxiv.org/abs/hep-ph/9605397}{[arXiv:hep-ph/9605397]}.
  
\bibitem{lat-1}J.A. Bailey {\it et al.}  [Fermilab Lattice and MILC Collaborations],
  \href{http://arxiv.org/abs/1503.07237}{arXiv:1503.07237 [hep-lat]}.
  
\bibitem{lat-2}H.~Na, C.~M.~Bouchard, G.~P.~Lepage, C.~Monahan and J.~Shigemitsu,
  \href{http://arxiv.org/abs/1505.03925}{arXiv:1505.03925 [hep-lat]}.
  
\bibitem{LHCb-Vub}
  R.~Aaij {\it et al.}  [LHCb Collaboration],
  \href{http://arxiv.org/abs/1504.01568}{arXiv:1504.01568 [hep-ex]}, submitted to Nature Physics.

\bibitem{Bevan:2014cya}
  A.~Bevan, {\it et al.} [UTfit Collaboration],
  \href{http://arxiv.org/abs/1411.7233}{arXiv:1411.7233 [hep-ph]}; for updates, see
  \href{http://www.utfit.org}{\tt http://www.utfit.org}.

\bibitem{LHCB-BDD-2013}R. Aaij {\it et al.} [LHCb Collaboration],
  Phys.\ Rev.\ D {\bf 87} (2013) 9, 092007
  \href{http://arxiv.org/abs/1302.5854}{[arXiv:1302.5854 [hep-ph]]}.

\bibitem{HMChiPT1}E.~Jenkins and M.J.~Savage,
  Phys.\ Lett.\ B {\bf 281} (1992) 331.
 
\bibitem{HMChiPT2}B.~Grinstein, E.~E.~Jenkins, A.~V.~Manohar, M.~J.~Savage and M.~B.~Wise,
  Nucl.\ Phys.\ B {\bf 380} (1992) 369
  \href{http://arxiv.org/abs/hep-ph/9204207}{[arXiv:hep-ph/9204207]}.

\bibitem{El-Khadra:2014sha}A.~X.~El-Khadra,
  PoS LATTICE {\bf 2013} (2014) 001
  \href{http://arxiv.org/abs/1302.5854}{[arXiv:1403.5252 [hep-lat]]}.

\bibitem{Bailey:2012rr}J.~A.~Bailey {\it et al.}, [Fermilab Lattice and MILC Collaborations]
  Phys.\ Rev.\ D {\bf 85} (2012) 114502
  [Erratum-ibid.\ D {\bf 86} (2012) 039904]
  \href{http://arxiv.org/abs/1302.5854}{[arXiv:1202.6346 [hep-lat]]}.

\bibitem{GHLR-94}M.~Gronau, O.~F.~Hernandez, D.~London and J.~L.~Rosner,
  Phys.\ Rev.\ D {\bf 50} (1994) 4529
  \href{http://arxiv.org/abs/hep-ph/9404283}{[arXiv:hep-ph/9404283]}.

\bibitem{BaBar-BDD} 
  B.~Aubert {\it et al.}  [BaBar Collaboration],
  Phys.\ Rev.\ D {\bf 79} (2009) 032002
  \href{http://arxiv.org/abs/0808.1866}{[arXiv:0808.1866 [hep-ex]]}.

\bibitem{Belle-BDD} 
  M.~R\"{o}hrken {\it et al.}  [Belle Collaboration],
  Phys.\ Rev.\ D {\bf 85} (2012) 091106
  \href{http://arxiv.org/abs/1203.6647}{[arXiv:1203.6647 [hep-ex]]}.

\bibitem{LHCb-BsDsDs} 
  R.~Aaij {\it et al.}  [LHCb Collaboration],
  Phys.\ Rev.\ Lett.\  {\bf 113} (2014) 21, 211801
  \href{http://arxiv.org/abs/1409.4619}{[arXiv:1409.4619 [hep-ex]]}.

\bibitem{Jung:2014jfa} 
  M.~Jung and S.~Schacht,
  Phys.\ Rev.\ D {\bf 91} (2015) 3, 034027
  \href{http://arxiv.org/abs/1410.8396}{[arXiv:1410.8396 [hep-ph]]}.

\bibitem{Belle-observation} 
  G.~Majumder {\it et al.}  [Belle Collaboration],
  Phys.\ Rev.\ Lett. {\bf 95} (2005) 041803
  \href{http://arxiv.org/abs/hep-ex/0502038}{[arXiv:hep-ex/0502038]}.

\bibitem{Belle-CPV-BDD-2008}I.~Adachi {\it et al.} [Belle Collaboration],
  Phys.\ Rev.\ D\  {\bf 77} (2008) 091101
  \href{http://arxiv.org/abs/0802.2988}{[arXiv:0802.2988 [hep-ph]]}.

\bibitem{Zupanc-Belle-2007}A. Zupanc {\it et al.} [Belle Collaboration],
  Phys.\ Rev.\ D\  {\bf 75} (2007) 091102
  \href{http://arxiv.org/abs/hep-ex/0703040}{[arXiv:hep-ex/0703040]}.

\bibitem{Huang-BESIII-2012}G.~Huang [BESIII Collaboration],
  \href{http://arxiv.org/abs/1209.4813}{arXiv:1209.4813 [hep-ph]}.

\bibitem{BESIII-2008} D.~M.~Asner {\it et al.} [BESIII Collaboration], 
  Int.\ J.\ Mod.\ Phys.\ A {\bf 24} (2009) S1
  \href{http://arxiv.org/abs/0809.1869}{[arXiv:0809.1869 [hep-ph]]}.

\bibitem{Zupanc-Belle-2013}A. Zupanc {\it et al.} [Belle Collaboration],
  JHEP {\bf 1309} (2013) 139
  \href{http://arxiv.org/abs/1307.6240}{[arXiv:1307.6240 [hep-ph]]}.

\bibitem{DF}R.~Fleischer and I.~Dunietz,
  Phys.\ Rev.\ D {\bf 55} (1997) 259
   \href{http://arxiv.org/abs/hep-ph/9605220}{[arXiv:hep-ph/9605220]}.

\bibitem{DDF}A.~S.~Dighe, I.~Dunietz and R.~Fleischer,
  Eur.\ Phys.\ J.\ C {\bf 6} (1999) 647
   \href{http://arxiv.org/abs/hep-ph/9804253}{[arXiv:hep-ph/9804253]}.

%
%
%
\end{thebibliography}
\end{document}